\documentclass{aa}

\newcommand{\cd}{$\mathrm{d}^{-1}$}

\usepackage{hyperref}
\usepackage{graphicx}
\usepackage{longtable}
\usepackage{multirow}
\usepackage{color}
\usepackage{soul}
\usepackage{txfonts}
\begin{document}

   \title{Pulsational instability of pre-main sequence models from accreting protostars}

   \subtitle{I. Constraining the input physics for accretion with spectroscopic parameters and stellar pulsations}

   \author{T. Steindl\inst{1}, K. Zwintz\inst{1}, T. G. Barnes\inst{2},  M. Müllner
          \inst{1}, E. I. Vorobyov\inst{3, 4}}
    
    \authorrunning{T. Steindl et al.}
    \titlerunning{Pulsational instability of pre-main sequence models from accreting protostars}
    
   \institute{\inst{1}Institut f\"ur Astro- und Teilchenphysik, Universit\"at Innsbruck, Technikerstra{\ss}e 25, A-6020 Innsbruck, Austria\\
              \email{thomas.steindl@uibk.ac.at}  \\
              \inst{2}The University of Texas at Austin, McDonald Observatory, 2515 Speedway, Stop C1402, Austin, TX 78712-1206, USA\\
              \inst{3} Department of Astrophysics, University of Vienna, Vienna 1180, Austria \\
              \inst{4} Research Institute of Physics, Southern Federal University, Rostov-on-Don 344090, Russia
              }

   \date{Received -; accepted -}

% \abstract{}{}{}{}{}      
% 5 {} token are mandatory
 
  \abstract
   {The pre-main sequence evolution is often simplified by choosing classical initial models. These have large initial radii and sufficient uniform contraction to make them fully convective. Contrary to that, real stars are born as small protostellar seeds in collapsing molecular clouds and obtain their final mass by means of accretion.}
   {We aim to constrain the input physics of accretion on protostellar seeds with observed spectroscopic parameters and stellar pulsations of young stellar objects and pre-main sequence stars.}
   {We conducted a literature search for spectroscopic samples of young stellar objects and pre-main sequence stars including all previously known pulsators. The sample size of pulsating pre-main sequence stars is increased by analysing TESS observations and presenting additional discoveries in CoRoT data. We employ  \texttt{MESA} and \texttt{GYRE} to calculate evolutionary tracks of accreting protostellar seeds in a constant accretion scenario, the subsequent pre-main sequence evolution, and their pulsation properties. The results are then compared with observations to constrain the input physics.}
   {We discuss 16 formerly unknown pulsating pre-main sequence stars and candidates that are either of SPB, $\delta$ Scuti,$\gamma$ Doradus, or $\delta$ Scuti - $\gamma$ Doradus hybrid type. We find that evolutionary tracks with a mass accretion rate of $5\times 10^{-6}\,M_\odot/{\rm yr}$ and fraction of injected accretion energy of $\beta = 0.1$ provide the best results in enveloping the spectroscopic parameters of pre-main sequence stars in a constant accretion scenario. The calculated instability regions constrain the atmospheric boundary conditions to Eddington Gray atmospheres; we discuss the future potential for additional constraints by instability regions that are dependent on radial order. Finally, we present a possible candidate for pulsations in M-type young stellar objects.}
   {We show that evolutionary calculations of accreting protostellar seeds match the observed spectroscopic parameters of pre-main sequence stars. Future observations that will allow the identification of radial orders in particular will present opportunities for additional constraints.}


   \maketitle
%
%-------------------------------------------------------------------

\section{Introduction}
Stellar structure and evolution is a colossal topic, both observationally and theoretically. Owing to the vastly different evolutionary stages of stars, it is increasingly difficult to retain a unified picture. Despite being a vital building block of stellar evolution, the pre-main sequence evolution often falls victim to this circumstance. Stars do not just pop up on the main sequence but are created as small stellar seeds \citep[or second hydrostatic cores, see][]{Larson1969, Masunaga2000} from the collapse of molecular clouds where the formation of stars is dependent on several parameters like metallicity \citep{Bate2019}. These seeds then continue to accrete material from their surroundings, most commonly from their embedded disks. Again, this process is dependent on the parameters of the surrounding material, leading to varying accretion rates for otherwise similar stellar seeds \citep{Vorobyov2010, Vorobyov2015, Vorobyov2020}.

Given that the physics governing early stellar evolution is much different, it is astonishing that  the discussion of the pre-main sequence in the non-specialised community often has not much changed since the pioneering works of \citet{Henyey1955}, \citet{Hayashi1961}, and \citet{Iben1965}. While the formation of stars from accreting protostellar seeds in one dimensional stellar models has been an ongoing area of research for many years \citep{Mercer1984, Palla1991, Hartmann1996, Baraffe2009, Hosokawa2011, Vorobyov2017, Kunitomo2017}, it seems that this insight is still far from reaching the entirety of the asteroseismic community. This series of papers aims to close the gap between these two areas of research by combining state-of-the-art one-dimensional pre-main sequence models with asteroseismology. 

Decades have passed since the first pre-main sequence pulsators have been discovered by \citet{Breger1972}. Today, it has become clear that Slowly Pulsating B-stars (SPBs), $\delta$ Scuti variables, and $\gamma$ Doradus stars also exist in the pre-main sequence phase of stellar evolution \cite[see e.g.][]{Ripepi2011, Gruber2012, Zwintz2013, Zwintz2014}. As such, both gravity modes (g-modes) and pressure modes (p-modes) are common in stars even before the main sequence. The existence of solar-like oscillations in pre-main sequence stars is theoretically predicted \citep{Samadi2005, pinheiro2008} and a first candidate has recently been suggested by \citet{muellner2020b}.  In addition, bursts in accretion rate might lead to short-lived radial oscillations \citep{Bastien2011}.

The advent of space telescopes like CoRoT \citep{Auvergne2009}, Kepler \citep{borucki2010, koch2010}, and TESS \citep{Ricker2015} and their stunning accuracy led to a multitude of discoveries and created the possibility for in-depth modelling of pulsation spectra \citep[e.g.][]{Aerts2018, Michielsen2019, Li2020}. Unfortunately, the main Kepler mission pointed away from star forming regions \citep{Zwintz2017b} and the latter are often too crowded for the pixel size of TESS. This explains why discoveries for pre-main sequence stars are comparable scarce, with a few bright examples \citep[e.g.][]{Steindl2020, Biddle2020, vanDam2020}. The situation will not much differ for the future space mission PLATO \citep{Rauer2018} calling in the need for a specialised mission. 

In this series of papers, we aim to investigate the influence of accretion on the pulsational properties of pre-main sequence stars. The work at hand concentrates on constraining the input physics for a constant accretion scenario by first enveloping a sample of pre-main sequence stars with known spectroscopic surface gravities and effective temperatures between the accreting evolutionary track and the zero age main sequence (ZAMS). This is especially desirable as luminosities of such objects in the main accretion phase are split into the star's produced radiation and the radiation of the accreted material's energy. The latter is both, hard to measure observationally and a direct consequence of the choice of input physics in one-dimensional stellar evolution models. 

Secondly, we calculate non-adiabatic pulsation models to obtain growth rates and, hence, pre-main sequence pulsation instability regions. The instability strip for the first three radial modes of A- and F-type pre-main sequence stars was first calculated by \citet{Marconi1998}. Instability regions for these stars were later also calculated by \citet{Grigahcene2006} and \citet{Bouabid2011}. The first discussion of the observational instability region for pre-main sequence $\delta$ Scuti stars was provided by \citet{zwintz2008}. \citet{Gruber2012} provide theoretical instability regions for pre-main sequence SPB stars. By comparing the resulting pre-main sequence instability regions with observed pulsating variables, we can further constrain the input physics for a constant accretion model. The second paper in this series will then investigate the influence of accretion on observed frequencies and compare them to expected frequency uncertainties for light curves obtained with different space telescopes. Hence, this aims to investigate the probing power of pre-main sequence pulsators for the accretion history of the protostar.

This paper is structured as follows: We discuss our sample of pre-main sequence and early ZAMS stars, including previously known pre-main sequence pulsators, in Section \ref{sec:sample}. Sections \ref{sec:Tess} and \ref{sec:Corot} present new pre-main sequence pulsators and candidates found in TESS and CoRoT data to expand upon the literature sample. We describe our stellar evolution models and pulsation analysis in Section \ref{sec:models} and present our results in Section \ref{sec:results}. Finally, we give concluding remarks in Section \ref{sec:conclusions}.

%--------------------------------------------------------------------
\section{Stellar Sample}
\label{sec:sample}
For our analysis, we searched the available literature for spectroscopic surveys of young clusters and star forming regions. Here, we describe the surveys used in this work and our selection process.

\subsection{Pre-main sequence and early ZAMS stars}
\label{sec:pre-ms-young}
\citet{Aidelman2018} obtained low resolution spectra of $68$ B-type stars in the four open clusters NGC 6087, NGC6250, NGC 6383, and NGC 6530. The respective ages are ${\sim} 55$ , ${\sim} 6$, ${\sim} 3{-}10$, and ${\sim} 4{-}7$~Myr \citep[see][for more detail and previous age estimates]{Aidelman2018}. Owing to the fast evolution towards the ZAMS of such intermediate mass stars, we omit NGC 6087 due to its age of ${\sim} 55$~Myr. The remaining $50$ stars are added to our sample of pre-main sequence and early ZAMS stars.

\citet{Alecian2013} performed a high resolution spectropolametric survey of Herbig AeBe stars. The characteristics of Herbig AeBe stars suggests that they remain surrounded by dust and gas \citep{Alecian2013} and are often still in their pre-main sequence phase of evolution. Hence, we add all stars, except the binaries, of this study to our sample. These are in total $65$ stars.

Similarly, \cite{Fairlamb2015} conducted a spectroscopic study of Herbig AeBe stars. We exclude stars that the authors could not model successfully with magnetospheric accretion. Thus, we add $81$ objects to our sample. The sample from \citet{Fairlamb2015} and \citet{Alecian2013} overlap, that is multiple stars are present in both surveys. In such a case, we use the stellar parameters provided in both studies as the measurements are obtained independently from each other.

\citet{frasca2017} obtained spectroscopy of young stellar objects in the ${\sim} 2$~Myr old Lupus star forming region. Their sample contains $104$ young stellar objects, $89$ of which are listed as members of Lupus. We therefore add these $89$ stars to our sample of pre-main sequence and early ZAMS stars. 

\citet{gutierrrez2020} derived cluster ages and memberships of open clusters as part of the Gaia-ESO survey. Of the $20$ clusters discussed in this work, three are young open clusters: $\rho$~Oph, Cha I, and $\gamma$~Vel with ages of $1{-}3$, $2$, and $10{-}20$~Myr, respectively. A total of $45$ stars are classified as members of one of these three young open clusters and we therefore add them to our sample.

\citet{Da2016} reported the results of infrared spectroscopy for pre-main sequence stars in the Orion A molecular cloud. They obtain $2700$ high resolution spectra and discover $383$ new sources. While the uncertainties provided in this study are mostly very small, some stars show very high uncertainties in surface gravity. Hence, we add 1859 stars to our sample that are members of Orion A and have an uncertainty in surface gravity smaller than $0.5$. 

\citet{Vioque2020} used a machine learning approach to find pre-main sequence stars. We queried the TESS Input Catalogue (TIC) \citep{Stassun2018} using the sample by \citet{Vioque2020} for entries that include values for the surface gravity and effective temperature. A total of $705$ stars have an entry for both which we added to our sample. 

Although the last two samples of \citet{Da2016} and \citet{Vioque2020} provide spectroscopic surface gravities and effective temperatures, we assign the other surveys higher weights. For \citet{Da2016}, the reasoning is mainly the different wavelength at which the spectra were obtained and the partially very small uncertainty (only a few Kelvin for a large subset of stars) that we interpret as formal and not astrophysical errors. For \citet{Vioque2020}, the parameters come from different sources and partly no uncertainties are available. We include both samples in our discussion with smaller weights and focus more on the stars selected from the other samples.

\subsection{Known pulsating pre-main sequence stars}
\label{sec:puls-pre-ms-young}
The pulsating pre-main sequence stars and candidates used in this work include SPB, $\delta$~Scuti, $\gamma$~Doradus, and  $\delta$~Scuti - $\gamma$~Doradus hybrids from different studies. In addition, we looked for pulsators among the stars in our sample of pre-main sequence and early ZAMS stars in the TESS Data (see Section \ref{sec:Tess}) and discuss previously unpublished pulsators found with CoRoT (see Section \ref{sec:Corot}). 

\subsubsection{Slowly pulsating B stars}
\citet{Gruber2012} reported the cases of GSC 00154-00785 and GSC 00154-01871 which are two probable members of the young open cluster NGC 2244. \citet{Zwintz2017} performed a photometric study of young B stars in the young open cluster NGC 2264, resulting in the discovery of four early ZAMS SPB stars. Ten new slowly pulsating B stars were discovered by \citet{Fedurco2020}. Two of them, HD 61712 and HD 36999 are potential pre-main sequence SPB candidates. \citet{Ketzer2020} reported the case of 2MASS J16260931-2434121, a member of the $\rho${-}Ophiuchus star{-}forming region embedded in the Upper Scorpius region. Given the young age of $\rho${-}Ophiuchus, 2MASS J16260931-2434121 is a prime candidate for a pre-main sequence SPB star. \citet{Labadie-Bartz2020} report g-mode frequency groups for HD 114981, a star in the sample by \citet{Alecian2013}, which makes it another potential pre-main sequence SPB star. In total, our literature sample includes ten pre-main sequence or early ZAMS SPB stars and candidates.

\subsubsection{$\delta$~Scuti stars}
\cite{Zwintz2014} showed the existence of a relationship between pulsation properties and evolutionary stage for pre-main sequence $\delta$~Scuti stars. Their work includes $34$ $\delta$~Scuti stars in their pre-main sequence to early ZAMS evolutionary phases. Additionally, \citet{Mellon2019} reported the discovery of $\delta$~Scuti-type pulsation in HD 156623. \citet{Forteza2020} proposed a scaling relation for $\delta$~Scuti stars. Their sample includes HD 287841 and HD 290764, two stars which are Herbig AeBe stars according to \citet{Fairlamb2015}. Recently, \citet{murphy2021} reported the case of HD 139614, a pre-main sequence candidate $\delta$~Scuti star. \cite{Murphy2020a} reported p-mode oscillations in HD 68695 and classify it as a $\lambda$ Bootis star. HD 68695 is included in the sample by \citet{Fairlamb2015} and added to our sample as pre-main sequence $\delta$~Scuti candidate. Our literature sample of pre-main sequence $\delta$~Scuti stars comprises of in total 39 stars.

\subsubsection{$\gamma$ Doradus stars and $\delta$~Scuti $\gamma$ Doradus hybrids}
Only two pre-main sequence $\gamma$ Doradus stars with known spectroscopic parameters are known from the literature. These are Cl$^{\star}$ NGC 2264 VAS 20 and  Cl$^{\star}$ NGC 2264 VAS 87 discovered in CoRoT data by \citet{Zwintz2013}. \cite{Ripepi2011} reported the only known $\delta$~Scuti $\gamma$ Doradus hybrid, CoRoT 102699796.

\subsection{Literature sample}
The literature sample for this study is shown in Fig. \ref{fig:sample}. Figure \ref{fig:sample} furthermore includes newly discovered pre-main sequence pulsators and candidates which we will discuss in Sections \ref{sec:Tess} and \ref{sec:Corot}.  Typical uncertainties have been calculated by taking the median plus one time the standard deviation of the uncertainties for all stars from a given source\footnote{Most of the stars in the study by \citet{Alecian2013} do not have uncertainties for the surface gravity. In such a case, we adopted an uncertainty of $0.5$~dex. The typical error for these stars does not include the added standard deviation for the values of surface gravities.}. 
%--------------------------------------------------------------------
\section{TESS observations}
\label{sec:Tess}
\subsection{The TESS Mission}
The TESS satellite \citep{Ricker2015} was launched on 2018, April 18, from Cape Canaveral on a SpaceX Falcon 9 Rocket, into a never-before-used lunar-resonant orbit known as P/2 \citep{Gangestad2013}. TESS carries four cameras,  each with a 24$\times$24 degree field of view used to survey almost the entire sky within two years in the primary mission. 
The research goal of the MIT-led NASA TESS mission is the discovery of transiting exoplanets. However, the high quality light curves provide much more research opportunities including the field of asteroseismology.
More information about the TESS mission can be found, for example, in \citet{Ricker2015}.

\subsection{Data reduction and analysis}
Multiple stars in our sample of pre-main sequence and early ZAMS stars have been observed by the TESS mission. For most of these stars, no short cadence data is available. In such a case we use the python package \texttt{lightkurve} \citep{lightkurve2018} to obtain the Full Frame Images (FFIs) and extract the TESS long cadence light curve. The latter is either a corrected light curve, where the scattered light is removed with the \texttt{RegressionCorrector}\footnote{see the tutorials of \texttt{lightkurve} at \url{https://docs.lightkurve.org}} or a background corrected simple aperture photometry (BCSAP) if the corrections do not work as anticipated. For the BCSAP, we use the flux in the aperture minus $n$ times the flux of a low intensity pixel, where $n$ is the number of pixels in the aperture. For all stars, both light curves are produced. We choose the corrected light curve if it resembles the BCSAP light curve with less scatter and the BCSAP light curve otherwise. We remove parts of the light curves if the corrections are unable to properly remove signal introduced by scattered light of the earth or moon. The resulting light curves are converted to magnitudes. We then used the python package \texttt{smurfs} \citep{mullner2020} to perform a Fourier analysis in an iterative process. At the start of every iteration, it calculates a Lomb-Scargle periodogram \citep{lomb1976, scargle1982} to detect the frequency and the amplitude of the highest peak. \texttt{smurfs} uses these and a phase of $0.5$ as initial values to perform a single sine fit to the residual light curve. The corresponding sine is subtracted from the residuals for the future iterations. We consider frequencies to be significant if their amplitudes exceed four times local noise level, e.g. SNR > 4 \citep{breger1993,kuschnig1997} and this procedure is continued until five consecutive frequencies are insignificant. With the exception of the data downlink gap at the  beginning of each orbit, TESS light curves are very close to evenly spaced time series. This results in an almost vanishing spectral window (see Fig. \ref{fig:dscuti_lcs_inc_sw}) that allows to ignore possible alias frequencies.

Stars often appear in multiple systems such as binaries. Some of the stars discussed in the following are known binary stars, while other stars might be multiple systems without our knowledge. Depending on stellar and orbital parameters, binarity can influence the frequencies extracted from Fourier analysis. This can range from the addition of frequency peaks at multiples of the orbital frequencies (eclipsing binaries, heartbeat stars, etc.) to the splitting of self driven pulsation modes into multiplets by different effects \citep{Steindl2020}. The first can in some cases be misinterpreted as signal of rotational variability, but is clearly distinguishable from a pulsation signal, in which the frequencies are not multiples of the orbital frequencies. Hence, the influence of binarity onto the newly discovered pre-main sequence pulsators presented in the following mostly affects the identification of the primary or secondary as the origin of the pulsation signal and has minor influence on the extracted frequencies an the detection of pulsation itself.

We use TESS FFIs, TIC, and GAIA \citep{gaia2016} data to clarify the source of the variability. In total, six new SPB, six new $\delta$~Scuti, and one new $\gamma$~Doradus pulsators were identified in the TESS data of pre-main sequence and early ZAMS stars which we add to our sample. Their stellar properties are listed in Table \ref{tab:newpulstable}.

\begin{figure}
   \centering
   \includegraphics[width=\linewidth]{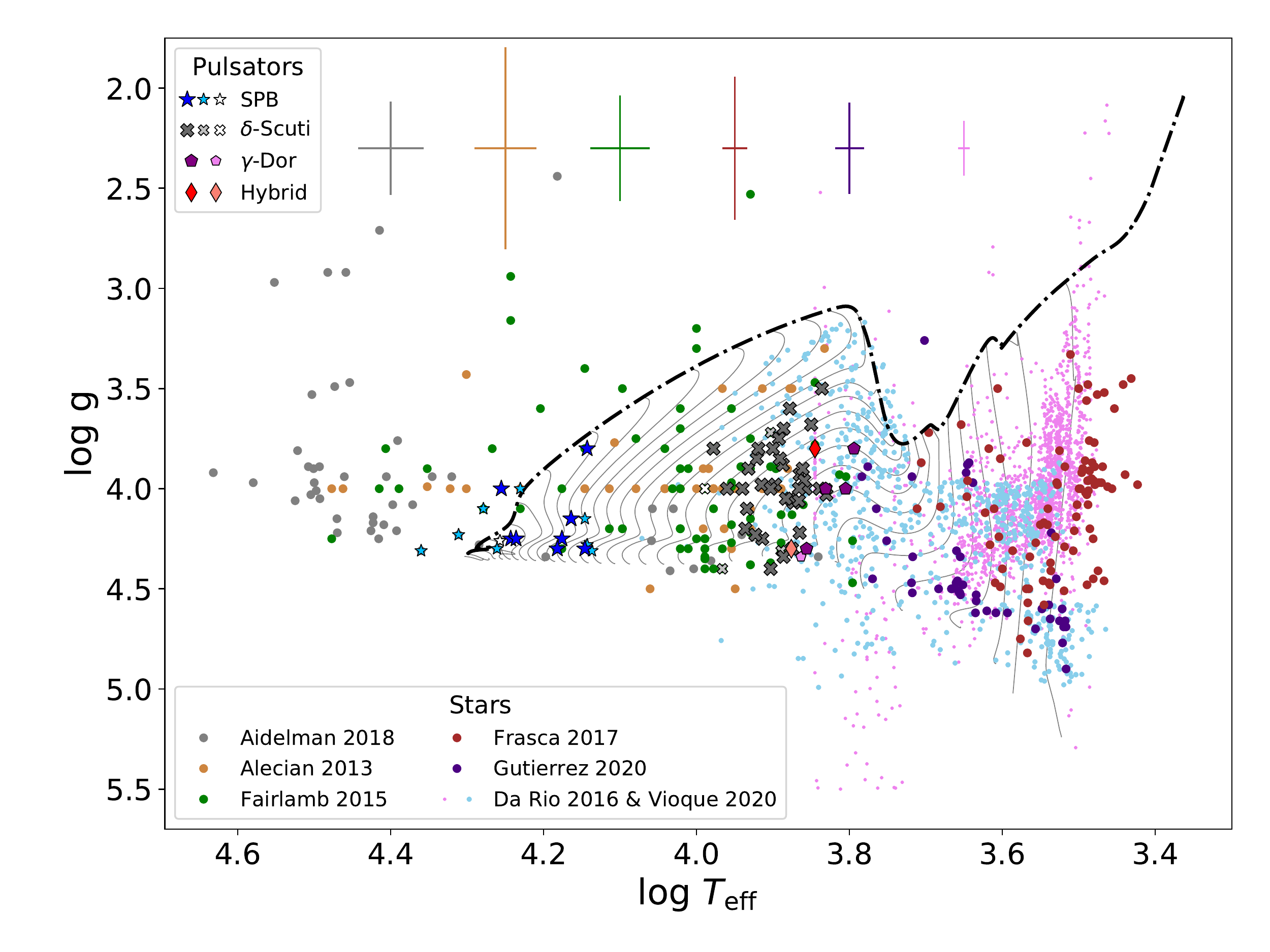}
      \caption{Stellar sample of pre-main sequence and early ZAMS stars and pulsators, including the newly discovered pulsators. Circles of different colours correspond to the different samples of stars discussed in Section \ref{sec:pre-ms-young}. Different symbols correspond to different pulsator classes described in Section \ref{sec:puls-pre-ms-young} and Sections \ref{sec:Tess} and \ref{sec:Corot}. Smaller and lighter coloured symbols correspond to candidates. The coloured error bars in the top give the typical uncertainty of the position of the stars from different sources. The black dash-dotted line illustrates an accreting evolutionary track as described in Section \ref{sec:results} and the thin black lines show the subsequent pre-main sequence tracks evolved from the latter.
              }
         \label{fig:sample}
\end{figure}

\subsection{Slowly Pulsating B stars}
\begin{figure}
   \centering
   \includegraphics[width=\linewidth]{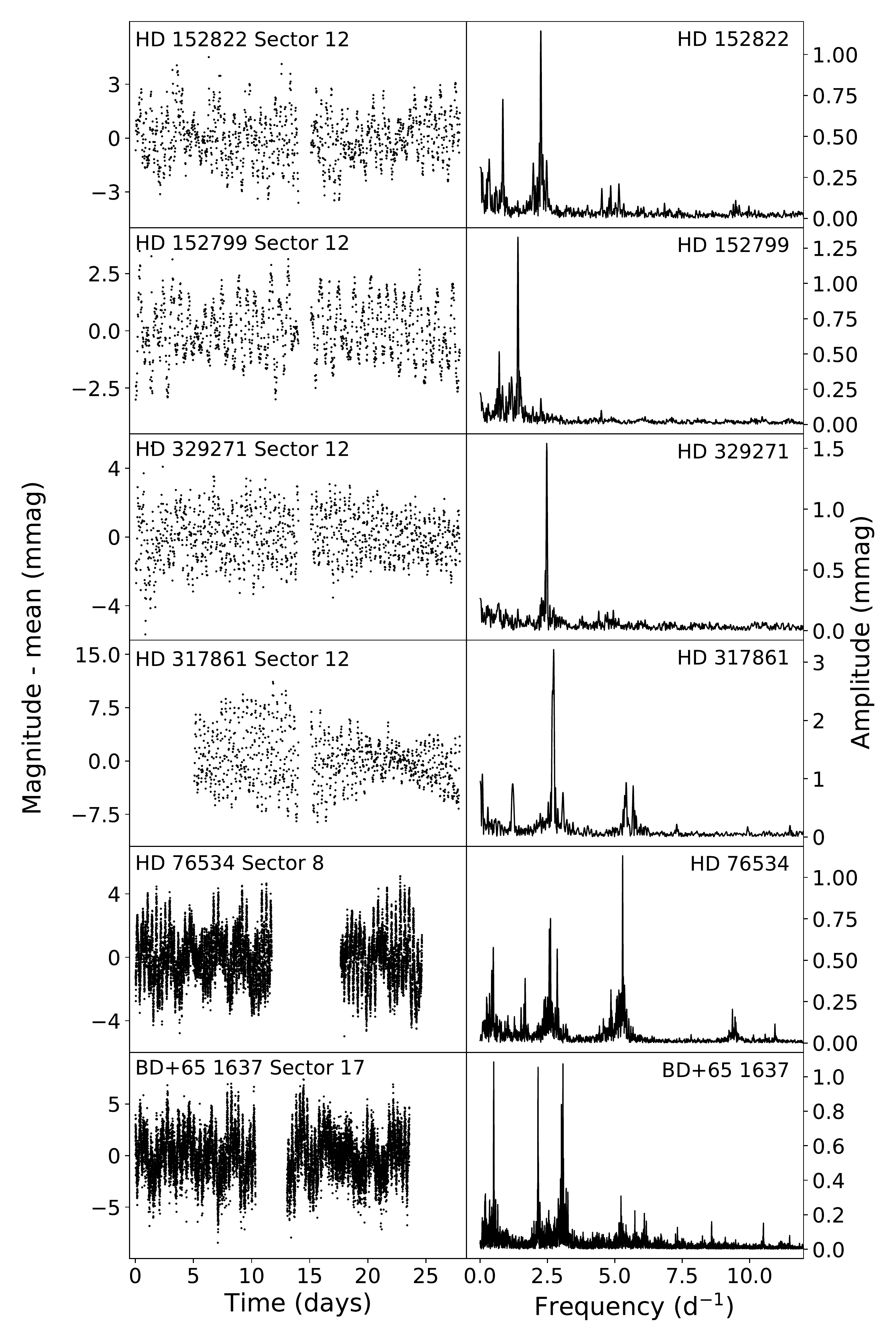}
      \caption{TESS observations of the newly discovered SPB stars and candidates. Left panel: one sector of the light curves. The time is relative to the start of the TESS sector. Right panel: corresponding amplitude spectra derived from the total light curve. 
              }
         \label{fig:spb_lcs}
\end{figure}

\textit{\underline{BD+65 1637}} is a B3IV-Vne star \citep{Aveni1972} that has been observed in sectors 16, 17, 18, and 24. Short cadence data is available for all of these sectors. BD+65 1637 is a member of the young open cluster NGC 7129 with an age smaller than $3$~Myr \citep{Straizys2014}. \citet{Straizys2014} furthermore give an age of $0.27$~Myr for BD+65 1637. We extract $19$ significant frequncies, mostly in the range $3$ and $5.5$~\cd, indicating g-mode pulsations. In addition, a few frequencies higher than $8$~\cd are significant, also indicating p-mode pulsations. Given the young age of NGC 7129 and especially of BD+65 1637, the latter is an excellent candidate to be a pre-main sequence SPB star. With $T_{\rm eff}  = 18000 \pm 1000 $~K and $\log(g) = 4.0 \pm 0.5 $ \citep{Alecian2013} we add it as young SPB star to our sample.

\textit{\underline{HD 152822}} is a B5/6IV star \citep{Houk1978} that has been observed in sector 12. No short cadence data is available for this star, hence, we use the light curve obtained using the \texttt{RegressionCorrector}. HD 152822 has a TESS magnitude of $8.844$ and is listed as a double or multiple star in SIMBAD. Indeed, two stars with TESS magnitudes of $9.589$ and $10.004$, TIC 1336947049 and TIC 1336947057, share almost the same position. None of the two are listed in SIMBAD. As a consequence, we lack the possibility to further constrain this assumed multiple system. In the aperture chosen by us, there are four more stars with relevant magnitudes (we chose a cut-off GAIA magnitude of HD 152822 minus $\sim6$). Three of them have an entry in SIMBAD, namely NGC 6250 7, NGC 6250 6, and NGC 6250 5 with respective TESS magnitudes of $11.7078$, $12.130$, and $10.966$. The last one, Gaia DR2 5963933729737076736, is by far the faintest of the four with a GAIA magnitude of $14.541$. We extract eight significant frequencies, of which five are in the frequency range of $0.8$ to $2.5$ \cd and an additional triplet around $4.8$ \cd. These frequencies clearly indicate g-mode pulsation. Given the involved magnitudes and the amplitude of the extracted frequencies, it is most likely that the pulsations originate in one of the three stars assumed to build up the multiple system. Since the spectroscopic parameters of HD 152822 are consistent with SPB stars, we view HD 152822 as an SBP candidate with $T_{\rm eff}  = 18236 \pm 625$~K and $\log(g) = 4.30 \pm 0.07)$ according to \citet{Aidelman2018}. The FFI and chosen aperture are shown in Fig. \ref{appfig:HD152822}.

\textit{\underline{HD 152799}} is a B2/3III star \citep{Houk1978} that has been observed in sector 12. No short cadence data is available for this star hence we use the light curve obtained using the \texttt{RegressionCorrector}. HD 152799 has a TESS magnitude of $8.674$ and only one relevant star is in the chosen aperture, that is GAIA DR2 5963932424051841536 with a TESS magnitude of $12.626$. Three fainter stars are just outside the aperture to the north. We extract five significant frequencies in the range from $0.7$ to $4.5$ \cd clearly indicating g-modes. In addition, two frequencies in the same range fall just short of our significance criterion. \citet{Aidelman2018} reports $T_{\rm eff}  = 22919 \pm 1577$~K and $\log(g) = 4.31 \pm 0.04$ which is hotter and thus presumably more massive compared to known young SBP stars. Given that no comparably bright stars are directly in the aperture, we view HD 152799 as SPB star, but given its effective temperature it is highly unlikely to be in its pre-main sequence phase. However, more observations, preferable with better angular resolution are needed to verify this classification and we hence add it as candidate to our sample. The FFI and chosen aperture are shown in Fig. \ref{appfig:HD152799}.

\textit{\underline{HD 329271}} is a B4V \citep{Aidelman2018} star that has been observed in sector 12. No short cadence data is available for this star hence we use the light curve obtained using the \texttt{RegressionCorrector}. HD 329271 has a TESS magnitude of $10.515$ and only one other relevant star is in the chosen aperture, that is GAIA DR2 5963932252267675392 with a TESS magnitude of $14.083$. We extract two close frequencies at $2.473508$ and $2.51648$ \cd. The frequency range indicates g-mode pulsation. However, the first frequency is very close to F5 extracted for HD 152822. Depending on the chosen aperture and the size of FFI cutoff, this frequency can sometimes also be found for HD 152799. Given that the three stars lie close to each other, this signal might be of instrumental origin. Therefore, we view HD 329271 with $T_{\rm eff}  = 18074 \pm 623$~K and $\log(g) = 4.26 \pm 0.06$ \citep{Aidelman2018} as less reliable SPB candidate compared to the two other cases. The FFI and chosen aperture are shown in Fig. \ref{appfig:HD329271}.

\textit{\underline{HD 317861 (CD-32 12908)}} is a B3Ve \citep{Aidelman2018} star that has been observed in sector 12. No short cadence data is available for this star hence we use the light curve obtained using the \texttt{RegressionCorrector}. CD-32 12908 has a TESS magnitude of $9.516$. Two other stars are in the chosen aperture, namely NGC 6383 75 and GAIA DR2 4054593854958545536 with TESS magnitudes $12.283$ and $14.798$. We extract twelve significant frequencies mainly in the range around $1.2$, $2.5$, and $5.2$~\cd and one at around $7.3$~\cd. The frequency list and light curve morphology strongly indicate g-mode pulsation. However, an additional frequency around $11.5$~\cd and some accessory signal in the amplitude spectrum suggest existence of possible p-modes. According to the TIC, NGC 6383 75 has $T_{\rm eff}  = 7530$~K and $\log(g) = 4.05796$, in the range of $\gamma$ Doradus stars. However, given the magnitude difference of the stars and the amplitude of the frequencies, we prefer the hypothesis that (at least the higher amplitude frequencies) originate from CD-32 12908. Additional observations, especially with better angular resolution is needed to verify any identification. We consider CD-32 12908 with $T_{\rm eff}  = 20473 \pm 822$~K and $\log(g) = 4.23 \pm 0.09$ \citep{Aidelman2018} an SPB candidate. The FFI and chosen aperture are shown in Fig. \ref{appfig:CD-3212908}.

\textit{\underline{HD 76534}} is a binary system where the primary is a B2Vn star \citep{Houk1978}. The binary has been observed in sector 8 and 9. With a TESS magnitude of $8.07$, the primary is significantly brighter than the secondary which has a magnitude of $9.409$. Short cadence data is available for both sectors, hence we downloaded the PDCSAP flux from the MAST portal. We extract $16$ significant frequencies, mostly around $2.5$, $5$, and $9$~\cd. Two additional frequencies are just below our significance criterion. The light curve of HD 76534 hence clearly indicates g-mode pulsations. However it is unclear whether the primary or the secondary is the source of the variability. For our study, we assume that the primary with $T_{\rm eff}  = 19000 \pm 500$~K and $\log(g) = 4.1 \pm 0.2$ \citep{Fairlamb2015} is a new SPB star candidate. 

Figure \ref{fig:spb_lcs} shows the light curves and amplitude spectra of all SBP stars and candidates found in this work and the extracted frequencies are given in Table \ref{tab:spbcan}.

\subsection{$\delta$ Scuti stars}

\begin{figure}
   \centering
   \includegraphics[width=\linewidth]{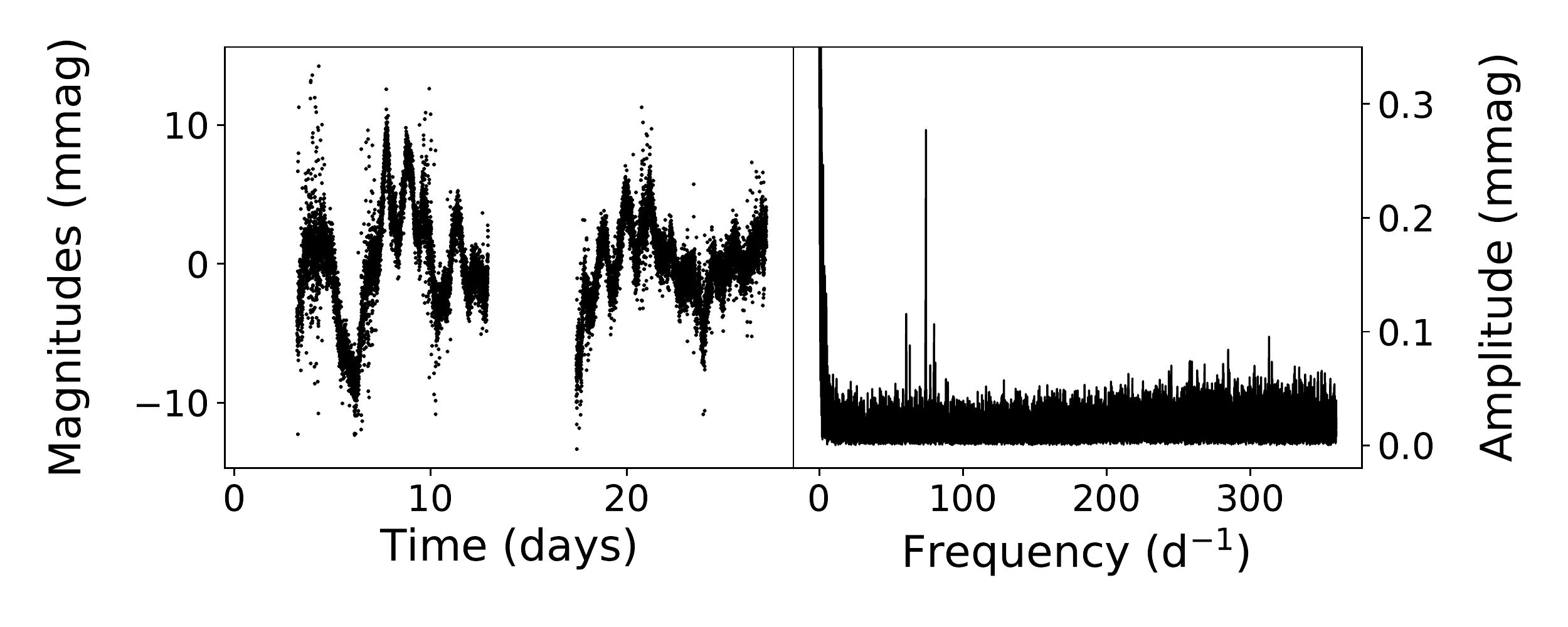}
      \caption{TESS observations of the $\delta$ Scuti candidate HD 135344A. Left panel: light curve minus the mean magnitude without removing the Savitzky-Golay filter. The time is relative to start of the TESS sector. Right panel: corresponding amplitude spectra of the light curve. 
              }
         \label{fig:HD135344}
\end{figure}
\begin{figure}
   \centering
   \includegraphics[width=\linewidth]{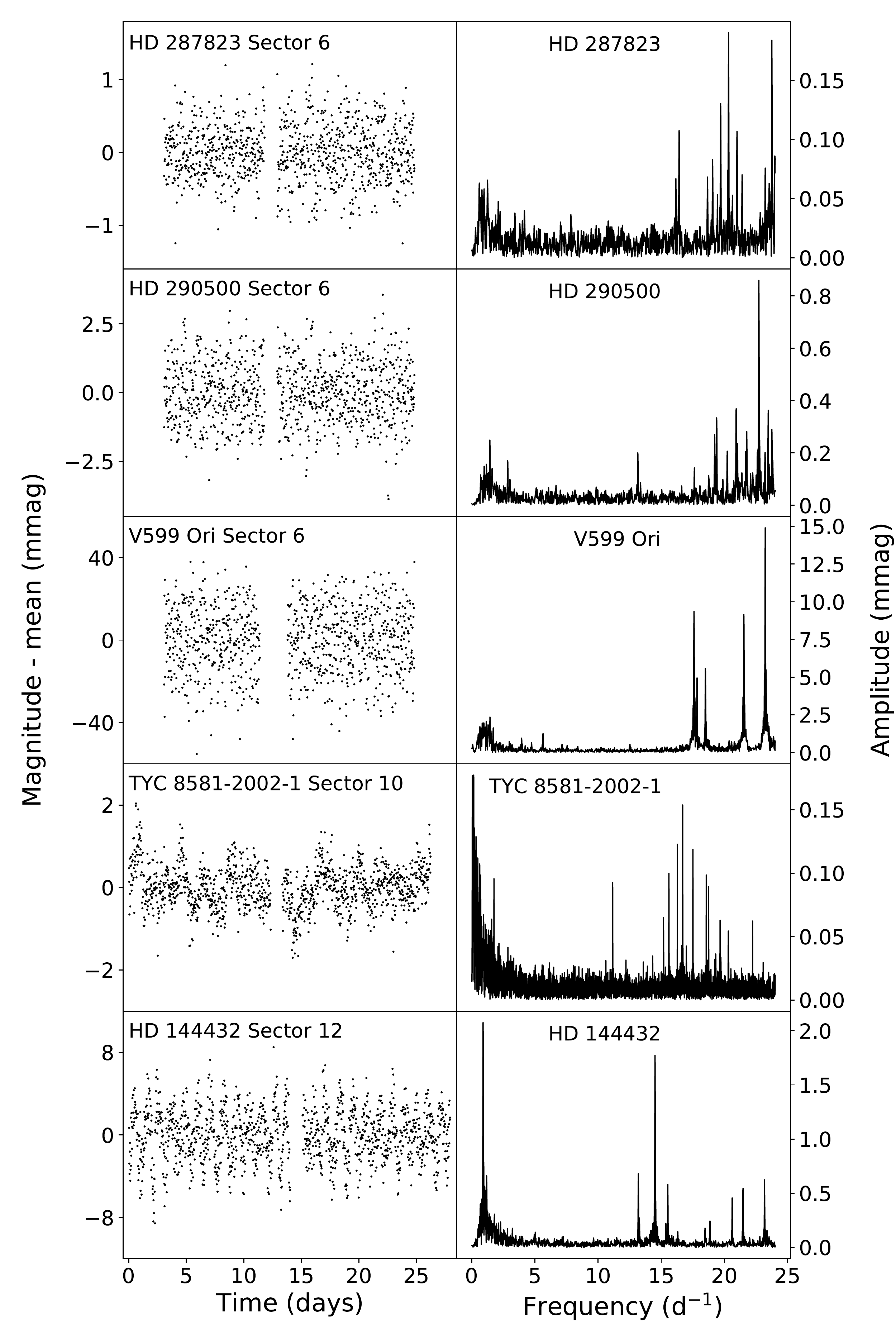}
      \caption{TESS observations of the newly discovered $\delta$ Scuti stars and candidates. Left panel: one sector of the (smoothed) light curves. The time is relative to start of the TESS sector. Right panel: corresponding amplitude spectra derived from the total light curve. 
              }
         \label{fig:dscuti_lcs}
\end{figure}
Amplitude spectra calculated from light curves extracted from TESS FFIs suffer from a Nyquist limit of $24$~\cd \citep{Murphy2015}. Real frequencies are reflected across the Nyquist frequency \citep{Murphy2012, Bell2020}. Given that pre-main sequence $\delta$ Scuti stars typically show oscillations below and above $24$~\cd \citep{Zwintz2014}, it is unclear whether the extracted frequencies are intrinsic to the star or Nyquist-aliases. However, the presence of the signal, independent of the intrinsic value, allows for the identification of the pulsation type. 
 
Figures \ref{fig:HD135344} and \ref{fig:dscuti_lcs} show the light curves and amplitude spectra of all six $\delta$ Scuti stars and candidates extracted from FFIs found in this work and the extracted frequencies are given in Tables \ref{tab:dscuti} and \ref{tab:dscutican}, respectively. 

\textit{\underline{HD 287823}} is an A0 star \citep{Vieira2003} that has been observed in sector 6. No short cadence data is available for this star, hence we use the light curve obtained using the \texttt{RegressionCorrector}. We remove some long term variability using a Savitzky-Golay filter with a window length of $101$, which removes intrinsic frequencies below $1$\cd\ but does not influence the p-mode frequencies. Most notably, this removes significant frequencies at $0.13334$, $0.35863$, and $0.32184$~\cd. With Gaia DR2 3234348869128541952, only one other star lies within the chosen aperture albeit the TESS magnitude is quite high ($14.344$) compared to the magnitude of $9.579$ for HD 287823. Three more stars lie in close vicinity, none of which are brighter than $12.08$ magnitudes.  Hence, we do not expect strong contamination from theses sources. We extract eleven significant frequencies in the range of $16.1$ to $23.75$~\cd (see Table \ref{tab:dscuti}), clearly indicating p-mode pulsation. More frequencies, above the Nyquist frequency ($24$~\cd) might be expected from shorter cadence data. Given its classification as Herbig AeBe star, we view HD 287823 as pre-main sequence $\delta$ Scuti star with $T_{\rm eff}  = 8375 \pm 125$~K and $\log(g) = 4.23 \pm 0.15$ \citep{Fairlamb2015}. The  FFI and chosen aperture are shown in Fig. \ref{appfig:HD 287823}.

\textit{\underline{HD 290500}} is an A2 star \citep{Vieira2003} that has been observed in sector 6. No short cadence data is available for this star, hence we use the light curve obtained using the \texttt{RegressionCorrector}. We remove some long term variability using a Savitzky-Golay filter with a window length of $75$, which removes intrinsic frequencies below $1$\cd\ but does not influence the p-mode frequencies. Two stars lie just at the outer edge of the chosen aperture but HD 290500 outshines them by at least four magnitudes. We extract $16$ significant frequencies in the range of $17.6$ to $23.76$~\cd (see Table \ref{tab:dscuti}), clearly indicating p-mode pulsation. More frequencies, above the Nyquist frequency ($24$~\cd) might be expected from shorter cadence observations. Given its classification as Herbig AeBe star, we view HD 290500 as a pre-main sequence $\delta$ Scuti star. With an effective temperature of $9500 \pm 500$~K \citep{Fairlamb2015}, it seems to be very hot for a pre-main sequence $\delta$ Scuti star. In comparison, \citet{Alecian2013} reported an effective temperature of $9000 \pm 500$~K indicating that it might actually be located at a lower effective temperature. Only a high-resolution, high signal-to-noise-ratio spectrum will provide a more reliable value in the future. A previous study performed by \citet{diaz-fraile2014} was inconclusive concerning the variability, but the TESS observations allow a clear identification. In addition, the frequencies F12 and F18 seem to be a signal of rational variability, indicating a surface rotation period of approximately $0.7$~d. The FFI and chosen aperture are shown in Fig. \ref{appfig:HD 290500}.

\textit{\underline{V599 Ori}} is as A8/9 star \citep{Hsu2013} that has been observed in sector 6. No short cadence data is available for this star and the \texttt{RegressionCorrector} fails for this target. Hence, we use the BCSAP flux light curve. We remove some long term variability using a Savitzky-Golay filter with a window length of $75$, which removes intrinsic frequencies below $1$\cd\, but does not influence the p-mode frequencies. A dark star lies just outside the chosen aperture and should not lead to strong contamination. However, the pulsating naked eye star 49 Ori lies in the same region of the sky. Given its brightness, some contamination might be expected. We extract seven significant frequencies in the range of $17.5$ to $23.3$~\cd (see Table \ref{tab:dscutican}), clearly indicating p-mode pulsation. More frequencies, above the Nyquist frequency ($24$~\cd) might be expected from short cadence observations. Some additional frequencies might indicate g-mode pulsations as well. This makes V599 Ori a fascinating object to study, especially as the spectroscopic parameters by \cite{Fairlamb2015} with $T_{\rm eff}  = 8000 \pm 250$~K and $\log(g) = 3.72 \pm 0.13$ would indicate that the effective temperature is too high for g-modes. However, the influence of 49 Ori on the extracted light curve of V599 Ori needs special attention. In addition, the fact that we are using a very simple background corrected aperture photometry does not enhance our trust in the extracted frequencies. A dedicated analysis as well additional observation will be needed to properly classify this object. We view V599 Ori as $\delta$ Scuti candidate. The FFI and chosen aperture are shown in Fig. \ref{appfig:V599 Ori}.

\textit{\underline{TYC 8581-2002-1}} is an A5V star \citep{Vieira2003} that has been observed in sector 8, 9, and 10. No short cadence data is available for this star, hence, we use the light curve obtained using the \texttt{RegressionCorrector}. Multiple stars lie in the vicinity of TYC 8581-2002-1; given that we have multiple sectors of FFI data, the stars that lie within the aperture vary slightly. TYC 8581-2002-1 has a TESS magnitude of $10.547$, while a very close star, GAIA DR2 5302462513641062144, has a TESS magnitude of $10.954$ and lies within the aperture in all sectors. GAIA DR2 5302462346139414656 lies to the East of TYC 8581-2002-1 and is either just inside (in sector 10) or outside the aperture ( in sectors 8 and 9). We extract 13 significant frequencies in the range from $11.1$ to $22.3$~\cd (see Table \ref{tab:dscutican}), clearly indicating p-mode pulsation. More frequencies, above the Nyquist frequency ($24$~\cd) might be expected. We find one additional frequency, possibly indicating a surface rotation period of approximately $0.57$~d. The TIC reports an effective temperature of $7145$~K for GAIA DR2 5302462513641062144, well within the range typical for $\delta$ Scuti-type pulsations. With the current observations, it is impossible to definitely attribute the pulsation signal to any of these stars and thus we choose TYC 8581-2002-1 with $T_{\rm eff}  = 9750 \pm 250$~K and $\log(g) = 4.0 \pm 0.1$ \citep{Fairlamb2015} as $\delta$ Scuti candidate. The FFI and chosen aperture for sector 10 are shown in Fig. \ref{appfig:TYC 8581-2002-1}.

\textit{\underline{HD 144432}} is an A9/F0V star \citep{Houk1982} that has been observed in sector 12. No short cadence data is available for this star, hence, we use the light curve obtained using the \texttt{RegressionCorrector}. We remove some long term variability using a Savitzky-Golay filter with a window length of $101$, which removes intrinsic frequencies below $1$~\cd\, but does not influence the p-mode frequencies. HD 144432 is a known multiple system \citep{mueller2011} and given their analysis, we assume that HD 144432A is mainly responsible for the signal in the chosen aperture and, hence, also the variability. We extract $12$ significant frequencies in the range of $13.1$ to $23.2$~\cd (see Table \ref{tab:dscuti}), clearly indicating p-mode pulsation. More frequencies, above the Nyquist frequency ($24$~\cd) might be revealed from short cadence data. In addition, F1 seems to be a signal of rational variability, indicating a surface rotation period of approximately $1.12$~d. Since the spectroscopic parameters of the other members of this system are unknown, we identify HD 144432 with $T_{\rm eff}  = 7500 \pm 250$~K and $\log(g) = 4.05 \pm 0.17$ \citep{Fairlamb2015} as $\delta$ Scuti candidate. The FFI and chosen aperture are shown in Fig. \ref{appfig:HD 144432}.

\textit{\underline{HD 135344A}} is an A0V \citep{Houk1982} primary of a binary system observed in sector 11. Short cadence observations are available for both, the primary and secondary component. We remove strong long term variability using a Savitzky-Golay filter with a window length of $201$, which removes intrinsic frequencies below $5$~\cd but does not influence the p-mode frequencies. We extract $7$ significant frequencies in the range of $60.5$ to $80.9$~\cd (see Table \ref{tab:dscuti}), clearly indicating p-mode pulsation. Additional, yet insignificant signals in the amplitude spectrum are observed at frequencies in the range of $250$ to $335$~\cd. Clearly, contamination from the secondary is expected in the light curve of the primary. However, the amplitude of the signal is evidently stronger in the light curve of HD 135344A and hence we identify it as $\delta$ Scuti candidate with $T_{\rm eff}  = 6750 \pm 250$~K and $\log(g) = 4$ \citep{Alecian2013}. Figure \ref{fig:HD135344} shows the un-smoothed light curve and corresponding amplitude spectrum of HD 135344A. The pulsation signal around $80$~\cd and the additional high frequency signal can clearly be seen.

\subsection{$\gamma$ Doradus stars}
\begin{figure}
   \centering
   \includegraphics[width=\linewidth]{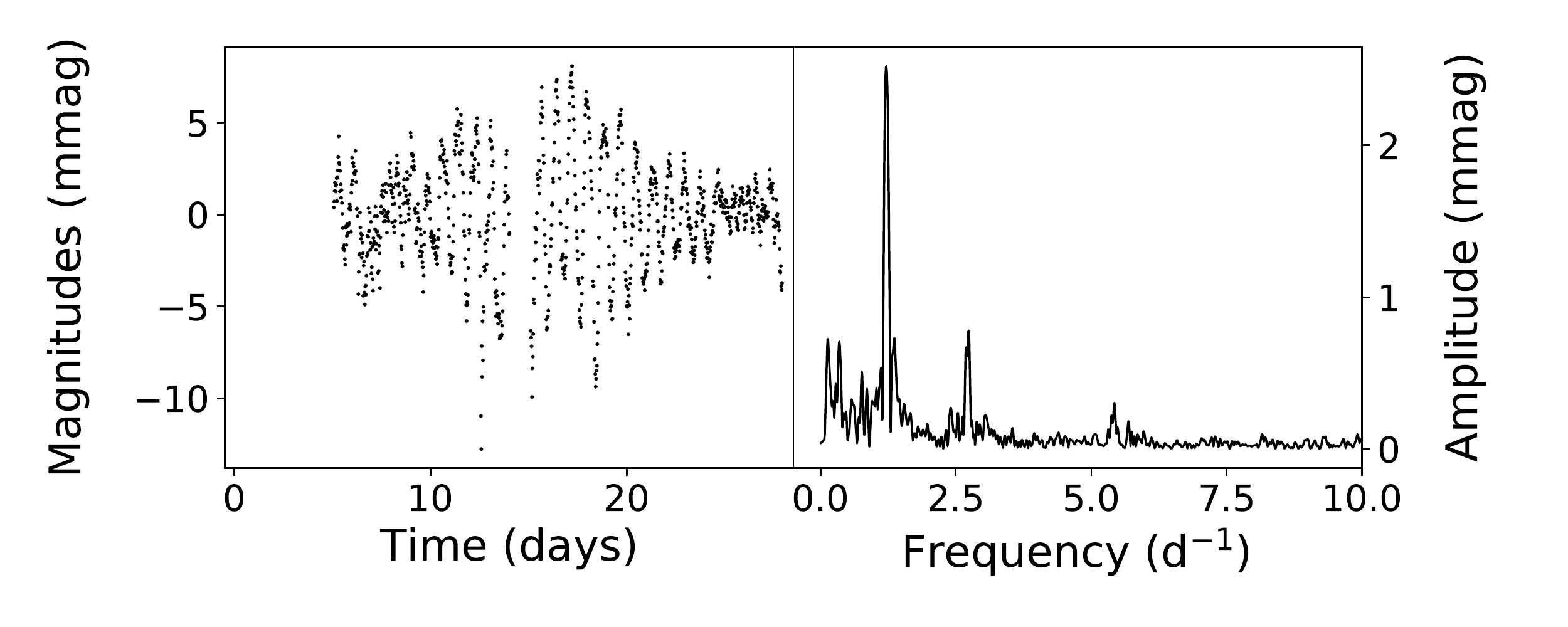}
      \caption{TESS observations of the $\gamma$ Doradus candidate HD 317859. Left panel: smoothed light curve extracted from the TESS FFIs. The time is relative to start of the TESS sector. Right panel: corresponding amplitude spectra of the light curve. 
              }
         \label{fig:CD-32 12910}
\end{figure}

\textit{\underline{HD 317859 (CD-32 12910)}} is a F3V \citep{Evans1978} star that has been observed in sector 12. No short cadence data is available for this star, hence, we use the light curve obtained using the \texttt{RegressionCorrector}. We remove some long term variability using a Savitzky-Golay filter with a window length of $601$, which removes intrinsic frequencies below $0.2$\cd. No other stars lie within the chosen aperture. We extract eight significant frequencies around $1.2$, $2.7$, and $5.4$~\cd (see Table \ref{tab:gamdor}), clearly indicating g-mode pulsation. \citet{Evans1978} reported HD 317859 to be a member of the young open cluster NGC 6383 with an age of $4{-}5$~Myr. At this estimated age, HD 317859 has a high probability to be a pre-main sequence $\gamma$ Doradus star with $T_{\rm eff}  = 7300 \pm 163$~K and $\log(g) = 4.34 \pm 0.01$ \citep{Aidelman2018}. However, a more reliable membership analysis is needed to verify this identification. We select HD 317859 as $\gamma$ Doradus candidate. The FFI and chosen aperture are shown in Fig. \ref{appfig:CD-3212910}.

\subsection{Other variable stars}
We found additional variability in the TESS observations of CD-49 11096, PDS 344, HD 36982, HD 329379, HD 96042, HD 53367, and HD 152743 of different origins. We discuss these stars in more detail in Appendix \ref{app:tess}. Especially HD 329379 and HD 96042 should be interesting objects for follow up studies.

    \begin{table*}
    \begin{footnotesize}
    \caption[]{Stellar and pulsational properties of the newly discovered pulsators with TESS and CoRoT.}
    \label{tab:newpulstable}
    \begin{tabular*}{\linewidth}{llrrrrrrlr}
        \hline
        \noalign{\smallskip}
         Designation & TIC ID  &\multicolumn{2}{c}{Magnitude}  &  \multicolumn{1}{c}{ST}    & \multicolumn{1}{c}{$T_{\rm eff}$} & \multicolumn{1}{c}{$\log g$}& \multicolumn{1}{c}{source} & type & frequency range(s)\\
        & &  \multicolumn{1}{c}{V} & \multicolumn{1}{c}{Tess} & & \multicolumn{1}{c}{(K)}   &  & && \multicolumn{1}{c}{(\cd)}\\
        \noalign{\smallskip}
        \hline
        \noalign{\smallskip}
        BD+65 1637 & 322312832 & $10.53$ & $9.53$ & B3 & $18000 \pm 1000$ & $4.0 \pm 0.5$ & A13 & SPB & $3 - 5.5$, $8$\\
        HD 152822 & 247980805 & $9.07$ & $8.84$ & B5/6 & $18236 \pm 625$ & $4.3 \pm 0.07$ & A18 & SPB & $0.8 - 2.5$, $4.8$\\
        HD 152799 & 247981153 & $8.74$ & $8.67$ & B2/3 & $22919 \pm 1577$ & $4.31 \pm 0.04$ & A18 & SPB & $0.7 - 4.5$\\
        HD 329271 & 247981406 & $10.65$ & $10.51$ & B4 & $18074 \pm 623$ & $4.26 \pm 0.06$ & A18 &  SPB & $2.5$\\
        HD 317861 & 102141335 & $9.83$ & $9.52$ & B3 & $20473 \pm 822$ & $4.23 \pm 0.09$ & A18 &  SPB & $1.2$, $2.5$, $5.2$, $7.3$\\
        HD 76534 & 30051402 & $7.5$ & $8.07$ & B2 & $19000 \pm 500$ & $4.1 \pm 0.2$ & F15 &  SPB &  $2.5$, $5$, $9$\\
        HD 287823 & 459990291 & $9.74$ & $9.58$ & A0 & $8375 \pm 125$ & $4.23 \pm 0.15$ & F15 &  $\delta$ Sct & $16.1 - 23.75$\\
        HD 290500 & 50656450 & $11.04$ & $10.75$ & A2 & $9500 \pm 500$ & $3.8 \pm 0.4$ & F15 &  $\delta$ Sct & $17.6 - 23.76$\\
        V599 Ori & 332914217 & $13.7$ & $11.66$ & A8/9 & $8000 \pm 250$ & $3.72 \pm 0.13$ & F15 &  $\delta$ Sct & $17.5 - 23.3$\\
        TYC 8581-2002-1 & 355320922 & $10.91$ & $10.55$ & A5 & $9750 \pm 250$ & $4.0 \pm 0.1$ & F15 &  $\delta$ Sct & $11.1 - 22.3$\\
        HD 144432 & 67671603 & $8.19$ & $7.8$ & A9/F0 & $7500\pm 250$ & $4.05 \pm 0.17$ & F15 &  $\delta$ Sct & $13.1 - 23.2$\\
        HD 135344A & 307606851 & $7.76$ & $7.71$ & A0 & $6750\pm 250$ & $4$ & A13 &  $\delta$ Sct & $60.5 - 80.9$\\
        HD 317859 & 102141942 & $9.52$ & $9.04$ & F3 & $7300\pm 163$ & $4.35 \pm 0.01$ & A18 &  $\gamma$ Dor & $1.2$, $2.7$, $5.4$\\
        Cl$^{\star}$ NGC 2264 VAS 196 & 231147340 & $10.07$ & $10.80$ & F0/2 & $6784\pm 150$ & $4.0 \pm 0.2$ & this work &  $\gamma$ Dor & $0.8 - 2.75$\\
        Cl$^{\star}$ NGC 2264 VAS 219 & 231147743 & $12.09$ & $11.85$ & A7 & $7170\pm 150$ & $4.3 \pm 0.3$ & this work &  $\gamma$ Dor & $0.5 - 3.8$\\
        Cl$^{\star}$ NGC 2264 VAS 230 &  231147556 & $12.37$ & $11.66$ & A5 & $7520\pm 150$ & $4.3 \pm 0.4$ & this work &  hybrid       & $2.1 - 3.3$, $8 - 12$\\

        \noalign{\smallskip}
        \hline

    \end{tabular*}    
    \tablefoot{
    The column source refers to the stellar parameter and the shorcuts refer to the following articles: A13: \citet{Alecian2013}, F15: \citet{Fairlamb2015}, A18: \citet{Aidelman2018}. Sources for the spectral types can be found in the main text. 
    }
    \end{footnotesize}
 \end{table*}
\section{COROT observations}
\label{sec:Corot}

\begin{figure}
   \centering
   \includegraphics[width=\linewidth]{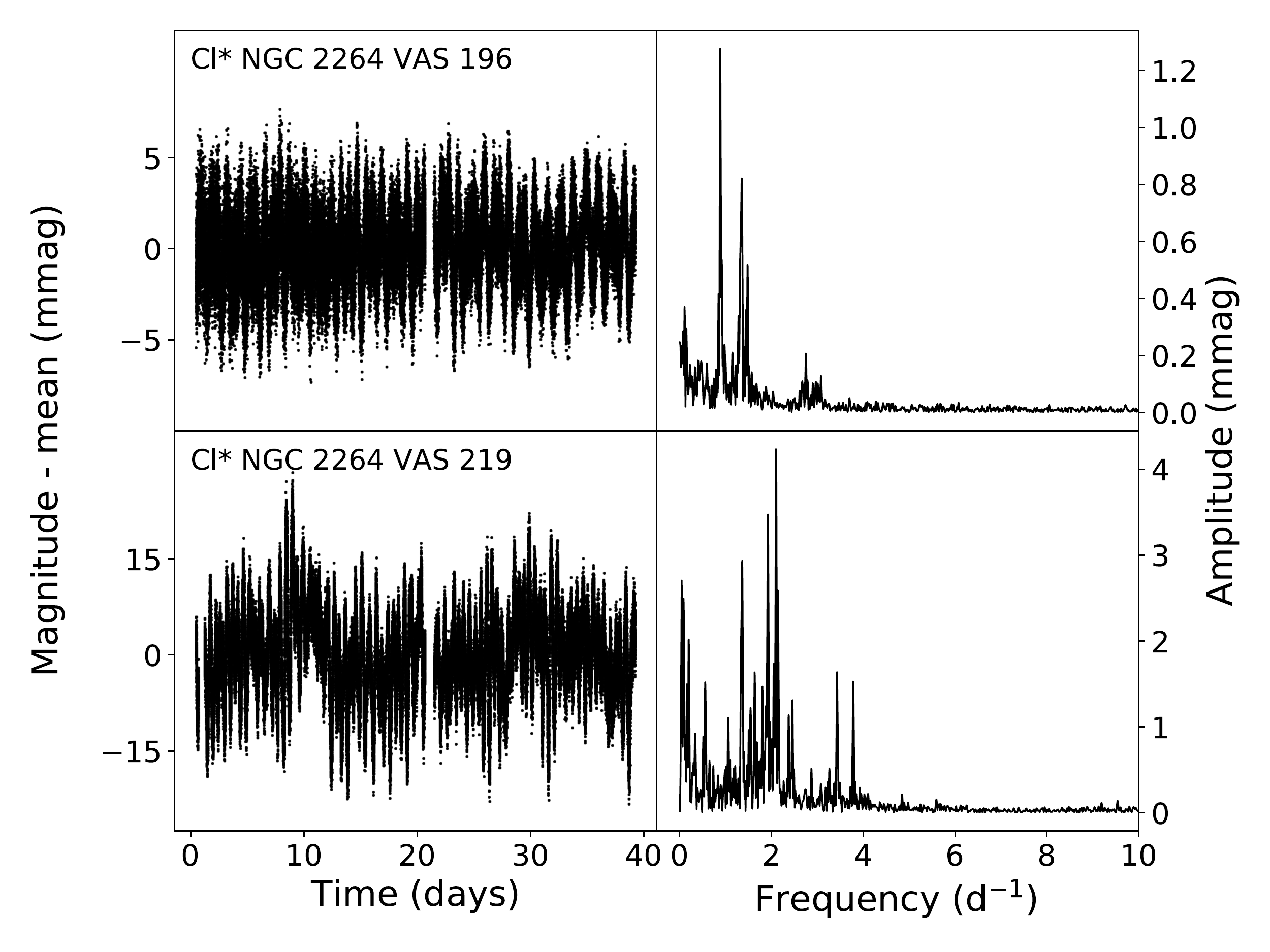}
      \caption{CoRoT observations of the newly discovered $\gamma$ Doradus stars. Left panel: light curves obtained in SRa05. The time is relative to the start of the observations. Right panel: corresponding amplitude spectra derived from the light curve of SRa05. 
              }
         \label{fig:gamma_dor_corot}
\end{figure}

\begin{figure}
   \centering
   \includegraphics[width=\linewidth]{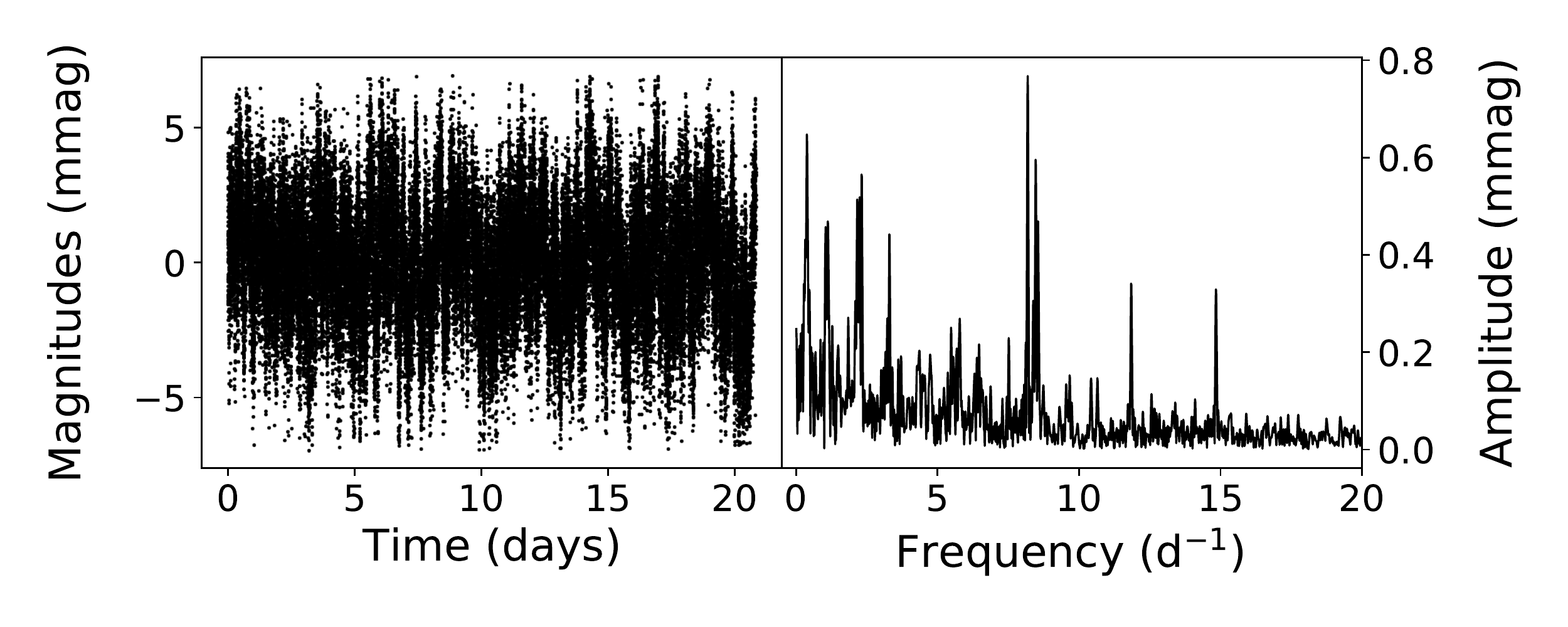}
      \caption{CoRoT observations of the newly discovered $\delta$ Scuti - $\gamma$ Doradus hybrid candidate Cl$^{\star}$ NGC 2264 VAS 230. Left panel: light curve obtained in SRa01. The time is relative to start of the observations. Right panel: corresponding amplitude spectra derived from the light curve of SRa01. 
              }
         \label{fig:VAS230}
\end{figure}
\subsection{The CoRoT mission}
The CoRoT satellite \citep{Auvergne2009} was launched on 2006, Dec. 27, from Baikonur onboard a Soyuz rocket into a polar, inertial circular orbit at an altitude of 896\,km. CoRoT carried a telescope with 27\,cm aperture and was able to observe stars inside two cones of $10^{\circ}$ radius, one at RA = 06:50 and the other at RA = 18:50 using a filter bandwidth from 370 to 1000 nm. The main research goals of the CoRoT mission was the search for extrasolar planets and asteroseismology of main sequence stars.

Three stars were originally discovered from CoRoT data to be candidate or bona-fide pre-main sequence $\gamma$ Doradus or $\delta$ Scuti -- $\gamma$ Doradus pulsators. Their frequency content was confirmed using recent TESS observations. Table \ref{tab:newpulstable} lists their stellar properties.

\subsection{Observations, Data reduction and analysis}
CoRoT obtained two short run observations on the young open cluster NGC 2264: Short Run SRa01 was conducted in March 2008 providing a time base of observations of 23.4 days and Short Run SRa05 observed the field from December 2011 to January 2012 for about 39 days. For both observations, the complete cluster NGC 2264 was placed in one of the Exofield CCDs and data were taken for all stars in the accessible magnitude range from 10 to 16 mag in $R$.

For the present analysis, the reduced N2 data for all stars were extracted from the CoRoT data archive and the standard procedures for the reduction of CoRoT data were applied \citep[see e.g.,][]{Zwintz2013}. The CoRoT data reduction pipeline \citep{Auvergne2009} flags those data points that were observed during the satellite's passages over the South Atlantic Anomaly (SAA) and replaces them with linearly interpolated values. In the present analysis, we omitted the linearly interpolated values, hence, used only the data points obtained outside of the SAA.
A detailed description of the CoRoT observations of NGC 2264 and the respective data reduction can be found for example in \citet{Zwintz2013}.

We perform the Fourier analysis of CoRot light curves in analogy to the TESS observations. Since CoRot observes from a low-earth orbit, aliasing issues from the spectral window might be expected. Data reduction for the light curves presented in this study mostly removed these effects (see Fig. \ref{fig:gamma_dor_corot_inc_sw} and Fig. \ref{fig:VAS230_inc_sw}), such that the spectral window has no influence in the identification of pulsation type.

\subsection{Pre-main sequence $\gamma$ Doradus stars in NGC 2264}
Additionally to the two previously reported pre-main sequence $\gamma$ Doradus stars \citep{Zwintz2013}, two further pre-main sequence cluster members observed in the two CoRoT Short Runs on NGC 2264 showed g-mode type pulsations. Their light curves can be seen in Fig. \ref{fig:gamma_dor_corot}. Only with high-resolution, high signal-to-noise spectra it was possible to clearly distinguish SPB from $\gamma$ Doradus type pulsators and add an additional piece of information on the membership of the individual stars to NGC 2264.
Appendix \ref{app:corot} describes the analysis of the spectroscopic data that led to the fundamental parameters used in the present study.

\textit{\underline{Cl$^{\star}$ NGC 2264 VAS 196 (CoID0223991870)}} is a F0/2 star \citep{Karlsson1972} that was observed by CoRoT in both Short Runs on NGC 2264 and by TESS in Sector 6 with a time base of 27 days. As the SRa05 data contain the longest continuous time base of $\sim$39 days (Fig. \ref{fig:gamma_dor_corot}), it also has the lowest noise level. Hence, we base our frequency analysis on this data set and compare to the shorter sets obtained by CoRoT in SRa01 and TESS in Sector 6. 
Nine frequencies were found to be significant in our analysis from the CoRoT SRa05 data (see Table \ref{app:tab_CoID1870}).
The membership probability of Cl$^{\star}$ NGC 2264 VAS 196 to the cluster is reported to be 94\% by \citet{Vasilevskis1965}. Hence, we add Cl$^{\star}$ NGC 2264 VAS 196 to our sample of pre-main sequence $\gamma$ Doradus pulsators.

\textit{\underline{Cl$^{\star}$ NGC 2264 VAS 219 (CoID0223997565)}} is an A7 star \citep{Voroshilov1985} that was observed by CoRoT in both short runs on NGC 2264 and by TESS in Sector 6. The CoRoT data obtained during SRa05 has the longest uninterrupted time base of $\sim$39 days (Fig. \ref{fig:gamma_dor_corot}), hence, yielding the lowest noise level. Consequently, we base our analysis on this data set and extract 24 significant g-mode frequencies in the range from 0.04 to 3.8\cd; 15 of those can also be found in the CoRoT SRa01 data set and five in the TESS data (see Table \ref{app:tab_CoID7565}). We interpret the presence of several emission lines in the spectrum (see Figure \ref{fig:spectrum_VAS219}) as an indicator for stellar youth, and, hence, add Cl$^{\star}$ NGC 2264 VAS 291 to our list of pre-main sequence $\gamma$ Doradus stars.
%CoRoT data
% McD 2010 spectrum

\subsection{Pre-main sequence $\delta$ Scuti -- $\gamma$ Doradus hybrid in NGC 2264}

\textit{\underline{Cl$^{\star}$ NGC 2264 VAS 230 (CoID0223999101)}} is an A5 star \citep{Voroshilov1985} that was observed by CoRoT in both short runs on NGC 2264 and by TESS in Sector 6. The CoRoT data obtained during SRa01 has the highest photometric precision, hence, yields the lowest noise level despite the fact that SRa05 has the longer time base. Therefore, we base our analysis on the SRa01 data set and compare to the SRa05 and TESS data. The light curve of SRa01 is shown in Fig. \ref{fig:VAS230}.  We find five frequencies between 8.1 and 12\,\cd\, that we identify as p-modes (F1 to F5 in Table \ref{app:tab_CoID9101}). 
The frequency at $\sim$14.8\,\cd\, which can easily be seen in Fig. \ref{fig:VAS230} is related to the orbital frequency of the CoRoT satellite, and, hence was omitted in our analysis. 
Additionally, we detect four frequencies in the range from 2.1 to 3.3\,\cd\, that might be g-modes (F6 to F9 in Table \ref{app:tab_CoID9101}). Hence, Cl$^{\star}$ NGC 2264 VAS 230 is a candidate pre-main sequence $\delta$ Scuti -- $\gamma$ Doradus hybrid pulsator.

%--------------------------------------------------------------------
\section{Stellar Models}
\label{sec:models}
\subsection{Stellar evolution models}
The stellar evolution models in this work are calculated with version v-12778 of  \textit{Modules for Experiments in Stellar Astrophysics} \citep[\texttt{MESA,}][]{paxton2011,paxton2013, paxton2015, paxton2018, paxton2019}. \texttt{MESA} is a software instrument that solves the fully coupled structure and composition equations simultaneously for a one dimensional spherically symmetric stellar model \citep{paxton2011}. We refer to the \texttt{MESA} instrument papers for a full description of the numerical methods. More details about the inputs for the microphysics is given in Appendix \ref{app:mesaphysics}.   \texttt{MESA}'s \textit{run\_star\_extras.f} file allows the extension of the otherwise very diverse code. We make use of this \textit{run\_star\_extras.f} file to implement extensions to \texttt{MESA} as discussed in Section \ref{subsec:treatment_of_accretion} \footnote{Our \texttt{MESA} input files will be publicly avaible on zenodo.}. 

\subsubsection{Input physics and chemical composition}
For our models we use the OPAL opacity tables \citep{seaton2005} and the standard \texttt{MESA} equation of state \citep{paxton2011}. We ignore stellar rotation and magnetic fields. We use the mixing length description of convection developed by \cite{Cox1968} together with the Ledoux criterion and the description of convective premixing developed by \citet{paxton2018} while not allowing for semiconvection. We use different values for the mixing length, $\alpha_{\rm MLT}$, and employ exponential overshooting with $f = 0.01$ and $f_0 = 0.005$ at the top and $f = 0.005$ and $f_0 = 0.0025$ at the bottom of any convective zone. 

The initial composition is subject to change in the models we calculate. Our standard model has an initial $^1\mathrm{H}$ abundance of $X_{^1\mathrm{H}} = 1 - X_{^2\mathrm{H}} - X_{^3\mathrm{He}} - X_{^4\mathrm{He}} - Z$, where $X_{^3\mathrm{He}} = 85$~ppm, $X_{^4\mathrm{He}} = 0.224 + 2*Z -  X_{^3\mathrm{He}}$, and an initial metallicity of $Z = 0.014$ according to the solar metallicity of \citet{asplund2009}. The mass fractions of metals are taken according to \citet{asplund2009} updated based on \citet{nieva2012} and \citet{Przybilla2013}. 

\subsubsection{Initial Models}
For the initial models of our calculations, we follow the description of \citet{Kunitomo2017} who in terms followed the works of \citet{Stahler1980} and \citet{Hosokawa2011}. In this approach, the initial mass and radius of a protostellar seed are set to $0.01\,M_\odot$ and $1.5\,R_\odot$, respectively. That is they correspond to slightly evolved protostars from stellar cores \citep{Kunitomo2017} and they possess a high value for the initial entropy of the star \citep{Baraffe2012}. We refer to \citet{Kunitomo2017} for more details of the initial model and the influence of different initial values on the models. 

\subsubsection{Treatment of accretion}
\label{subsec:treatment_of_accretion}
We follow the description of \citet{Baraffe2009} to model the influence of accretion in a one-dimensional stellar model. Hence, we adopt similar simplifications as \citet{Hartmann1997} and \cite{Siess1997}. In this picture, the accretion onto the protostar proceeds non-spherically. Thus, the object is free to radiate its energy over most of the photosphere \citep{Hartmann1997}. The accretion of material from the surrounding cloud introduces gravitational energy per unit of accreted mass, $-G M/R$, as well as internal energy, $+\epsilon G M/R$, to the system, leading to an energy rate of
\begin{align}
    \frac{\mathrm{d} E_{\rm acc}}{\mathrm{d} t} = (\epsilon -1)\frac{G M \Dot{M}}{R}.
\end{align}
Here, $G$ is the gravitational constant, $M$ and $R$ are the mass and radius of the star in solar units, and $\Dot{M}$ is the accretion rate in solar masses per year. The value $\epsilon$ describes the geometry of the accretion process. Accretion from gravitationally bound material is described by $\epsilon \leq 1$,  while $\epsilon \leq 0.5$ describes the accretion from a thin disk at the object's equator \citep{Hartmann1997, Baraffe2009}. In this study, we follow the approaches of previous authors \citep[e.g.][]{Baraffe2009, Jensen2018} and choose $\epsilon = 0.5$.

The energy of the accreted material is either added to the star ($L_{\rm add}$) or radiated away as accretion luminosity ($L_{\rm acc}$). The amount of each is controlled by a factor $\beta$, leading to 
\begin{align}
    L_{\rm add} = \beta\epsilon \frac{G M \Dot{M}}{R}
\end{align}
and
\begin{align}
    L_{\rm acc} = (1-\beta)\epsilon \frac{G M \Dot{M}}{R}.
\end{align}
The case of $\beta = 0$ corresponds to cold accretion where all of the energy is radiated away, while $0 < \beta \leq 1$ corresponds to hot accretion. However, $\beta$ is expected to vary with accretion rate: a higher accretion rate corresponds to higher values of $\beta$ \citep[see e.g.][and references therein]{Baraffe2012, Vorobyov2017, Jensen2018, Elbakyan2019}. In this study, we use a fixed accretion rate. Therefore, $\beta$ is fixed to values in the range $0{-}0.5$. 

The energy added to the star is injected as extra heat. However, it is unclear how this extra heat is distributed inside the star. Descriptions vary from uniform distribution \citep[e.g.][]{Baraffe2010} to heat injection only in the outer zones with a step function \citep{Jensen2018}. In this study we follow the linear distribution of \citet{Kunitomo2017}. Hence, the heat is distributed only in an outer region of fractional mass $M_{\rm outer}$ and increases linearly with the mass coordinate $m_r$ according to
\begin{align}
    l_{\rm extra} = \frac{L_{\rm add}}{M} {\rm max} \left\{0, \frac{2}{M_{\rm outer}^2} \left(\frac{m_r}{M} -1 + M_{\rm outer} \right)\right\}.
\end{align}
In this study, we vary $M_{\rm outer}$ in the range of $0.025{-}0.5$. 
    
The rate of accretion is most likely unique to any star{-}disk system. Numerical hydrodynamics simulations of such systems show that the accretion indeed varies substantially on a case to case level \citep{Vorobyov2010, Vorobyov2015, Vorobyov2021}. Such episodic accretion has been modelled with one dimensional stellar evolution codes with assumed accretion rates \citep[e.g.][]{Baraffe2009, Baraffe2010, Baraffe2012} as well as accretion rates stemming from such simulations \citep[e.g.][]{Vorobyov2017, Jensen2018, Elbakyan2019}. However, this is computationally expensive and complicates the calculations of controlled stellar grids. Hence, we choose a constant accretion rate for this study.

Accreting evolutionary tracks are calculated until the central hydrogen abundance has dropped by $0.001$ compared to the initial value. During the evolution of the accreting protostellar seed, we save \texttt{MESA} models at specific masses. These are saved once the mass of the accreting evolution is $0.001\,M_\odot$ below the mass wanted at the ZAMS. A different \texttt{MESA} run then reads in these models to calculate the subsequent pre-main sequence (post-accretion) evolution towards the ZAMS. At the start of this subsequent evolution, we continuously lower the accretion rate over a time of $t_{\rm acc} = 0.003/\Dot{M}$ for $500$ equal timesteps according to
\begin{align}
    \Dot{M}(t_i) = \Dot{M}\left(1-\frac{t_i - t_0}{t_{\rm acc}}\right)^2,
\end{align}
where, $t_i$ is the star age at timestep $i$ and $\Dot{M}$ is the constant accretion of the protostellar evolution. Currently, we do not have a good understanding of the processes that lead to a halt of stellar accretion. As a consequence, we are unable to choose the accretion rate towards the end of the accretion process in a fully physically motivated sense. It is often a convenient choice to abruptly stop the mass accretion rate, once the desired stellar mass is reached \citep[see e.g.][]{Kunitomo2017}, while other approaches often assume an exponential drop over a few million years \citep[see e.g.][]{Elbakyan2019}. As our approach needs to be applicable for both low and intermediate mass stars with vastly different time scales of stellar evolution, we choose a short timescale to decrease the accretion rate and allow the stellar model to relax to the non-accreting setup this way. This sharp termination of the accretion may be caused by photoevaporative winds creating holes in the inner disk regions and effectively cutting out the outer disk mass reservoir \citep{Owen2011, Owen2012}.

\subsubsection{Atmospheric boundary condition}
For the calculation of theoretical pulsation spectra, it is a necessity to build the atmosphere from a temperature-opacity ($T{-}\tau$) relation. We initially follow the work of \citet{Amard2019} and use the $T{-}\tau$ relation described by \citet{Krishna1966} with varying opacities but also calculate models using the relation by \cite{Eddington1926}. However, the use of OPAL opacity tables \citep{seaton2005} lead to convergence problems for low mass stars. Therefore, the initial models are constructed using \textit{tau\_100} tables \citep{paxton2011} based on model atmospheres by \citet{Castelli2003} and \citet{Allard2001}. For a solution of these convergence problems we furthermore relax the error tolerance of the integration routine at the start of the evolution.

\subsubsection{Shortcomings of our accreting models}
\label{sec:short_acc_model}
Starting from a small protostellar seed of $0.01\,M_\odot$ poses a problem to \texttt{MESA} which manifests itself in convergence issues. Hence, our evolutionary calculations need multiple retries and backups as the solver runs into problems. In addition, we are not able to perform a convergence study due to problems arising from smaller time steps and more zones. A typical protostellar evolution takes around $\sim9000$ time steps, while the amount of zones rises from $\sim3000$ in the initial model to $\sim9000$ at the end of the evolution. 

We are unable to produce accreting evolutionary tracks for any choice of input physics and further discuss this for the accretion rate. High accretion rates onto our $0.01\,M_\odot$ starting model will not converge using a  $T{-}\tau$ atmosphere. However, starting from higher mass seeds allows to explore accretion rates up to $10^{-2}\,M_\odot/{\rm yr}$ for integrated atmospheric boundary conditions. Low accretion rates pose a different challenge. In this case, the calculation of the atmospheric boundary conditions fails due to problems with the opacity tables and the equation of state. This might be overcome by using  \textit{tau\_100} tables at the start of the evolution, and jumping to the $T{-}\tau$ once a higher mass is reached (i.e. $0.04\,M_\odot$). While the influence of this approach on the accreting evolutionary track on the Hertzsprung-Russel or Kiel diagram is marginal, it influences seismic properties and, hence, is not an option for this study. However, we use the \textit{tau\_100} tables to show results with different values of accretion rate. A different solution would be the use of different opacity tables at lower temperatures such as tables based on \citet{Freedman2008}. However, we prefer the other opacity tables especially for the calculation of higher mass stars and thus decline this option.

In a similar manner, other input physics lead to convergence issues. These include for example very low values of $\beta$, $M_{\rm outer}$ and the mixing length.

Emphasis has to lie on the physical interpretation of the evolutionary models calculated in this work. As discussed in Section \ref{subsec:treatment_of_accretion}, using a constant accretion rate is most probably not physical. But overall we do not aim to model individual stellar parameters but rather find an envelope to the combined stellar parameters of our sample.

In addition, multiple assumptions have been made concerning the treatment of accretion. Currently, we miss the full physical description of how the energy of the accreted material is deposited inside the star. We expect that different descriptions will influence the outcome of our calculations to some degree. 

%--------------------------------------------------------------------
\subsection{Stellar Pulsations}
\label{subsec:stellar_pulsations}
We use version 5.2 of the stellar oscillation code \texttt{GYRE} \citep{Townsend2013, Townsend2018} to calculate non{-}adiabatic pulsations of our stellar models. We refer to the above mentioned instrument papers for details on the numerical methods. 

We calculate dipole modes (i.e. harmonic degree $l=1$) with azimuthal order $m=0$ for g-modes in the range $n = -1, \cdots, -50$ and p-modes in the range $n = 1, \cdots, 50$. These have to be calculated separately as they call for different frequency scan parameters. That is, we use frequency points distributed uniformly in inverse frequency for g-modes and distributed uniformly in frequency for p-modes. We look for frequencies in the range $0.05 {-} 10$~\cd\, for g-modes and $2 {-} 100$~\cd\, for p-modes. We choose regularity{-}enforcing inner boundary conditions and outer boundary conditions according to the formulation following \citet{Dziembowski1971}.

The time dependence of perturbations is assumed to be proportional to $\exp(-i\sigma t)$, where $t$ is the time, $i$ the imaginary unit, and $\sigma$ the eigenfrequency. In non{-}adiabatic calculations, $\sigma$ is imaginary. Hence, the real part $\sigma_{\rm R}$ is associated with the period of oscillation
\begin{align}
\Pi = \frac{2\pi}{\sigma_{\rm R}}    
\end{align}
and the imaginary part $\sigma_{\rm I}$ is associated with the growth rate
\begin{align}
    \tau = \frac{1}{\sigma_{\rm I}}.
\end{align}
Given an eigenmode, the sign of the growth decides whether the mode is excited ($\tau > 0$) or stable  ($\tau < 0$). In other words, an eigenmode is unstable whenever the work, $W$, of this mode is positive. The work is given by
\begin{align}
    W = 4 \pi \frac{\sigma_{\rm I}}{\sigma_{\rm R}} E,
\end{align}
and describes the amount of energy change in the system performed by one cycle \citep{Unno1989}. Here $E$ is the total energy of oscillation. It is often convenient to explore the differential work ${\rm d}W/{\rm d}x$ such that
\begin{align}
    W = \int_0^1\frac{{\rm d}W}{{\rm d}x}{\rm d}x,
\end{align}
where $x = r/R$ is the coordinate of fractional radius. Hence, the differential work describes the driving and damping throughout the star. The normalised growth rate
\begin{align}
    \eta = \frac{\int_0^1\frac{{\rm d}W}{{\rm d}x}{\rm d}x}{\int_0^1\left \vert \frac{{\rm d}W}{{\rm d}x}\right \vert{\rm d}x}
\end{align}
as proposed by \citet{Stellingwerf1978} gives viable hints to the power of the driving. For $\eta = 1$ the entire stellar envelope contributes to the driving of the oscillation, while the envelope is entirely damping for $\eta = -1$. Typical values for excited modes in main-sequence SBP stars are values up to $\sim 0.3$ \cite[see e.g.][]{Townsend2018}.

\subsubsection{Shortcomings of our pulsation analysis}
The description of convection is a known problem for one-dimensional stellar evolution codes \citep[see e.g.][]{Aerts2019}. The mixing length theory, first proposed by \citet{bohm1958}, uses averages of equations over lengths exceeding the typical travel length of convective motions \citep{Aerts2019}. Hence, this description misses the time-dependence of stellar convection. As a consequence, we are missing the influence of time dependent convection on the pulsation properties. \citet{Dupret2005} showed that this time dependence is needed to produce the red edge of the p-mode instability strip. Similarly, this influences the obtained growth rates for $\gamma$ Doradus stars. \texttt{GYRE} allows to freeze convective heating altogether \citep[case 1 of][]{Pesnell1990} or freeze the Lagrangian perturbation of convective luminosity \citep[case 4 of][]{Pesnell1990}. We explore the effect of choice in our results. Similarly, the linear nature of the pulsation equation prevents accurate calculation of the red edge of the pre-main sequence p-mode instability strip as has been shown by \cite{Marconi1998}. Non-linear analysis and time dependent convection for pre-main sequence stars are hence suspect for future projects.

\begin{figure}
   \centering
   \includegraphics[width=\linewidth]{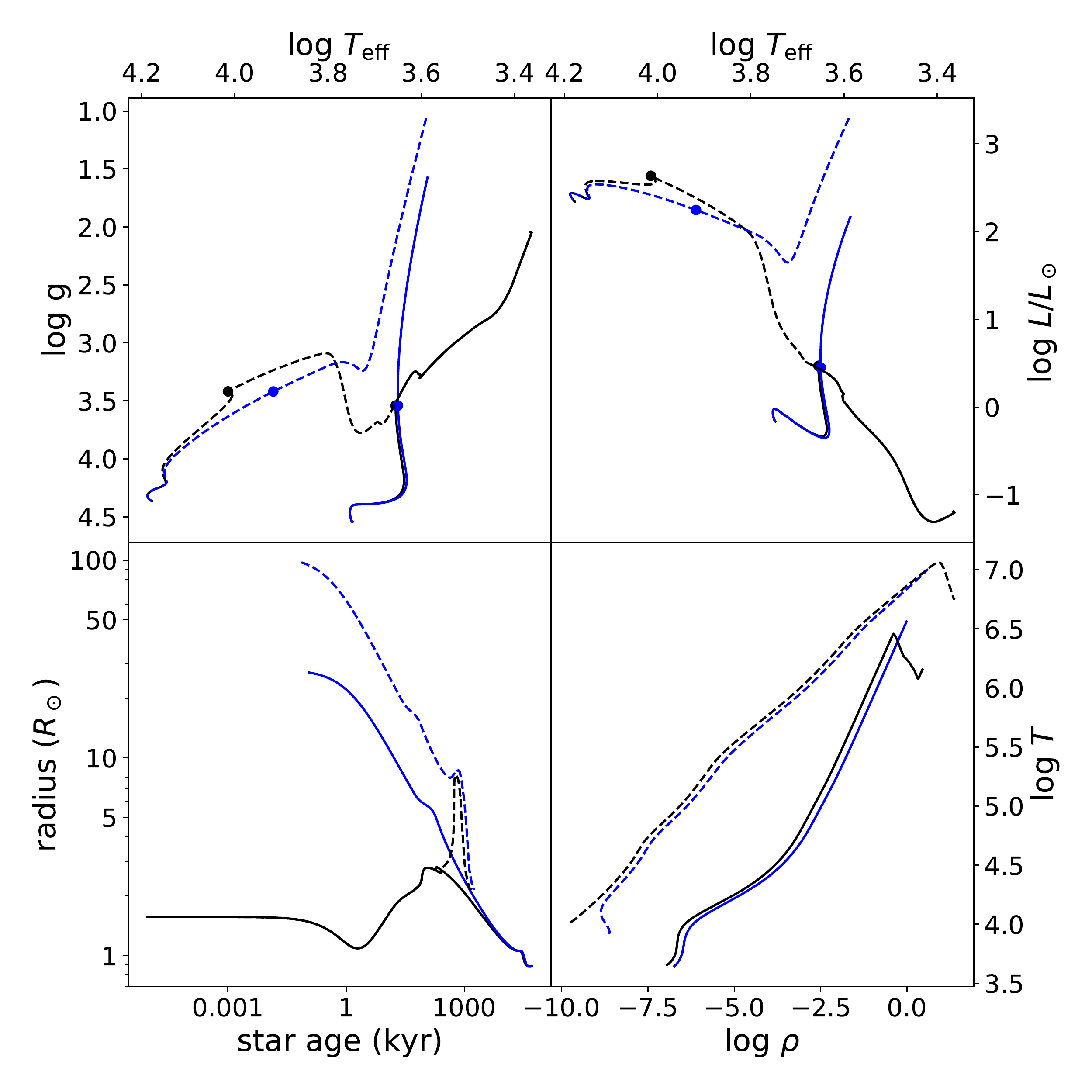}
      \caption{Stellar models from accreting protostars compared to classical models. Black lines show the models originating from the accreting initial model, while blue lines show the evolution from classical initial models. The full lines correspond to a one solar mass model while the dashed lines correspond to a four solar mass model. Top left panel: Kiel diagram. Top right panel: Hertzprung-Russel diagram, where $L$ is the photospheric luminosity. Bottom left panel: time evolution of the stellar radius. Bottom right: temperature -- density (cgs units) profiles of selected models which are marked with a circle in the Kiel and Hertzsprung-Russel diagrams. }
         \label{fig:models_from_protostars}
\end{figure}
%--------------------------------------------------------------------
\begin{figure*}
   \centering
   \includegraphics[width=\textwidth]{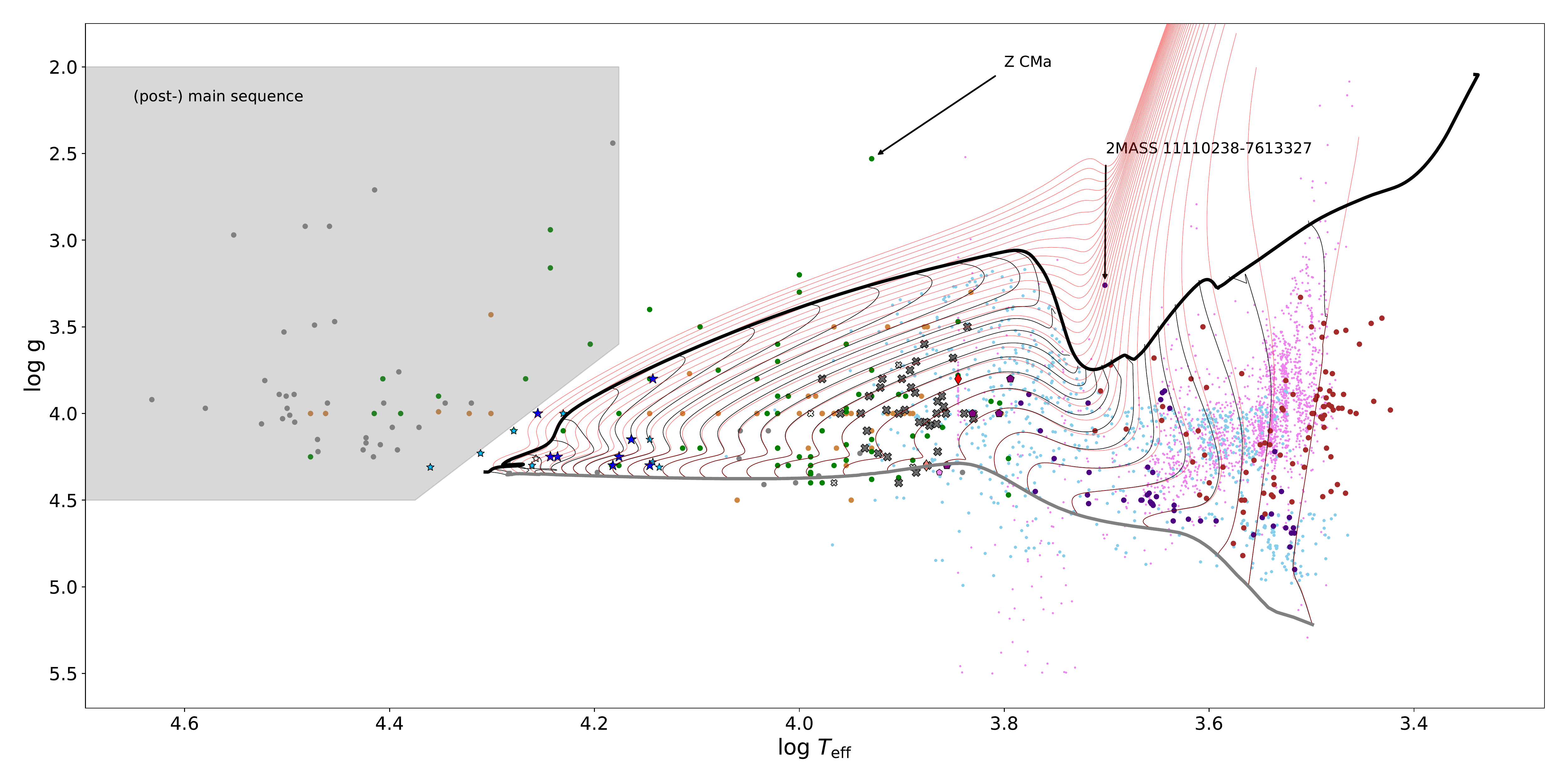}
      \caption{Comparison of the accreting evolutionary tracks from protostars to the classical pre-main sequence evolution. The thick black line shows our accreting standard model and the thin black line the pre-main sequence models evolved from this accreting model. Red lines show classical pre-man sequence tracks with the same masses at the ZAMS as the pre-main sequence models from the accretion protostar The ZAMS is shown as grey line. The grey shaded area at higher temperatures marks the position of (post-) main sequence stars according to our interpretation. The colour code of the stars is the same as in Fig. \ref{fig:sample}.
              }
         \label{fig:standard_model}
\end{figure*}

\begin{figure*}
   \centering
   \includegraphics[width=\linewidth]{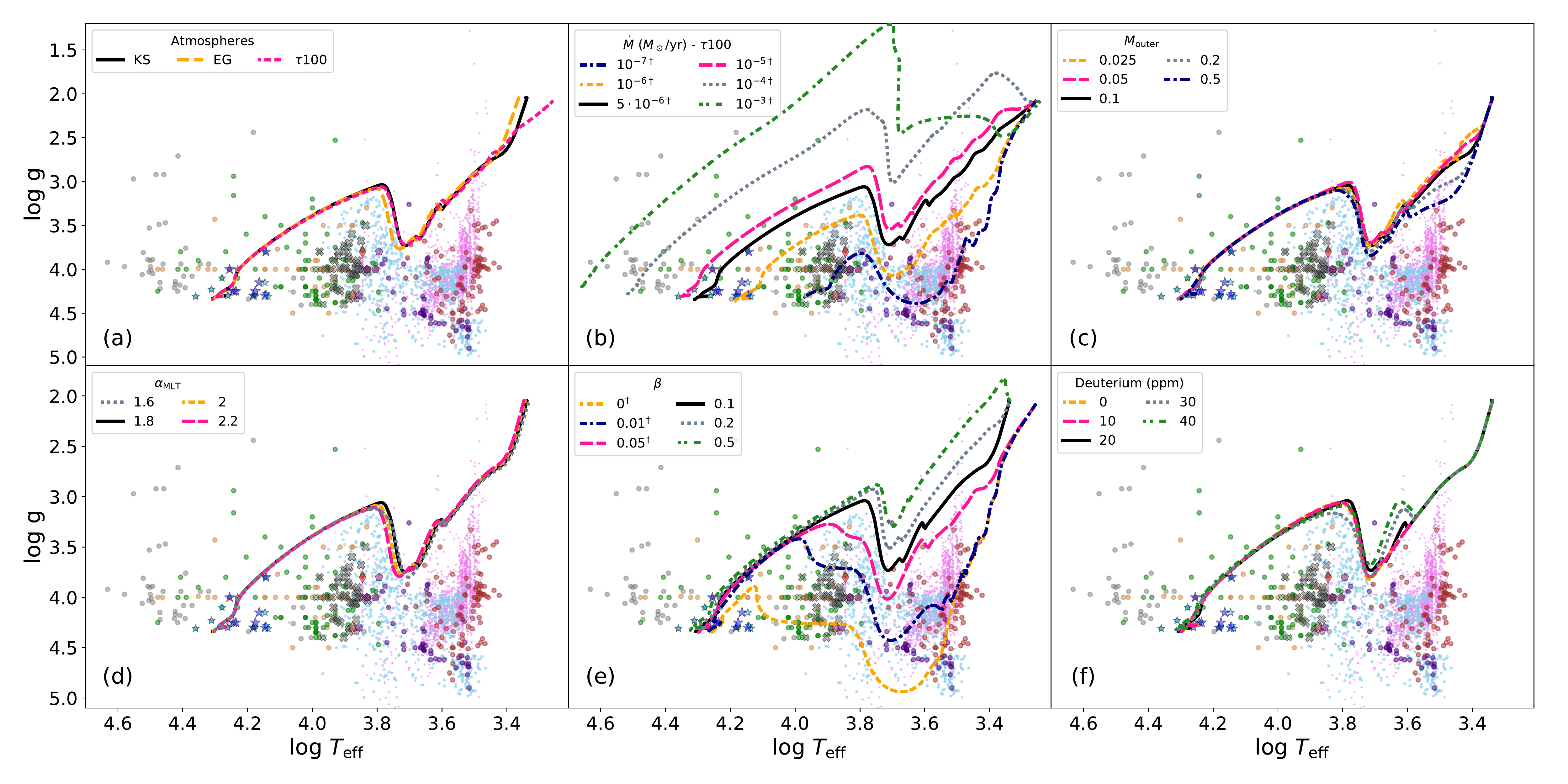}
      \caption{Influence of chosen input physics on the accreting evolutionary track. (a) Different atmospheric boundary conditions. (b) Different constant accretion rates. (c) Different regions of heat injection(d) Different values for the mixing length. (e) Different values for $\beta$, controlling the amount of energy injected into the star. (f) Different initial abundances of Deuterium. Our standard model is shown as black lines in all panels. Lines appended with an dagger in panels (b) and (e) are calculated using the \textit{tau\_100} instead of Krishna-Swamy atmosphere. The colour code of the stars is the same as in Fig. \ref{fig:sample}.}
         \label{fig:input_physics}
\end{figure*}
\section{Results}
\label{sec:results}

\begin{figure*}
   \centering
   \includegraphics[width=\linewidth]{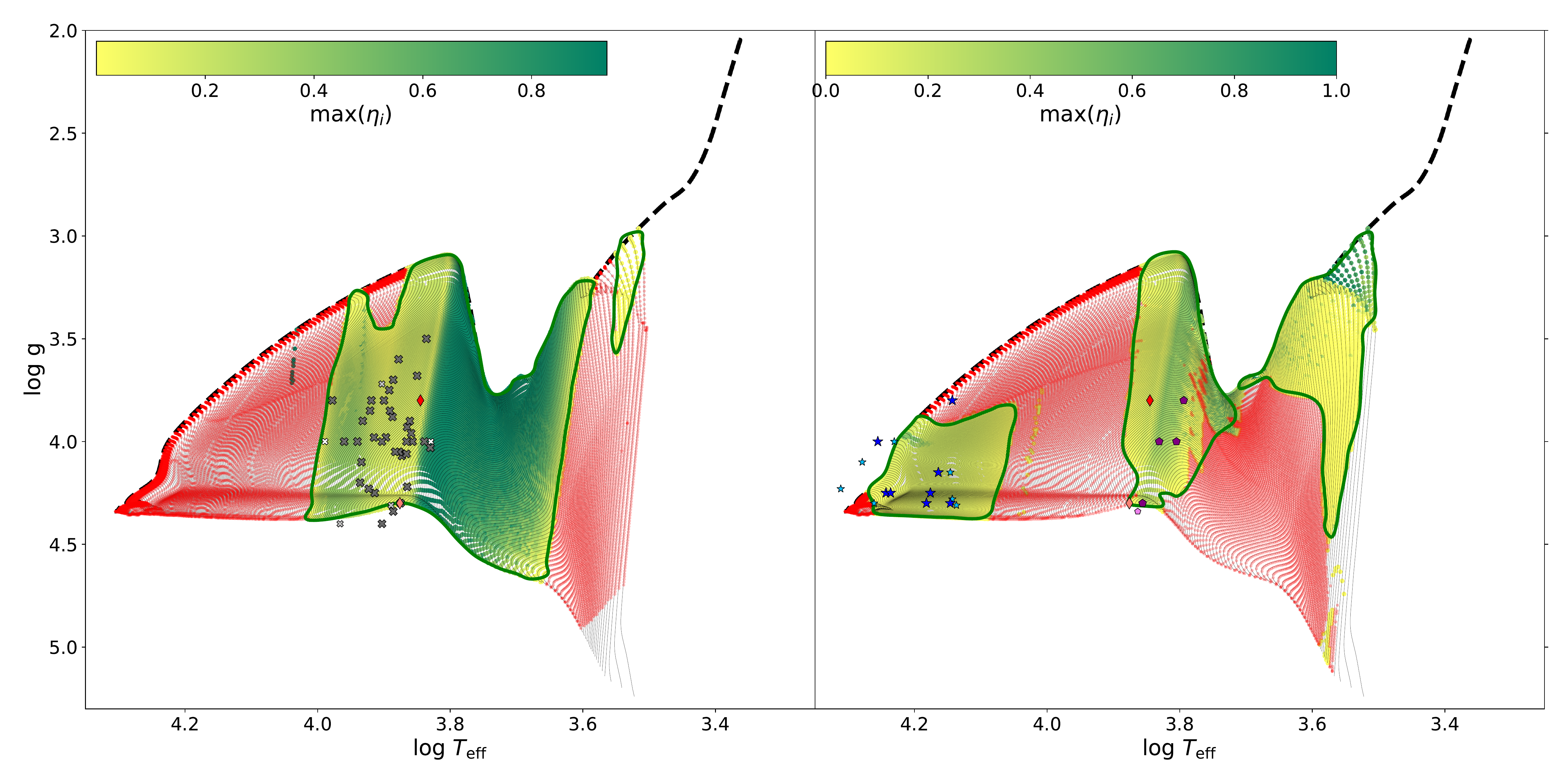}
      \caption{Pre-main sequence instability regions from normalised growth rates. The black dotted line is our final accretion model. Thin black lines show the pre-main seqeunce tracks for different masses from the accreting evolutationary track to the ZAMS. The colour code for green dots shows the maximum of the normalised growth rate of a given model and hence the instability strips. Models without unstable modes are shown as red dots. The green line follows the borders of the instability region resulting from our calculations. Left panel: p-modes. Right panel: g-modes. The colour code of the shown pulsators is the same as in Fig. \ref{fig:sample}.}
         \label{fig:instab_main}
\end{figure*}
\begin{figure}
   \centering
   \includegraphics[width=\linewidth]{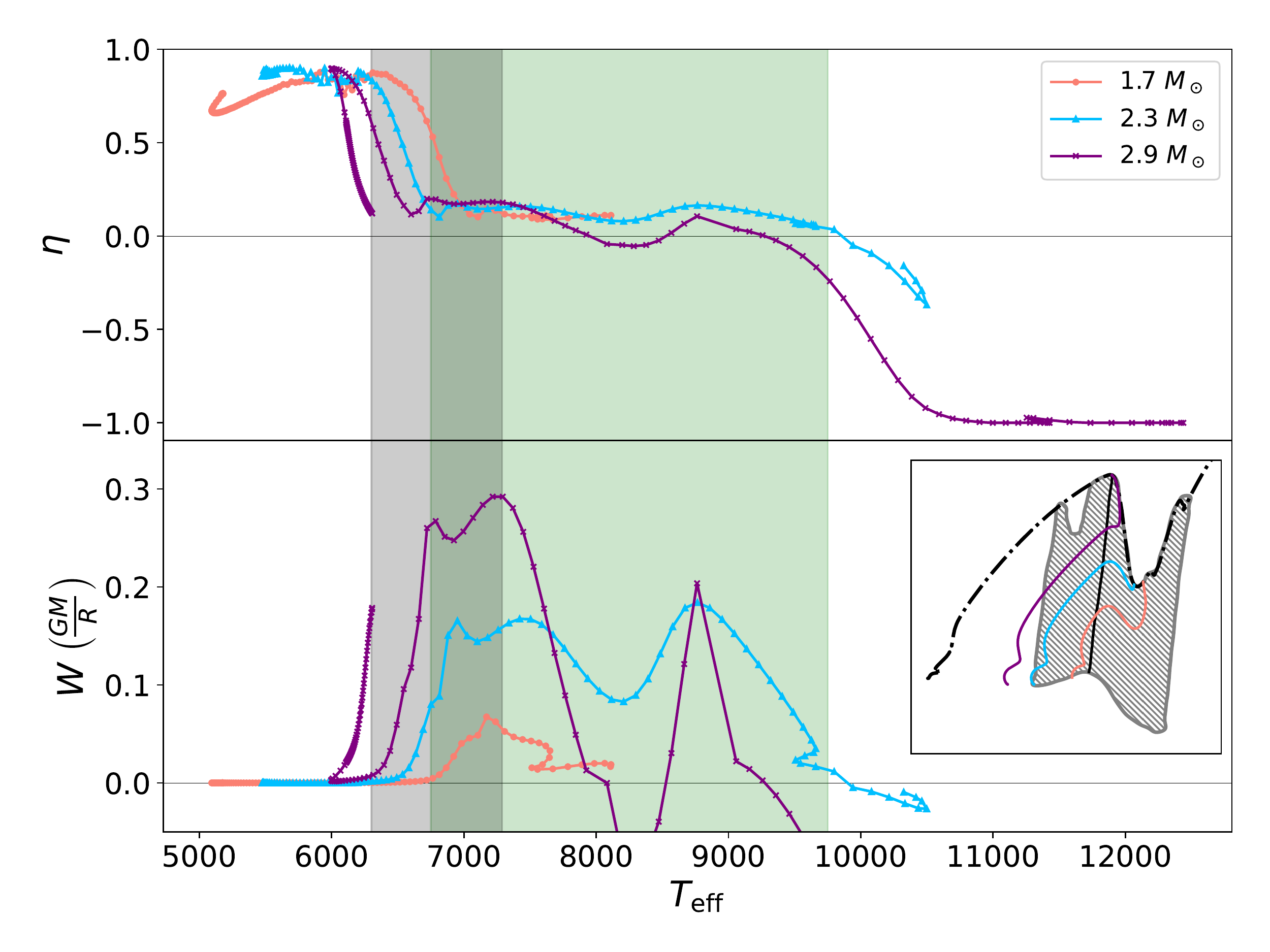}
      \caption{Pulsation parameters for the pressure mode of radial order $n= 4$ for selected models. Upper panel: normalised growth rate. Lower panel: work performed by one pulsation cycle. The inset in the lower panel shows the Kiel diagram of the three corresponding evolutionary tracks (coloured lines) and the accreting evolutionary track (black dash dotted line). Also shown is the p-mode instability region for any modes (gray area) and the boundary for high $\eta$ as in Fig. \ref{fig:instab_main}. The upper and lower panel also include the temperature ranges of the boundary (grey area) and the effective temperatures of observed pre-main sequence stars (green area) which overlap.
              }
         \label{fig:eta_work}
\end{figure}
Before calculating instability regions for the subsequent pre-main sequence evolution, it is important to try and constrain the input physics for the accreting protostellar evolutionary track (see Section \ref{subsec:treatment_of_accretion}) . We aim to find a set of input physics that results in an envelope (between the accreting evolutionary track and the ZAMS) which includes most pre-main sequence stars in our sample. We find that the best values are given by an accretion rate of  $5\times 10^{-6}\,M_\odot/{\rm yr}$ and $\beta = 0.1$. Additional parameters are only constrained marginally, as described in detail in Sections \ref{subsec:result_tracks} and \ref{subsec:result_tracks2}.

The resulting instability strips provide constraints for the atmospheric boundary conditions of pre-main sequence stars. We find that the Krishna-Swamy atmosphere does not reproduce the observed spectroscopic parameters for pre-main sequence $\delta$ Scuti stars while the Eddington-Grey atmosphere works successfully (see Sections \ref{sec:res:puls} and \ref{sec:res:puls2}). 

\subsection{Stellar models from accreting protostars}
The evolution of a star beyond the main sequence is primarily controlled by the stellar mass at which it arrives at the ZAMS. However, the evolution prior to the main sequence is inherently dependent on how the stellar model aquires said mass. The classical pre-main sequence evolution assumes an (unrealistic) initial model that has already obtained the mass of the main sequence star. In contrast, accreting evolutionary tracks start as small stellar seeds and obtain their final ZAMS mass by means of accretion. In this section, we shortly summarise the differences in the pre-main sequence evolutionary tracks for two distinct models that will, at some point, arrive at the ZAMS with the same stellar mass.

Figure \ref{fig:models_from_protostars} compares the evolution of a one and a four solar mass model for both, a classical pre-main sequence evolution and the evolution from an accreting protostar. The initial accreting evolutionary track for both masses is the same for the latter (black lines in Fig. \ref{fig:models_from_protostars}) since they both originate from the same stellar seed. From the evolution of the radius, it is evident that these models never reach the huge radii of the classical initial models. Once they have reached their final mass, they leave the accreting evolutionary track and continue their subsequent evolution towards the ZAMS. At this time, the position in the Kiel- and Hertzsprung-Russel diagram of the one solar mass model is already similar to the classical model, while the stellar parameter of the four solar mass models converge only later.

Figure \ref{fig:models_from_protostars} also illustrates the internal structure of the stellar models. We searched for the \texttt{MESA} timestep where the surface gravity of the classical evolution models reaches the value of the models from accreting protostars after their accretion has stopped (black and blue dots in the top panels of Fig. \ref{fig:models_from_protostars}). The models from accreting protostars show a temperature inversion between the radiative core and the convective outer regions in the high density region towards the centre of the star. This is due to the heating of the stellar atmosphere from hot accretion \citep[see e.g.][]{Stahler1980, Jensen2018}.

In the following we will refer to the evolution from the initial stellar seed with constant mass accretion as accreting evolutionary track. The term pre-main sequence tracks is used for the subsequent evolution towards the ZAMS for models of different masses after the accretion has stopped.

\subsection{Enveloping our stellar sample}
\label{subsec:result_tracks}
We have calculated multiple accreting evolutionary tracks for different choices of input physics. The constraints from spectroscopic parameters of our sample are sufficient to limit some free parameters of the accretion picture adopted by us (see Section \ref{subsec:treatment_of_accretion}). 

We have chosen the following input physics as our standard model: an accretion rate $\Dot{M}= 5\times 10^{-6}\,M_\odot/{\rm yr}$, fractional mass of the outer region $M_{\rm outer} = 0.1$, mixing length parameter $\alpha_{\rm MLT}$, and  factor $\beta = 0.1$ with Krishna-Swamy atmospheres and initial Deuterium abundance of $20$~ppm. The accreting evolutionary track of this standard model is shown in Figure \ref{fig:standard_model}, together with the resulting pre-main sequence tracks and classical pre-main sequence evolution. The accreting evolutionary track, together with the ZAMS, beautifully envelopes our stellar sample for lower temperatures and explains the `missing' stars in the area between $\log(T_{\rm eff}) = 3.6 - 3.8$ and $\log(g) = 3.2 - 3.75$.

However, this model does not envelop our complete stellar sample. Most notably, it reaches the ZAMS at approximately $6\, M_\odot$ and cuts out multiple stars with effective temperatures higher than $15000$~K. This is not surprising since higher effective temperatures correspond to higher stellar mass. This in turn leads to shorter evolutionary timescales. Stars with O spectral types only have pre-main sequence lifetimes of tens to hundreds of thousand years according to classical evolution models, making it increasingly difficult to observe such stars in their pre-main sequence phase. At this timescale, they might still be embedded in their birth cloud when the stellar core already burns hydrogen in equilibrium. Furthermore, our calculations do not include input physics that would put them within the envelope of our accreting evolutionary track without straying far away from well fitting envelope for lower temperatures. Therefore, we believe that most of these stars have arrived on the main sequence, or are even already in the post-main sequence phase of their evolution.

A few more stars do not lie within the region enveloped by our accreting model and the ZAMS. Some stars of the sample by \citet{Fairlamb2015} lie just above the accretion model. This might be due to multiple reasons. While they might be, similar to the hotter stars, in their post main sequence phases, this is less probable at the respective temperture. 
A more logical explanation might be that they undergo phases of strong accretion. As our model assumes constant accretion rates, excursions from this accreting evolutionary track are expected due to episodic accretion \citep{Elbakyan2019}. Given the uncertainties of the spectroscopic parameters (see Fig. \ref{fig:sample}) it is also  possible that some even lie within the envelope.

A few low temperature stars of the sample by \citet[pink symbols in Figure \ref{fig:standard_model}]{Da2016} seem to have surface gravities above $3$. We suspect that these are either red giant branch stars or outliers in the analysis by \citet{Da2016}.

Two stars need special attention. The first one, Z CMa is part of the sample by \citet{Fairlamb2015}. With $T_{\rm eff} = 8500 \pm 500$ and $\log(g) = 2.53 \pm 0.17$ it lies far off our accreting model. This star is a well known FU Orionis object \citep{Hartmann1996}, meaning it undergoes very strong accretion, leading to high luminosities. This leads to heating of the stellar envelope and increase in stellar radius \citep{Elbakyan2019}. This explains that the surface gravity is very low and the star does not fit within our envelope.The second one, 2MASS 11110238-7613327 is part of the sample by \citet{gutierrrez2020} as a member of Cha I. With $T_{\rm eff} = 5032 \pm 56$ and $\log(g) = 3.26 \pm 0.09)$ it lies in the area of the Kiel diagram where the accreting evolutionary track takes a dip in surface gravity. There is no literature available for this star and the TESS lightcurve does not show any variability. At this stage we are unable to explain the reason for the star's location in the Kiel diagram and we strongly encourage follow up observations.

The samples by \citet{Da2016} and \citet{Vioque2020} show some stars below the zero age main sequence. These might be dwarfs, but their existence does not influence our results for the accreting models.

\subsection{Constraining the input physics with spectroscopic parameters}
\label{subsec:result_tracks2}
We calculated models that differ from the standard model by one parameter at a time. From the discussion in Section \ref{subsec:result_tracks} it is clear that a significant portion of the stars in our sample of pre-main sequence and early ZAMS stars might be in later evolutionary stages even tough they are of comparably young age (several million years old) because of the short evolutionary time scales of high mass stars. Without a distinction between these stars and genuine pre-main sequence stars, it is inconvenient to define a likelihood or merit function and statistically infer a best fitting model. As discussed in Section \ref{sec:short_acc_model}, the accretion history is most probably unique for each individual star. Furthermore, more massive stars have most likely undergone an early accretion phase with higher mass accretion rates than their lower mass counterparts. Hence, we rely on a subjective choice of the models, whenever the difference in the track is big enough to do so, without investigating a two dimensional parameter space and hence ignoring possible degeneracy. While different combinations may yield similar results (e.g. higher mass accretion rate and lower $\beta$-value) the values for our standard model are chosen to reflect what is typically used in the literature. Estimates for the mass accretion rates for molecular clouds temperatures of $\sim30$\,K lead to a mass accretion rate of $\sim10^{-5}\,M_\odot/{\rm yr}$ \citep{Kunitomo2017}. As such we are inclined to stay near these estimates instead of straying far from them. The results are shown in Figure \ref{fig:input_physics} and discussed in the following in detail. 

\subsubsection{Atmospheric boundary conditions}
Panel (a) shows the influence of the choice for atmospheric boundary conditions. We find that this choice only marginally changes the position of the accreting evolutionary track and, hence, the observationally determined atmospheric parameters cannot be used to constrain them. At the same time, this allows us to use \textit{tau\_100} table atmospheres to calculate evolutionary tracks for those cases that would otherwise lead to strong convergence problems.

\subsubsection{Mass accretion rate}
Panel (b) shows the influence of the mass accretion rate. It is evident, that the lower accretion rates of $\Dot{M}= 10^{-6}$ and $10^{-7}\,M_\odot/{\rm yr}$ cut in our stellar sample. In these cases, a lot of stars would need to undergo phases of variable accretion rates to explain their positions above the respective accreting evolutionary tracks.
For the very high mass acretion rates of $\Dot{M}= 10^{-3}$ and $10^{-4}\,M_\odot/{\rm yr}$, the tracks cut the ZAMS at significantly higher masses. This would lead to the O-type stars in our sample still being within the area enveloped between the accreting evolutionary track and the ZAMS. For these accretion rates, the increase in stellar radius leads to lower surface gravities. As a consequence, these accreting evolutionary tracks fail to describe the location of lower mass stars at temperatures below $16 000$~K.
Finally, the location of the track with $\Dot{M}=  10^{-5}\,M_\odot/{\rm yr}$ is comparably close to the one with an accretion rate of $\Dot{M}=5\times 10^{-6}\,M_\odot/{\rm yr}$ which we adopted for our standard model. Comparison of these tracks in the region below $10000$ K shows that the track with an accretion rate of $\Dot{M}=5\times 10^{-6}\,M_\odot/{\rm yr}$ follows the observations more closely and is to be prefered in our analysis.

\subsubsection{Fractional mass of heat injection}
In the description of accretion adopted in this work, $M_{\rm outer}$ controls the position up until the extra heat from the accreted material is injected. Hence, for higher $M_{\rm outer}$, larger parts of the stars are being heated. Intuitively, one would expect the accreting material to be able to penetrate only the very outer parts of the star. Panel (c) shows the influence of different choices of $M_{\rm outer}$ onto the accreting evolutionary tracks which is largely confined to very early stages. We find that the descriptions using $M_{\rm outer} \geq 0.2$ may be the only ones that might be considered unphysical as cannot explain a small part of the sample. Additionally we do not expect the accreted material to penetrate that far into the star. Lower values of $M_{\rm outer}$ lead to marginal differences that cannot be constrained with our spectroscopic sample.

\subsubsection{Mixing length of convection}
Panel (d) shows the influence of different mixing lengths. Mixing lengths below $\leq 1.5$ lead to convergence issues and are hence omitted from this study. The influence of the mixing length on the accreting evolutionary track is marginal and therefore we can not employ our sample to constrain it. However, we can observe that higher mixing lengths tend to cut into the sample obtained from \citet{Vioque2020}.

\subsubsection{Fraction of injected energy from accreted material}
Panel (e) shows the influence of the parameter $\beta$ that controls how much of the accreted material's energy is injected into the star or radiated away. The case of $\beta = 0$ corresponds to cold accretion and definitely fails to explain our sample as it cuts through it at very high surface gravities. The story is similar for the small values $\beta = 0.01$ and $0.05$. Values higher than the value adopted by us give accretion evolutionary tracks that have too low surface gravities for small effective temperatures. 

\subsubsection{Deuterium abundance}
Panel (e) shows the influence of the mass fraction of Deuterium in the initial composition of the stellar seed and the accreted material. Given the description of accretion adopted for the standard model, the Deuterium abundance only marginally influences the accreting evolutionary track. Most notably, increasing the Deuterium abundance leads to a smaller surface gravity after Deuterium ignition at approximately $\log T_{\rm eff} = 3.6$. Hence, we cannot constrain the Deuterium abundance with our stellar sample.
  \begin{table}
    \caption[]{Comparison of the temperature regions to previously calculated pre-main sequence instability strips.}
    \label{tab:instab_compare}
    \tabcolsep=0.1cm
    \begin{tabular*}{\linewidth}{llllll}
        \hline
        \noalign{\smallskip}
        & MP98  &  G06\ & B11  & G12 & this work \\
        \noalign{\smallskip}
        \hline
        \noalign{\smallskip}
         type  & non-linear & linear & linear & linear & linear \\
         note & - & ${\rm TDC}^a$ & TDC & - & ${\rm ASE}^b$ \\
         radial order & 1-3 & 1-7 & $1$-$60^c$ & ? & 1-50  \\
         \multirow{2}{*}{$\delta$ Sct}  & $6500-$ & $6450-$& \multirow{2}{*}{-} &\multirow{2}{*}{-} & $6300^d-$  \\
         & $7500$~K & $8400$~K &    &  & $10300$~${\rm K}$ \\
         \multirow{2}{*}{$\gamma$ Dor}  & \multirow{2}{*}{-} & \multirow{2}{*}{-} & $6550-$ &\multirow{2}{*}{-} & $5200-$  \\
           &  &  & $7400$~K &  & $7650$~${\rm K^e}$ \\
         \multirow{2}{*}{SPB}  & \multirow{2}{*}{-} & \multirow{2}{*}{-} & \multirow{2}{*}{-} &  $10500-$ & $11100-$  \\
           &  &  &  & $19200$~K & $18700$~K  \\

        \noalign{\smallskip}
        \hline

    \end{tabular*}    
    \tablefoot{
    Sources: MP98: \citet{Marconi1998}, G06: \citet{Grigahcene2006},B11: \cite{Bouabid2011} ,  G12: \citet{Gruber2012}
    ($a$) time dependent convection, ($b$) accretion from protostellar seed, ($c$) assumed from results, ($d$)the red edge of the p-mode instability region is taken to be the boundary to high normalised growth rates as discussed in the text. (e) the difference at the blue edge is an accretion effect for modes with low radial order only. For radial orders $\geq 15$, we obtain the temperature range $6450-7550$~K.
    }
 \end{table}
\begin{figure}
   \centering
   \includegraphics[width=\linewidth]{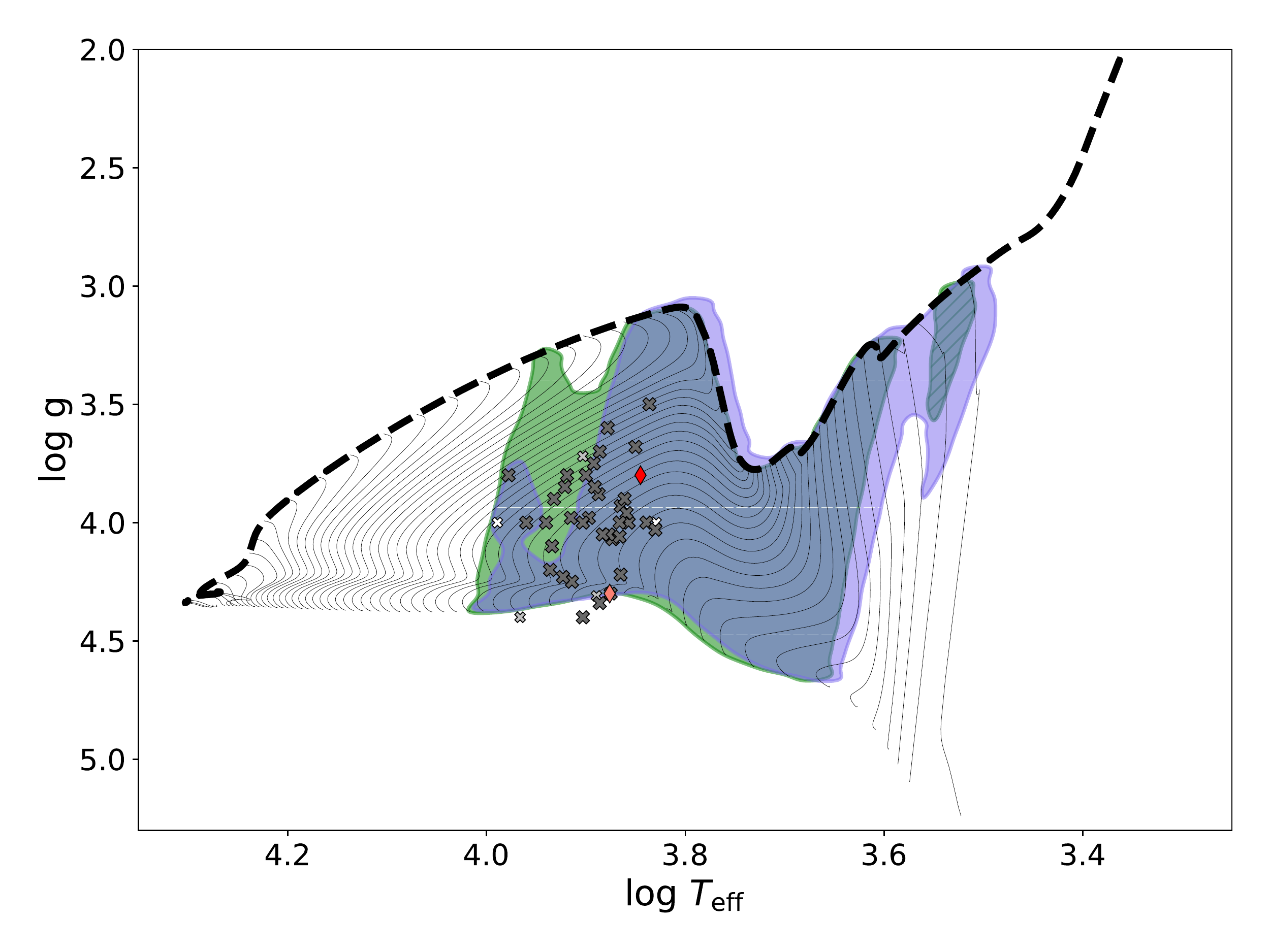}
      \caption{p-mode instability strips for different atmospheres. The green shaded area shows the result for Eddington Gray atmosphere, while the blue shaded area shows the result for the Krishna-Swamy atmosphere. The colour code of the shown pulsators is the same as in Fig. \ref{fig:sample}.
              }
         \label{fig:atmospheres}
\end{figure}

\subsection{Exploring the instability regions for pre-main sequence stars}
\label{sec:res:puls}
The calculations performed in this paper include the growth rates for dipole modes across the Kiel diagram. Figure \ref{fig:instab_main} shows the resulting instability strips as the area where any mode is unstable. We discuss the implications of these results, starting with the p-mode instability region (left panel in Fig. \ref{fig:instab_main}). 

All p-mode pulsating pre-main sequence $\delta$ Scuti stars within our sample lie within the calculatedinstability region. As expected, the red edge of the instability region does not agree to the observational red edge as implied by our sample. The non-inclusion of time dependent convection leads to unstable modes in stars with effective temperatures way below $6500$ K. The normalised growth rates indicate that the driving is very strong in this part of the instability region. Figure \ref{fig:eta_work} shows that the high values for the normalised growth rate come with low values of actual work performed by one pulsation cycle. Furthermore, the change towards higher values for the work is abrupt, leading to a clear boundary in Figure \ref{fig:instab_main}. The temperature region of this boundary overlaps with observed effective temperatures of pre-main sequence $\delta$ Scuti stars in our sample. Hence, we propose to interpret only the hotter part (higher temperatures than the boundary) as the $\delta$ Scuti instability region. The remaining temperature region is similar to the one obtained by \citet{Dupret2005} and, hence, the unstable modes towards lower temperatures can be attributed to missing stabilisation effects from convection.

For g-modes, we obtain three distinct instability regions with our calculations (right panel of Fig. \ref{fig:instab_main}). With the exception of GSC 00154-01871, all SPB stars and candidates from our sample (that fall within our pre-main sequence tracks) also lie in the SPB instability region. GSC 00154-01871 is a spectroscopic binary and the real physical parameters of the pulsating star may differ from the adopted ones. \citet{handler2011} obtained $\log g = 4.4 \pm 0.2$ and $T_{\rm eff} = 14300 \pm 600$ in contrast to the adopted values from \citet{Gruber2012} which are $\log(g) = 3.80 \pm 0.15$ and $T_{\rm eff} = 13900 \pm 350$. However, $\log(g)$ is primarily obtained from the H$\alpha$ line, which is strongly affected by the companion. We conclude that the position of GSC 00154-01871 in our Kiel diagram is most likely not correct. Hence, all SPB stars with well defined parameters lie within our calculated pre-main sequence instability region. 

Our results show that it is possible to obtain the $\gamma$ Doradus instability region even without time dependent convection. This can only be accomplished by employing case 4 of \citet{Pesnell1990} in the calculations with \texttt{GYRE}. Case 1 will return stable modes throughout this region of the Kiel diagram. Hence, we strongly recommend using the description of case 4 when calculating g-modes with \texttt{GYRE}. While the blue edge of our instability region is well defined, the red edge seems a bit problematic. The models in small regions seem to host stable modes only. We tried to increase the resolution of \texttt{GYRE}'s spatial grid but without success. We choose to draw the border to include the low temperature region ($3.7 \leq \log(T_{\rm eff}) \leq 3.75$) of this instability region as these are the early parts of the pre-main sequence $\gamma$ Doradus tracks. However, it is most likely that the instability in this region is due to the influence of accretion (see discussion below). The resulting instability region includes all $\gamma$ Doradus stars and  $\gamma$ Doradus -- $\delta$ Scuti hybrids. Two $\gamma$ Doradus candidates are just outside the instability region, which is not concerning given the uncertainties on their positions.

We find an additional instability region for $\log(T_{\rm eff}) \leq 3.7$. We are currently not aware of any known pulsators in this region of K- and M-type stars. Pulsation in these types of stars is expected \citep{Baran2011, rodriguez2012, rodriguez2014} and multiple projects look for observational discoveries \cite[see][for a review]{rodriguez2019}. We looked for possible pulsations in the TESS observations of K- and M-type stars and present our results in Section \ref{sec:mtypepuls}.

Table \ref{tab:instab_compare} compares the temperature ranges of our instability regions with the published instability regions of pre-main sequence stars by \citet{Marconi1998}, \citet{Grigahcene2006}, \citet{Bouabid2011}, and \citet{Gruber2012}. Besides the blue edge of the p-mode instability region, our calculations compare well with these previous results. Our sample includes five pre-main sequence $\delta$ Scuti stars and two candidates with effective temperatures above the blue edge of \citet{Grigahcene2006} at temperatures up to $8400$~K. Based on our sample, the observational blue edge for pre-main sequence stars lies around $8900$~K. In later evolutionary phases, $\delta$ Scuti stars are observed up to effective temperatures of $10000$~K, but much less frequently \citep{Murphy2019}. Given the size of our current sample, it is impossible to decide whether the red edge resulting from our calculations is supported by real stars. Only a future increase in the number of observed pre-main sequence p-mode pulsators can put this hypothesis to the test.

\begin{figure}
   \centering
   \includegraphics[width=\linewidth]{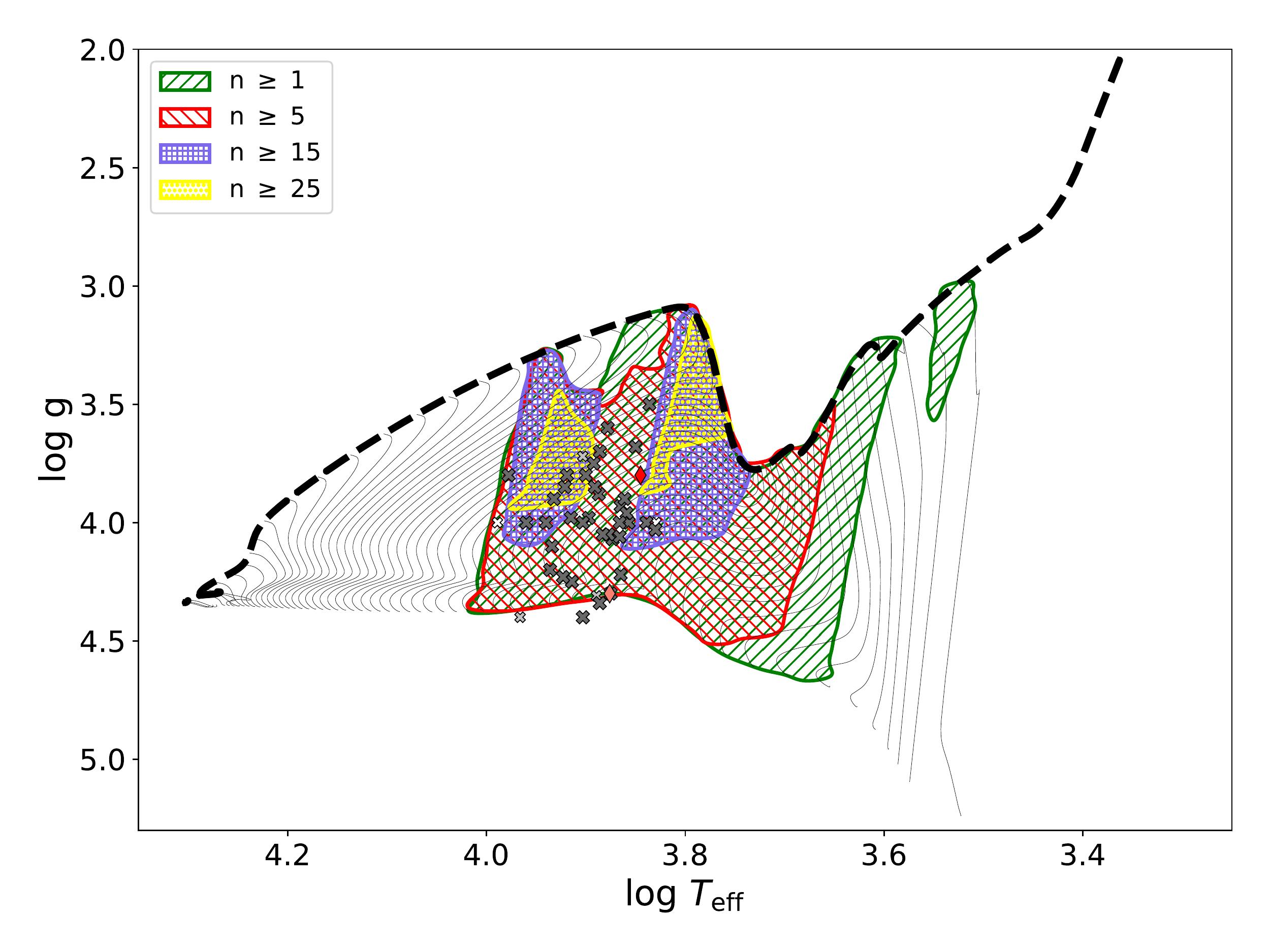}
      \caption{p-mode instability regions dependent on the excited orders. The black lines show the accreting evolutionary track (dashed) and the pre-main sequence evolution from this models (thin). Different coloured areas show the regions where modes with radial orders higher than $n\geq 1, 5, 15,$ and $25$ are unstable. The colour code of the shown pulsators is the same as in Fig. \ref{fig:sample}.
              }
         \label{fig:ho_pmodes}
\end{figure}
\begin{figure}
   \centering
   \includegraphics[width=\linewidth]{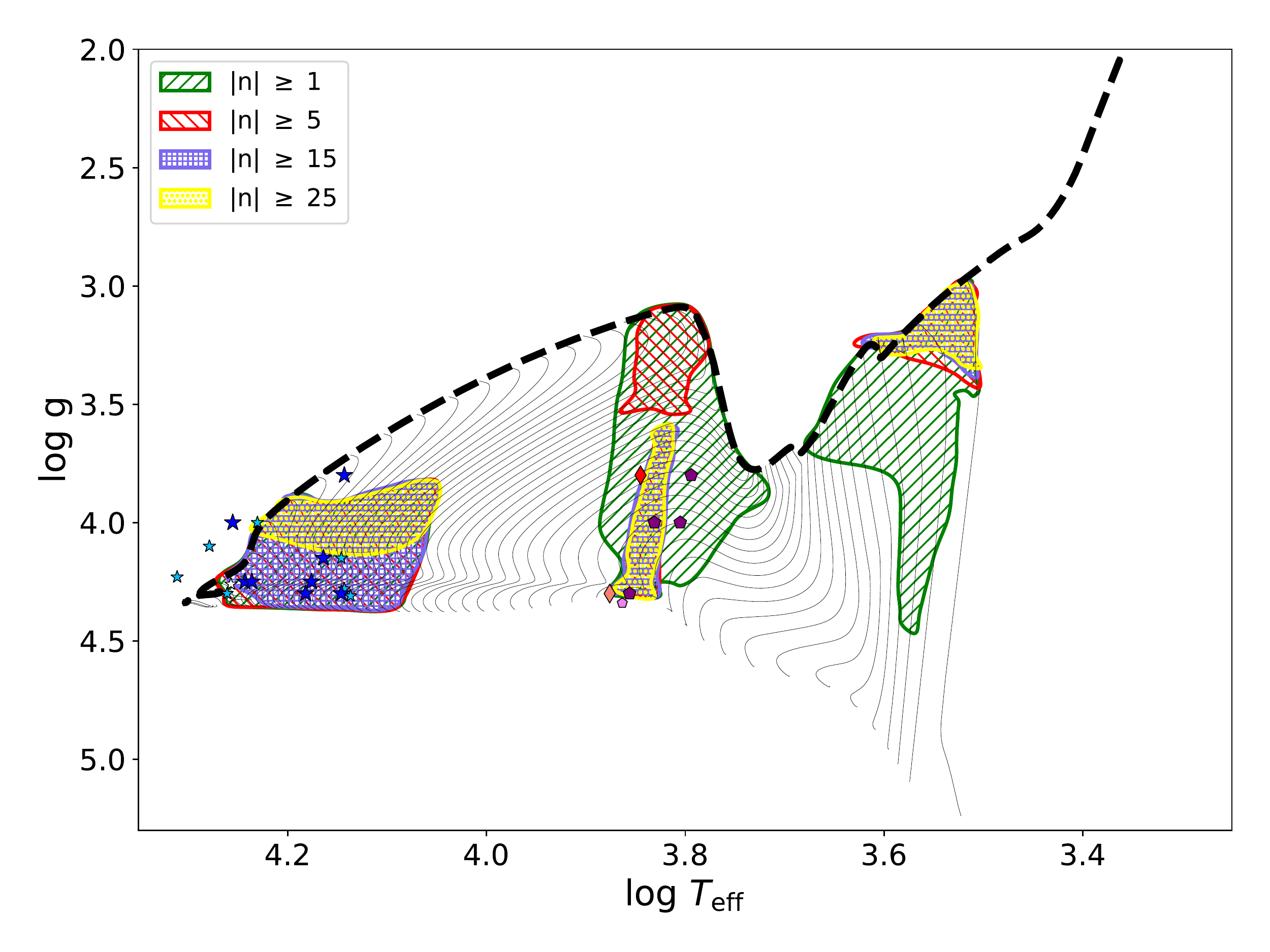}
      \caption{g-mode instability regions dependent on the excited orders. The black lines show the accreting evolutionary track (dashed) and the pre-main sequence evolution from this models (thin). Different coloured areas show the regions where modes with radial orders higher than $\vert n\vert \geq 1, 5, 15,$ and $25$ are unstable. The colour code of the shown pulsators is the same as in Fig. \ref{fig:sample}.
              }
         \label{fig:ho_gmodes}
\end{figure}
\subsection{Constraining the input physics with pre-main sequence pulsators}
\label{sec:res:puls2}
We calculated instability regions for models with different atmospheres and mixing length parameters. In addition, we compare the obtained results with instability regions obtained from classical models. The results are discussed in the following.

\subsubsection{Atmospheric boundary conditions}
Figure \ref{fig:atmospheres} shows the p-mode instability region for Eddington-Gray and Krishna-Swamy atmospheres. Models calculated with Krishna-Swamy atmospheres have no excited modes in a region of the Kiel diagram populated with known pre-main sequence $\delta$ Scuti stars. In comparison, the observed $\delta$ Scuti stars are fully embedded in the instability region resulting from models using Eddington Gray atmosphere. Hence, it is evident that Krishna-Swamy atmospheres are not suitable for the modelling of pre-main sequence stars.

\subsubsection{Mixing length of convection}
We have calculated instability regions for models with $\alpha_{\rm MLT} = 1.8, 2.0, 2.2$ (see Fig. \ref{fig:mix_pmodes} and Fig. \ref{fig:mix_gamma}). Both, the p-mode and $\gamma$ Doradus instability region are influenced by the choice of input physics. Given the uncertainties of the stellar parameters we are not able to constrain the mixing length with pre-main sequence pulsators. Since higher mixing lengths cut into the stellar sample of \citet{Vioque2020}, we note that $\alpha_{\rm MLT} = 1.8$ fits best.

\subsubsection{Comparison to classical evolution}
Figure \ref{fig:instab_compare_classic} shows the comparison between classical models and the models from accreting protostars. The p-mode instability region is very similar, with the exception that lower surface gravity is allowed due to different initial models.

For SPB stars, the early phases off the accreting evolutionary track is stable to g-modes. This is different to the classical pre-main sequence tracks, which show an instability at this location. The $\gamma$ Doradus instability region is located at  similar positions in both calculations, but the borders look significantly different. For the classical models, unstable modes are located in a smaller temperature region and especially for low temperatures ($3.7 \leq \log(T_{\rm eff}) \leq 3.75$) no modes are excited. Hence, we attribute the instability region at these low temperatures and the widening of the temperature range found in our models of accreting protostars to an effect related to the accretion history.

\subsection{Instability of higher order modes}
Long time base observations obtained with space telescopes allow the identification of successive radial orders \citep[see e.g.][for $\gamma$ Doradus stars observed with Kepler]{Li2020}. To investigate the potential of performing similar analyses for pre-main sequence stars, we performed our calculations such that radial orders up to $n=\pm 50$ are included. A distinction of the observed pre-main sequence instability regions including the information of which radials orders are excited will create more constraints on the input physics once sufficient observations of pulsating pre-main sequence stars are available. Here, we report our results for the instability regions of higher orders for both p- and g-modes.

Figure \ref{fig:ho_pmodes} shows order-dependent pre-main sequence instability regions for p-modes\footnote{The areas of excited higher orders may be incomplete due to our chosen frequency limits. Higher order modes with frequencies above $100$~\cd might be unstable in regions near the ZAMS. Especially the areas with $n\geq 25$ are cut off by this limit. However, the lower boundary of the left area with $n\geq 15$ is resolved.}. There are two separate regions with unstable modes of radial order $\geq 15$: region I at lower temperatures ($3.75 \leq \log(T_{\rm eff}) \leq 3.85$) and region II at hotter temperatures ($3.9 \leq \log(T_{\rm eff}) \leq 4$). The blue boundary of region I coincides with the red edge of the observed instability deduced from our sample of pre-main sequence and early ZAMS stars. Region I is most likely introduced by missing stabilisation effects due to convection, and, hence most likely non-physical. 

In comparison, region II matches the location of observed pre-main sequence $\delta$ Scuti stars and candidates in our sample. As such, there are multiple known pulsators that fall within this part of the Kiel diagram. We strongly encourage the search for more pre-main sequence pulsators to further populate the p-mode instability region.

Figure \ref{fig:ho_gmodes} shows order-dependent instability regions for g-modes with even more distinct features. For SPB stars, the results indicate the modes with radial orders $\vert n \vert \geq 25$ are confined to lower surface gravity. Astonishingly, all of the observed pre-main sequence and early ZAMS SPB stars and candidates with well defined spectroscopic parameters fall within the region without unstable modes of radial order higher than $25$. 
The currently available photometric time series of pre-main sequence and early ZAMS SPB stars are sufficient for an identification of the pulsation type. The identification of 
g-mode period spacings is often difficult due to the inadequate frequency resolution stemming from short observations with time bases well below $100$ days. However, \citet{Zwintz2017} showed that it is possible to extract such information even with shorter observation runs. Unfortunately, this only includes a few radial orders, which is insufficient to compare it with the results shown in Fig. \ref{fig:ho_gmodes}. Only with observational data of sufficiently long time bases (i.e., on the order of 100 days and more) obtained from space it will be possible to extract the necessary information about excited frequencies to test this result of our calculations. 

The results for the pre-main sequence $\gamma$ Doradus instability region indicate that mostly low order modes are excited. However, in a small region around $6750$~K, higher order modes are expected. Hence, the situation is similar the to the case of SPB stars. Future observations of higher accuracy should be able to deduce whether this is  a real physical or a modelling effect. Our results furthermore indicate a region with excited modes with orders $1 \leq \vert n \vert \leq 15$ just below the accreting evolutionary track shown as the red area in Fig. \ref{fig:ho_gmodes}. 

The instability region for M-type stars described in Sect. \ref{sec:res:puls} includes higher order modes at very early evolutionary stages. As such, we would expect to be able to observe period spacings at very low frequencies for such stars.

  \begin{table}
    \caption[]{Extracted frequencies for the possible pulsating M-type young stellar object 2MASS J11120327-7637034.}
    \label{tab:mstar_fs}
    \begin{tabular*}{\linewidth}{lrrrr}
        \hline
        \noalign{\smallskip}
        Designation &\multicolumn{1}{c}{$f$}  &  \multicolumn{1}{c}{$A$}    & \multicolumn{1}{c}{$\phi$} & \multicolumn{1}{c}{SNR}\\
        &\multicolumn{1}{c}{(\cd)} &  \multicolumn{1}{c}{(mmag)}   & \multicolumn{1}{c}{($\frac{\rm rad}{2 \pi}$)} & \\
        \noalign{\smallskip}
        \hline
        \noalign{\smallskip}
         F1  & $1.27260(2)$  & $2.32(10)$ & $0.881(7)$  & $9.25$  \\
         F2  & $0.19716(4)$  & $1.28(10)$ & $0.503(13)$ & $4.31$  \\
         F3 = $f_{\rm rot}$ & $0.93922(5)$  & $0.99(10)$ & $0.363(16)$ & $5.12$  \\
         F4  & $0.24018(6)$  & $0.87(10)$ & $0.567(19)$ & $4.43$  \\
         F5 = $2f_{\rm rot}$ & $1.88202(10)$ & $0.50(10)$ & $0.22(3)$   & $5.89$  \\
        \noalign{\smallskip}
        \hline

    \end{tabular*}    
    \tablefoot{
    The values in parentheses give the $1\sigma$ uncertainty as reported by the standard error estimates formulated by \citet{montgomery1999}. The uncertainties are most probably underestimated given that they reflect only the uncertainty of the extracted light curve and do not include uncertainties stemming from the extraction process.
    }
 \end{table}

 \begin{figure}
   \centering
   \includegraphics[width=\linewidth]{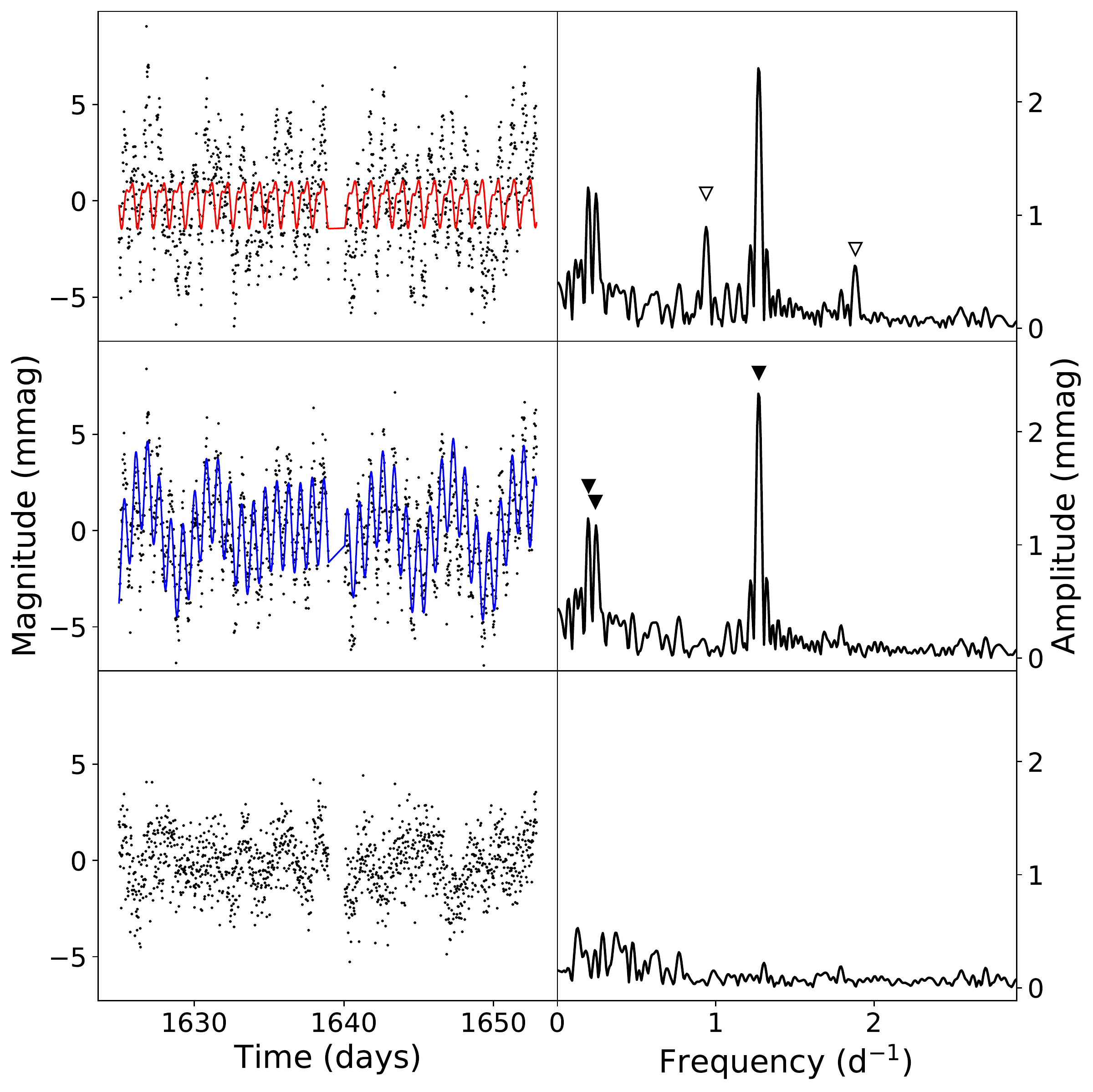}
      \caption{Light curve and amplitude spectra of 2MASS J11120327-7637034. Left panels show the corresponding light curve (black points) and a model (coloured lines), while right panels show the corresponding amplitude spectra. The frequencies used in the model are marked with a filled (pulsation) or open (rotation) triangle. Top panel: TESS light curve and the model for rotational variability (red line) calculated from F3 and F5. Middle Panel: TESS light curve after removing the rotational signal. Also shown is the pulsation model (blue line) calculated from F1, F3 and F4. Bottom Panel: Residuals after subtracting all five frequencies.
              }
         \label{fig:mstar_lc}
\end{figure}

\begin{figure}
   \centering
   \includegraphics[width=\linewidth]{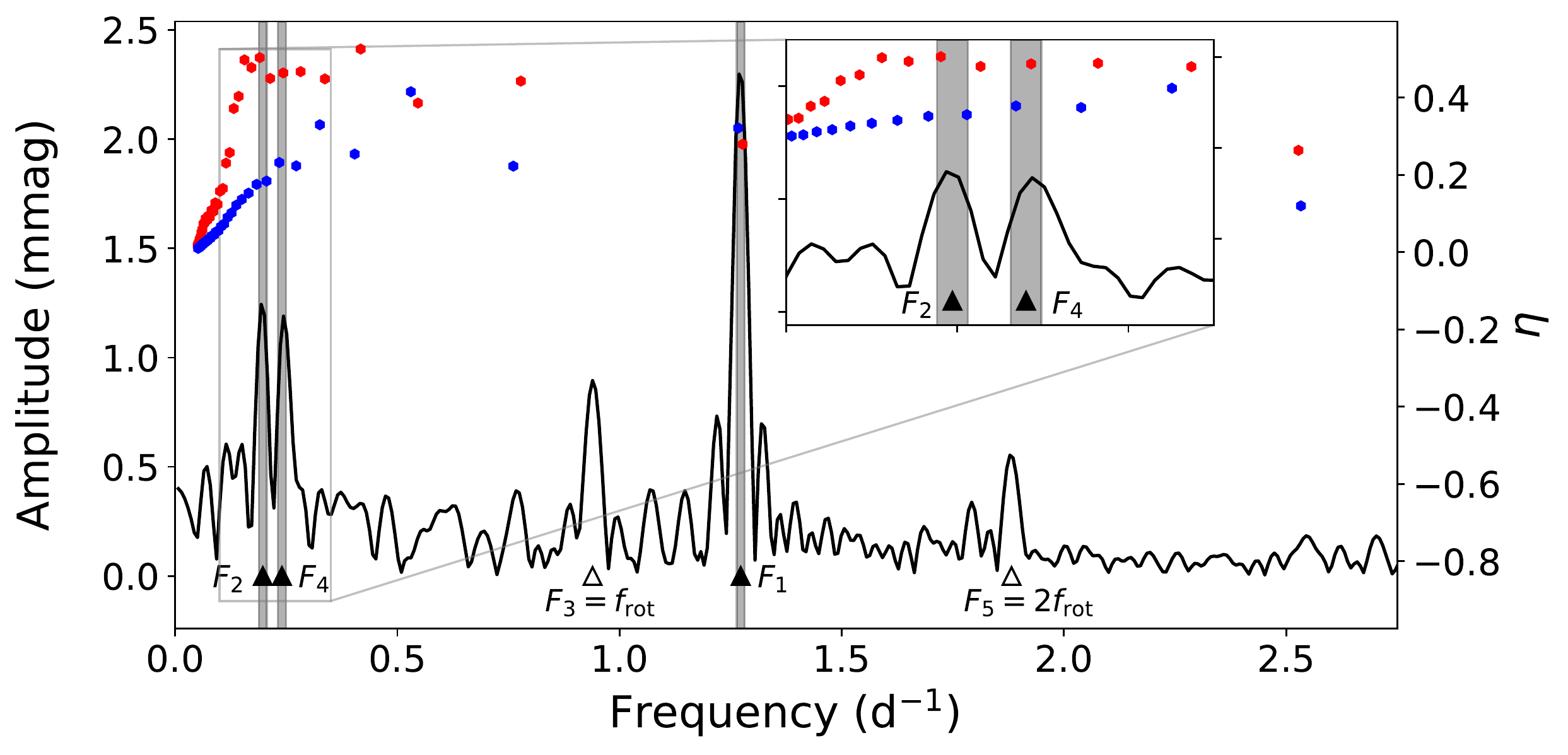}
      \caption{Amplitude spectra and excited frequencies of models A and models B (see Table \ref{tab:mstar_vals}). Observed frequencies are marked with triangles at the bottom and the frequencies attributed to pulsations are covered in a gray shaded area with the size of a fourth of the Rayleigh limit. Open triangles mark frequencies attributed to rotation. The excited frequencies of two chosen models are shown as red and blue hexagons. The inset shows a zoom in in the low frequency region of F2 and F4. 
              }
         \label{fig:mstar}
\end{figure}
\subsection{2MASS J11120327-7637034: a possible pulsating M-type young stellar object}
\label{sec:mtypepuls}
2MASS J11120327-7637034 (also CHXR 84, Hn 16, TIC 454364154, etc.) is a M5.5 \citep{Luhman2004} type star and a member of the Chameleon I region with an estimated age of $2$~Myr \citep{gutierrrez2020}. TESS observed 2MASS J11120327-7637034 (TESS magnitude 14.062) in sector 14. No short cadence data is available for this star and hence we obtained the light curve from the FFIs using the \texttt{RegressionCorrector}. The FFIs are very crowded, but the chosen aperture includes only three more stars (Gaia DR2 5201302259854227968, Gaia DR2 5201302255559778304, and Gaia DR2 5201302259855347072) with magnitudes (19.326, 19.346, and 20.108\,mag) that are significantly lower than the magnitude of 2MASS J11120327-7637034 from which no contamination is expected. Multiple other stars lie in the vicinity of 2MASS J11120327-7637034; given their magnitude, they might contaminate the extracted light curve to some degree. The FFI and chosen aperture are shown in Fig. \ref{appfig:2MASS 11120327-7637034}.

We extracted five significant frequencies (see Table \ref{tab:mstar_fs}). We attribute the frequencies F3 and F5 to rotational signal from stellar spots. \citet{Flaherty2016} found evidence for cold spots for 2MASS J11120327-7637034 consistent with this attribution. We furthermore interpret the other three frequencies as possible pulsation frequencies. Figure \ref{fig:mstar_lc} shows the light curve and amplitude spectra extracted from the TESS FFIs. We removed the rotation signal (top panel) and the suspected pulsation signal (middle panel) to obtain the residuals in the bottom panel. The amplitude spectrum of the residuals shows evidence for more variability, but no more significant frequencies could be extracted. 

\citet{gutierrrez2020} reported an effective temperature of $3304 \pm 103$~K, which is higher than the temperature expected from the spectral type, that is $3058$~K \citep{Luhman2003, Luhman2004}. We looked for models with temperatures in the range of the values adopted by \citet{gutierrrez2020} in our grid. While the grid is sparse in the corresponding region of the Kiel diagram, we find two models for which dipolar g-modes are excited that match F1, F2, and F4 within a fourth of the Rayleigh limit which is $0.009$~\cd. Figure \ref{fig:mstar} compares the excited frequencies in these models with the observations. The specifics of the two models are given in Table \ref{tab:mstar_vals}. The stellar mass is higher than estimates by \citet{Kirk2011} and \cite{Lafreniere2008}. Both used the temperature estimate by \cite{Luhman2004} and hence some discrepancy is expected. \citet{Feigelson2004} estimated the age of 2MASS J11120327-7637034 to be $1$~Myr. This estimate of course relies on models that are not originating from accreting stellar seeds and hence this discrepancy is also expected.

Given the relatively poor observations at hand, it is clear that no full asteroseismic modelling can be done. However, it is promising that even such a sparse grid includes two models that fit the observed frequencies.
A closer look at the amplitude spectrum in Fig. \ref{fig:mstar} shows additional signal at around $0.8$~\cd and $2.6$~\cd where both models predict excited frequencies (radial orders $-1$ and $-3$), but the observational amplitudes are not significant. 2MASS J11120327-7637034 will be observed again in TESS sectors 38 and 39. With additional data, it might be possible to further constrain whether the variability originates from 2MASS J11120327-7637034 or any other star in the vicinity and what its origin is. A longer baseline should also give better estimates for the observed frequencies and produce better constraints for modelling. We plan on re-evaluating the status of 2MASS J11120327-7637034, once the new TESS data is available.

  \begin{table}
    \caption[]{Frequencies and stellar parameters for the models describing the observed frequencies of 2MASS J11120327-7637034.}
    \label{tab:mstar_vals}
    \begin{tabular*}{\linewidth}{lrrrr}
        \hline
        \noalign{\smallskip}
        parameter & observed & model A  &  model B \\
        \noalign{\smallskip}
        \hline
        \noalign{\smallskip}
         F1 (\cd)  & $1.27260(2)$  & $1.26706$ &  $1.27699$ \\
         radial order  &  - & -2 & -2 \\
         F2 (\cd) & $0.19716(4)$  & $0.205598$ &  $0.190394$ \\
         radial order  &  - & -9 & -10 \\
         F4 (\cd)  & $0.24018(6)$  & $0.234298$ & $0.243129$  \\
         radial order  &  - & -8 & -8 \\
         $T_{\rm eff}$ (K) & $3304(103)^a$  & $3317.3$ & $3368.7$ \\
         $\log(g)$ &  - & $3.299$ & $3.289$ \\
         radius ($R_\odot$)&  - & $1.484$ & $1.592$ \\
         mass ($M_\odot$)&  $0.137^b\, /\, 0.15^c$ & $0.16$ & $0.18$ \\
         age (Myr)& $1^d$  & $0.176$ & $0.160$ \\
        \noalign{\smallskip}
        \hline

    \end{tabular*}    
    \tablefoot{
    Sources: ($a$) \citet{gutierrrez2020}, ($b$) \citet{Kirk2011}, ($c$) \citet{Lafreniere2008}, ($d$) \citet{Feigelson2004}.
    }
 \end{table}

%--------------------------------------------------------------------
\section{Conclusions}
\label{sec:conclusions}

This series of paper aims to combine the power of asteroseismic tools with state-of-the-art pre-main sequence models. In this first paper, we present the first calculations of pulsation instability regions for pre-main sequence models originating from small stellar seeds. For this, we first gathered a spectroscopic sample of pre-main sequence stars and early ZAMS stars and constrained the input physics of accretion to obtain an envelope for these stars in a constant accretion scenario. This can be considered as a first step towards more realistic scenarios provided by episodic accretion rates either obtained from numerical hydrodynamics simulations or other more sophisticated approaches than this constant accretion scenario. While this effort was able to restrict the constant accretion rate and fraction of injected heat from accreted material, other free parameters (i.e., the atmospheric boundary conditions, $\alpha_{\rm MLT}$, fractional mass of heat injection, and initial Deuterium abundance) cannot be constrained. The resulting accreting evolutionary track presents a good envelope to the stellar sample even though not all stars in the latter end up being interpreted as pre-main sequence stars. However, besides one star, 2MASS 11110238-7613327, the reason for not fitting within the envelope is discussed and of not much concern. Not much is known about the star 2MASS 11110238-7613327 and we strongly encourage follow up investigation to explain its position on the Kiel diagram.

The location of observed pre-main sequence pulsators compared to the calculated instability regions provides further constraints on the theoretical models. Hence, we additionally gathered a sample of known pulsating pre-main sequence and early ZAMS stars with spectroscopic parameters and looked for new pulsators in the TESS observation of our stellar sample. We presented light curves and pulsation spectra of new pre-main sequence SPB stars (1), SPB candidates (5), $\gamma$ Doradus candidates (1), $\delta$ Scuti stars (3), and $\delta$ Scuti candidates (3). In addition we also present three new pulsators discovered in CoRoT data.

We constructed theoretical pre-main sequence instability regions by calculating non-adiabatic pulsation spectra of models originating from a small stellar seed in a constant accretion scenario. The resulting pulsation regions follow the position of the observed pulsators, with few exceptions. The p-mode instability region indicates unstable modes in low temperature regions, where no pulsating star is known. We attributed this to missing stabilisation effects from stellar convection, as our models do not include time dependent treatment of convection. A closer inspection of the results shows that the pulsation modes in the corresponding models have high normalised growth rates but the work performed by one pulsation cycle is almost vanishing. Furthermore, there is a clear border in normalised growth rate that coincides with the observed red edge of the pre-main sequence instability region deduced from our sample of pre-main sequence and early ZAMS stars. We hence propose to interpret only the part to the hotter regions of this boundary as pre-main sequence p-mode instability region. Both, the pre-main sequence SPB und $\gamma$ Doradus instability regions, explain the observed pulsators. We have shown that is essential to use the description of case 4 of \citet{Pesnell1990} when calculating the pulsation spectra of pre-main sequence $\gamma$ Doradus stars with \texttt{GYRE}.

While the instability regions obtained by our calculations are not able to constrain the mixing length, we have shown that Krishna-Swamy atmospheres are not able to produce the unstable p-modes in populated regions of the Kiel diagram. Hence, Eddington Gray atmospheres are to be preferred. 

In addition, we discussed the instability of pre-main sequence stars depending on the radial order of unstable modes. Future observations, with baselines long enough to identify g-mode period spacings, will allow to test these obtained instability regions and as such provide more constraints. However, current and currently planned space missions are unfortunately not able to provide such observations. This calls in the need for a specialised space mission to observe stars in their early evolutionary stages. 

Our calculations also result in a g-mode instability region for K- and M- type stars, where long period spacings would be expected for M-type stars in their earliest evolutionary phases. Such pulsations are expected for some time \citep{rodriguez2019}, but to our knowledge no candidate has been detected from observations yet. We present the case of 2MASS J11120327-7637034, a young stellar object with a spectral type of M5.5. The light curve obtained from the crowded TESS FFIs shows three frequencies attributed to pulsations and additional variability most likely originating from stellar spots. While our grid is sparse in this area of the Kiel diagram we looked for models fitting both, the effective temperature constraints of \citet{gutierrrez2020} and the observed frequencies. We found two models with unstable modes within a fourth of the Rayleigh limit of each of the three frequencies. Because the FFIs are crowded and the obtained light curve is comparably short, we are not able to simply attribute the observed variability to 2MASS J11120327-7637034 and, hence, can only suggest 2MASS J11120327-7637034 to be the first potential candidate pulsator of this class. TESS will revisit 2MASS J11120327-7637034 in sectors 38 and 39 which should improve our analysis. Once this data is available, we hope to obtain a better picture of the status of this star.

\begin{acknowledgements}
    We are grateful to the anonymous referee whose comments helped to improve the clarity of this work. We thank Masanobu Kunitomo for providing valuable help in the early stages of this project. We are grateful to Bill Paxton and his collaborators for their valuable work on the stellar evolution code MESA. 
    EV acknowledges financial support by Southern Federal University, 2020 (Ministry of Science and Higher Education of the Russian Federation). The TESS data presented in this paper were partly obtained from the Mikulski Archive for Space Telescopes (MAST) at the Space Telescope Science Institute (STScI). Funding for the TESS mission is provided by the NASA Explorer Program. STScI is operated by the Association of Universities for Research in Astronomy, Inc., under NASA contract NAS5-26555. Support for MAST for non-HST data is provided by the NASA Office of Space Science via grant NNX13AC07G and by other grants and contracts. 
    The CoRoT space mission was developed and operated by the French space agency CNES, with participation of ESA's RSSD and Science Programmes, Austria, Belgium, Brazil, Germany, and Spain.
    This work has made use of data from the European Space Agency (ESA) mission {\it Gaia} (\url{https://www.cosmos.esa.int/gaia}), processed by the {\it Gaia} Data Processing and Analysis Consortium (DPAC, \url{https://www.cosmos.esa.int/web/gaia/dpac/consortium}). Funding for the DPAC has been provided by national institutions, in particular the institutions participating in the {\it Gaia} Multilateral Agreement. 
    Spectroscopic data were obtained with the 2.7-m telescope at Mc Donald Observatory, Texas, US.
    This research has made use of the SIMBAD database, operated at CDS, Strasbourg, France; NASA's Astrophysics Data System; matplotlib, a Python library for publication quality graphics \citep{Hunter:2007}; SciPy \citep{Virtanen_2020}; Lightkurve, a Python package for Kepler and TESS data analysis \citep{lightkurve2018};  Astrocut, tools for creating cutouts of TESS images \citep{Brasseur2019}; Astropy, a community-developed core Python package for Astronomy \citep{2018AJ....156..123A, 2013A&A...558A..33A}; NumPy \citep{van2011numpy}; MESA SDK for Linux (Version 20.3.1) \citep{townsend2020}, WebPlotDigitizer (\url{http://arohatgi.info/WebPlotDigitizer/}) by Ankit Rohatgi, \texttt{tpfplotter} by J. Lillo-Box (publicly available in www.github.com/jlillo/tpfplotter) \citep{Aller2020}.
\end{acknowledgements}

\bibliographystyle{aa} % style aa.bst
\bibliography{bib} % your references Yourfile.bib

\appendix

\section{TESS observations}
\label{app:tess}

\subsection{Spectral window of TESS and CoRot observation}
\begin{figure}
   \centering
   \includegraphics[width=\linewidth]{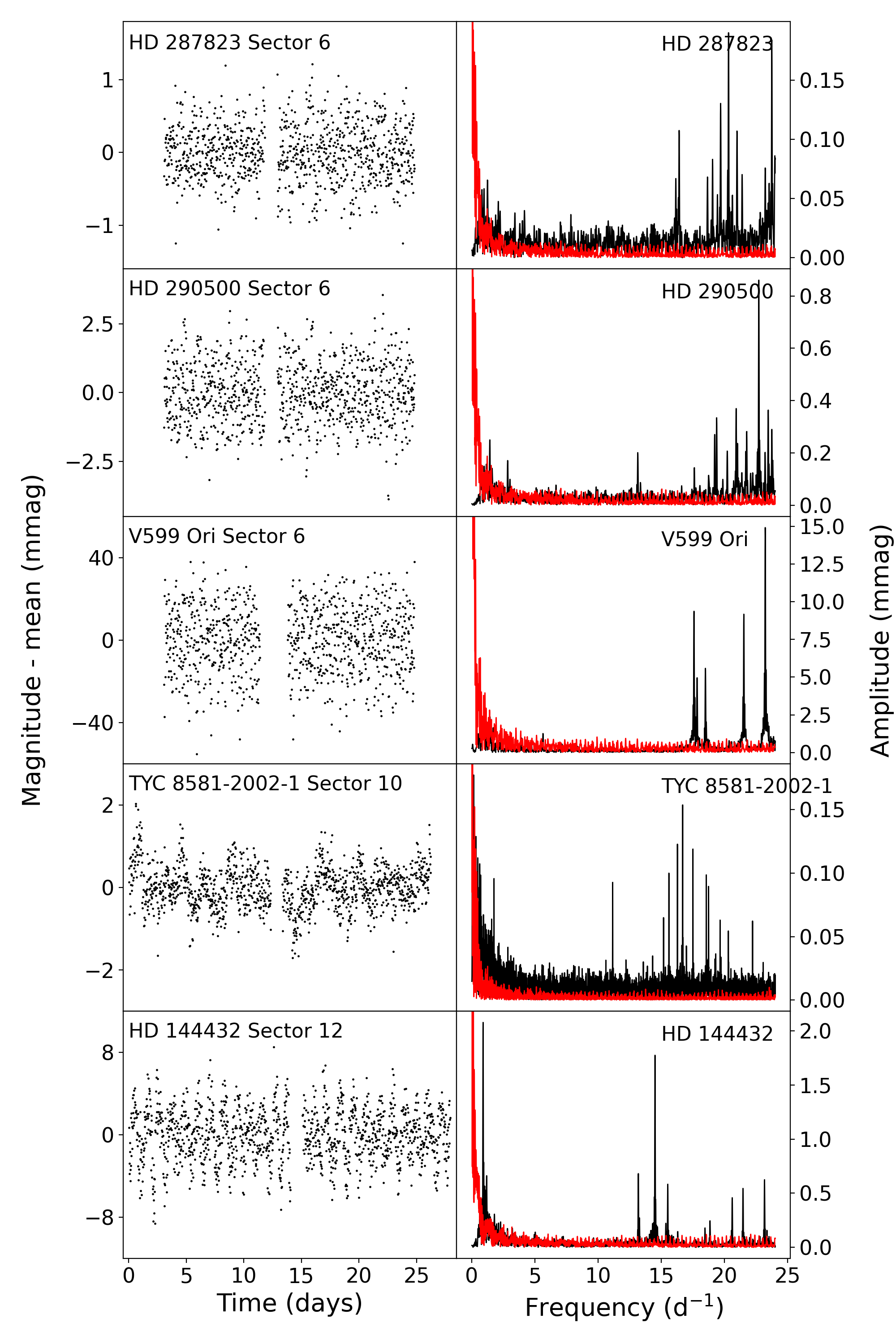}
      \caption{Same as Fig. \ref{fig:dscuti_lcs} but including the spectral window. (red line) of the light curves in the right panels. The spectral window is normalised to the highest peak in the amplitude spectrum and magnified by a factor of $100$.
              }
         \label{fig:dscuti_lcs_inc_sw}
\end{figure}

\begin{figure}
   \centering
   \includegraphics[width=\linewidth]{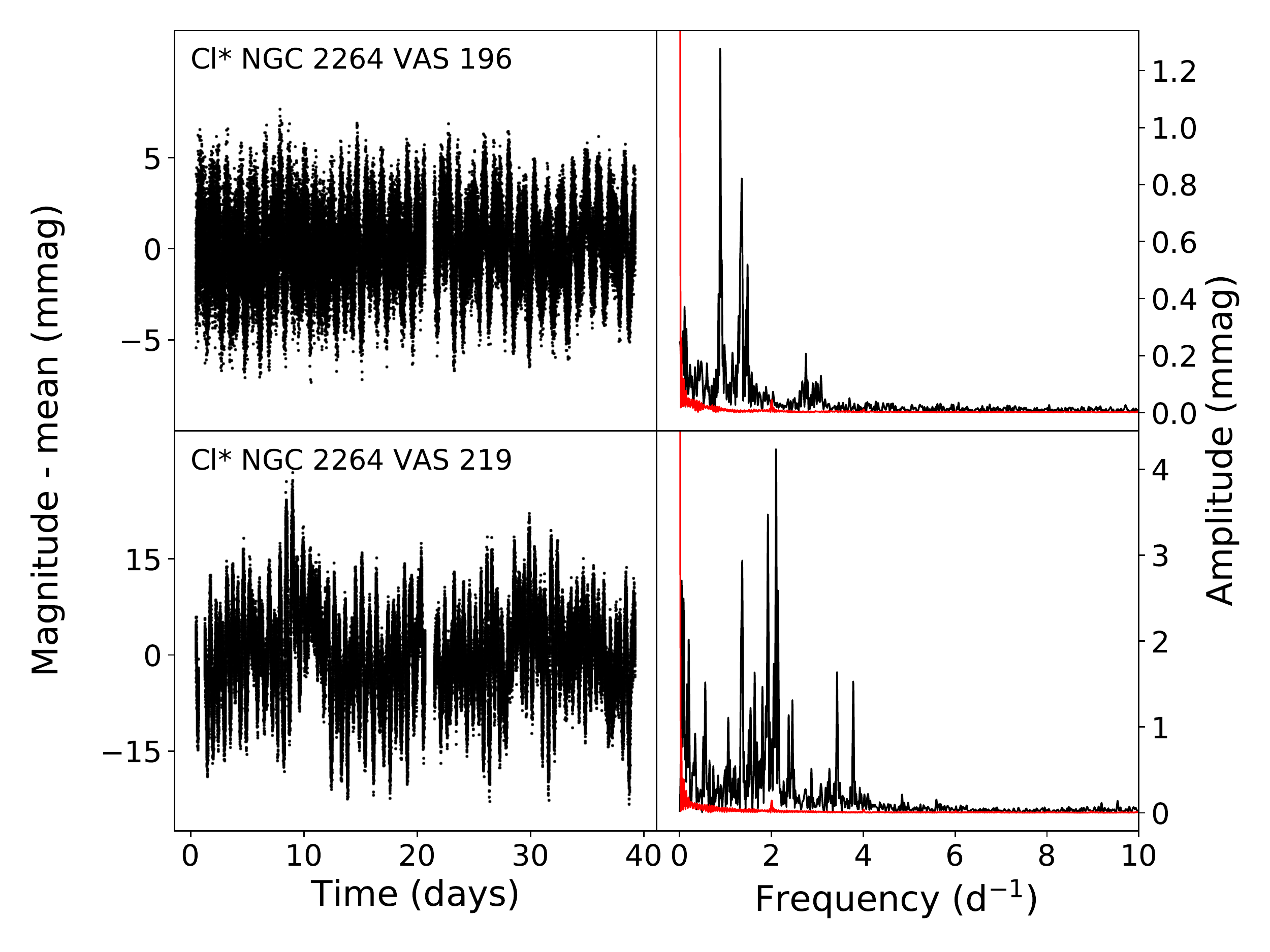}
      \caption{Same as Fig. \ref{fig:gamma_dor_corot} but including the spectral window. (red line) of the light curves in the right panels. The spectral window is normalised to the highest peak in the amplitude spectrum and magnified by a factor of $10$.
              }
         \label{fig:gamma_dor_corot_inc_sw}
\end{figure}

\begin{figure}
   \centering
   \includegraphics[width=\linewidth]{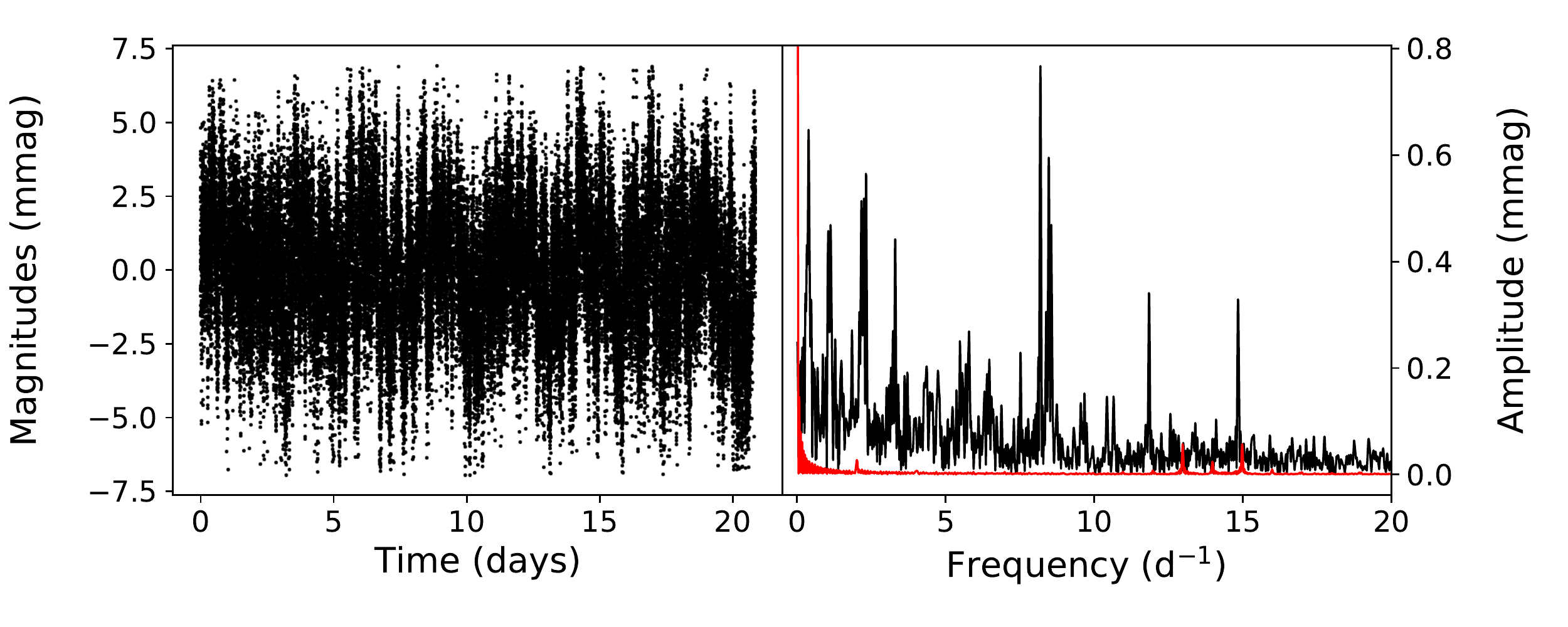}
      \caption{Same as Fig. \ref{fig:VAS230} but including the spectral window. (red line) of the light curves in the right panels. The spectral window is normalised to the highest peak in the amplitude spectrum and magnified by a factor of $10$.
              }
         \label{fig:VAS230_inc_sw}
\end{figure}

In the Fourier analysis of time series photometry, the time stamps of observations often introduce alias frequencies due to the convolution of the amplitude spectrum with the spectral window. The spectral window of different light curves analysed in this work are shown in Fig. \ref{fig:dscuti_lcs_inc_sw} - \ref{fig:VAS230_inc_sw}. While there is no signal whatsoever in the spectral window of the TESS observations, there are some small peaks in the CoRot observations. However, their amplitude is small enough that this does not lead to serious aliasing problems in our analysis. 

\subsection{Frequency tables and FFI images of the newly discovered pre-main sequence pulsators}

\begin{figure}
   \centering
   \includegraphics[width=\linewidth]{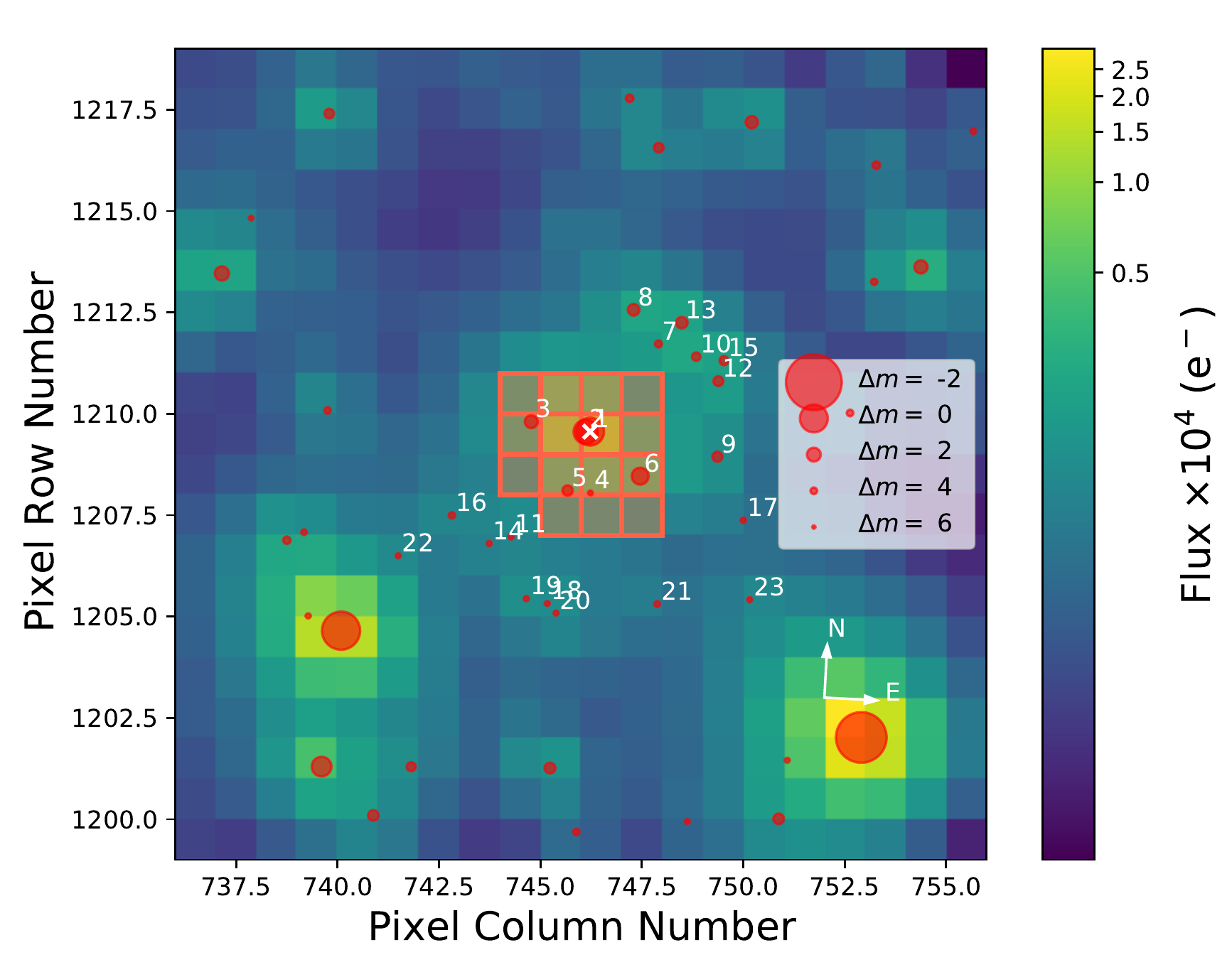}
      \caption{TESS FFI image in a $20 \times 20$ pixel cutout around the position of HD 152822 in sector 12. Red filled area shows the chosen aperture. The red dots show stars within a GAIA magnitude limit $\propto 6$ with varying size according to their GAIA magnitude. The GAIA query failed for HD 152822 and hence the magnitudes are relative to  TYC 8327-955-1, which is marked by the white cross.
              }
         \label{appfig:HD152822}
\end{figure}

\begin{figure}
   \centering
   \includegraphics[width=\linewidth]{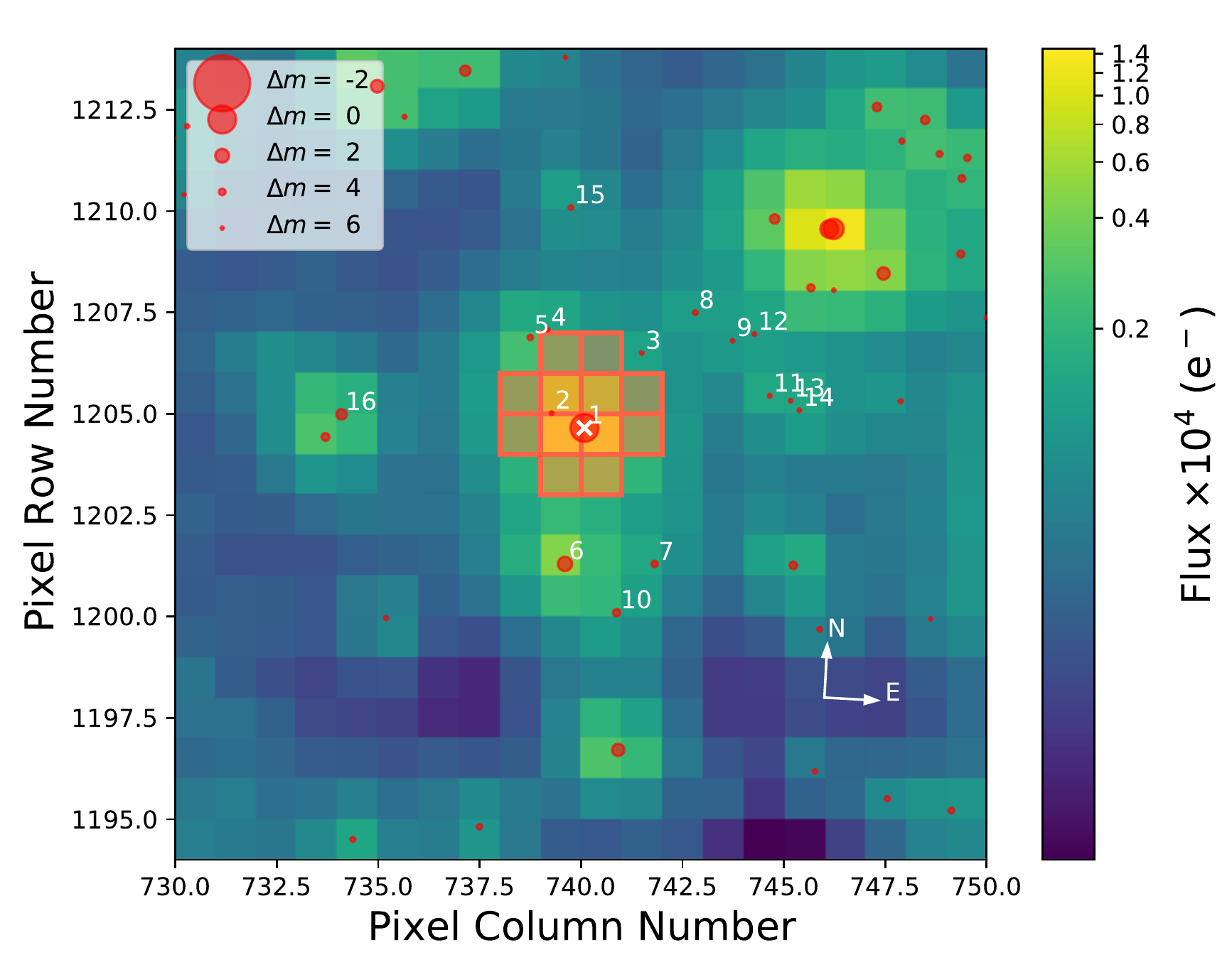}
      \caption{TESS FFI image in a $20 \times 20$ pixel cutout around the position of HD 152799 in sector 12. Red filled area shows the chosen aperture. The red dots show stars within a GAIA magnitude limit $\propto 6$ with varying size according to their GAIA magnitude and the white cross marks the position of HD 152799.
              }
         \label{appfig:HD152799}
\end{figure}

\begin{figure}
   \centering
   \includegraphics[width=\linewidth]{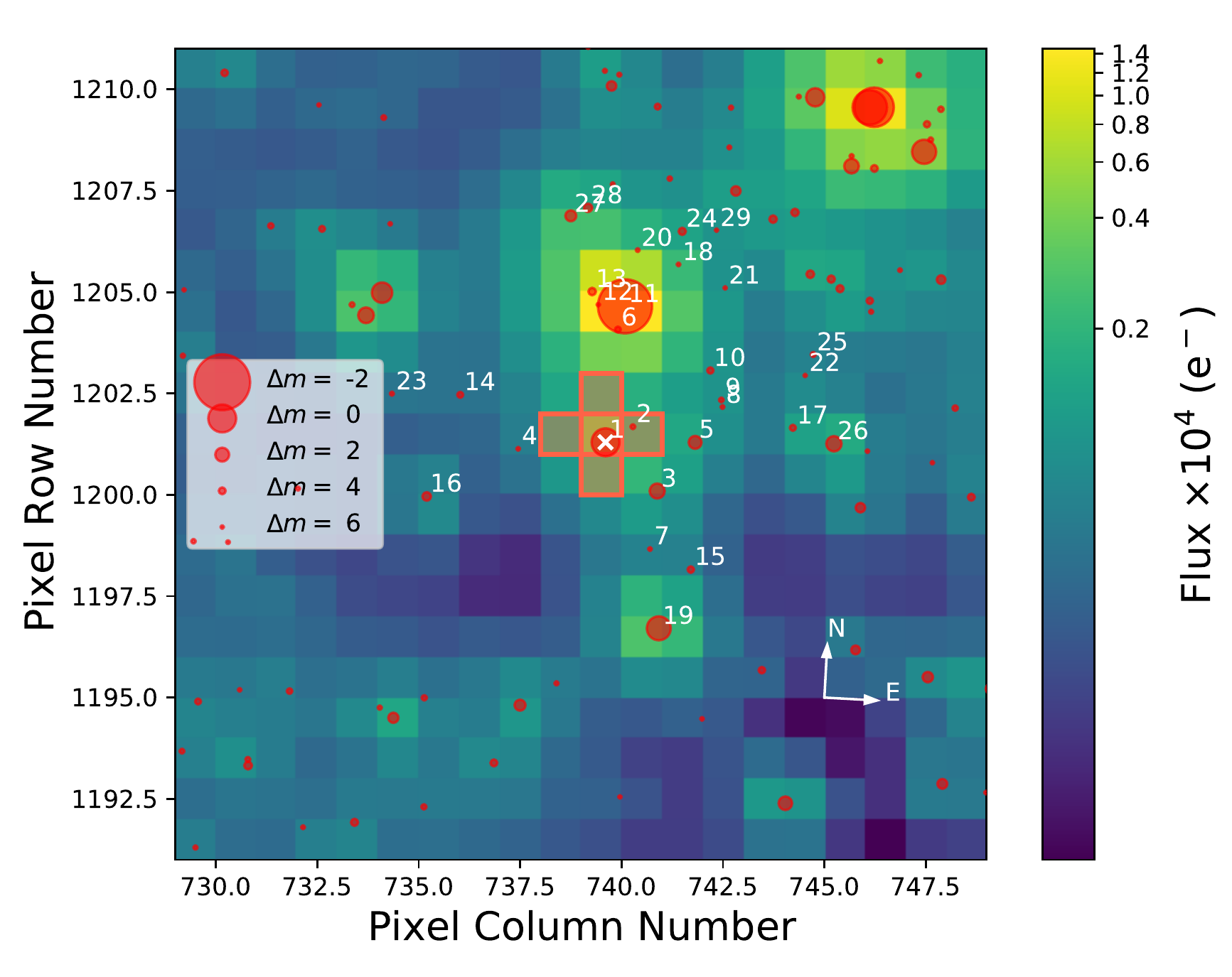}
      \caption{Same as fig. \ref{appfig:HD152799} but for HD 329271 in sector 12.
              }
         \label{appfig:HD329271}
\end{figure}

\begin{figure}
   \centering
   \includegraphics[width=\linewidth]{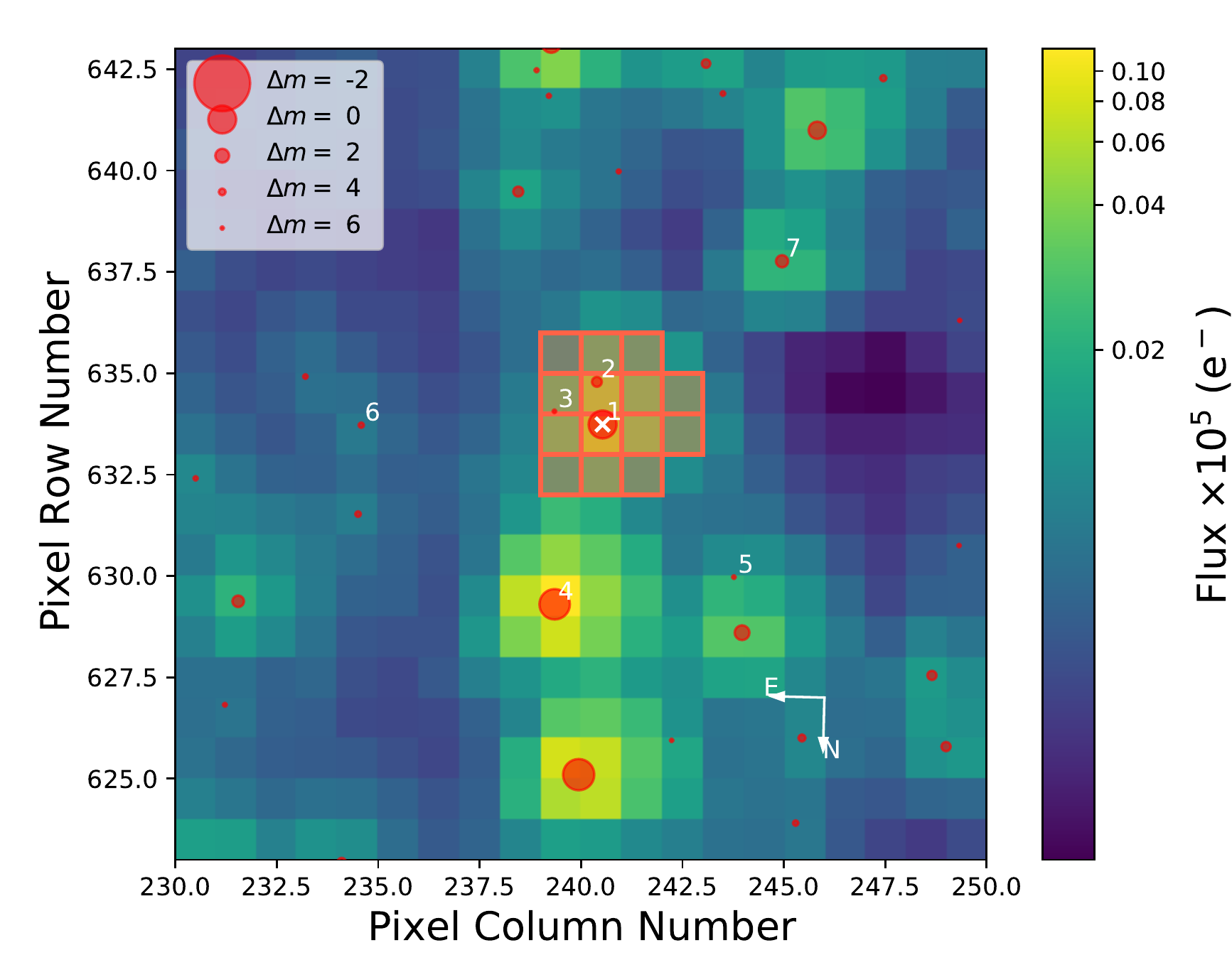}
      \caption{Same as fig. \ref{appfig:HD152799} but for CD-32 12908 in sector 12.
              }
         \label{appfig:CD-3212908}
\end{figure}

\begin{figure}
   \centering
   \includegraphics[width=\linewidth]{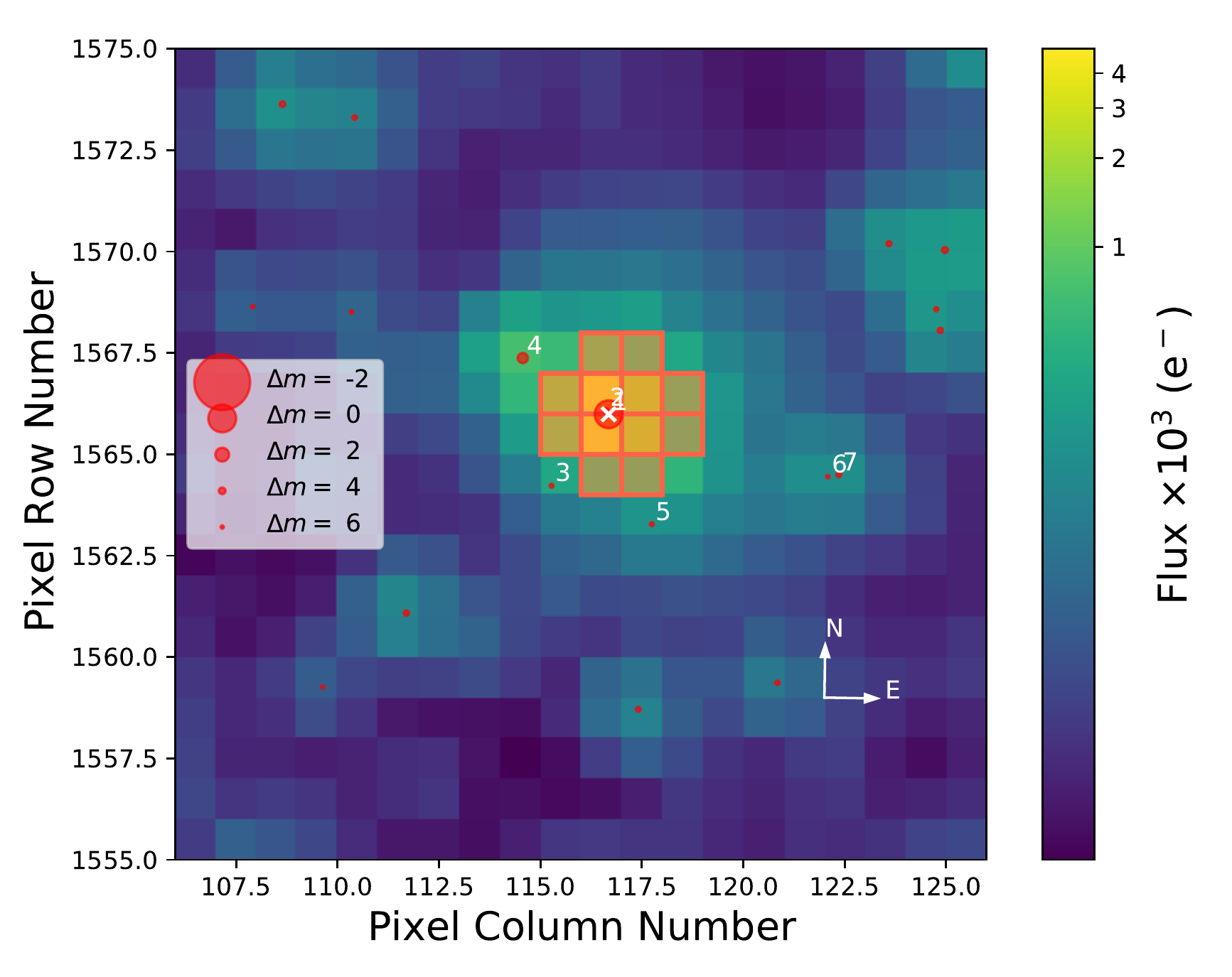}
      \caption{Same as fig. \ref{appfig:HD152799} but for HD 287823 in sector 6. 
              }
         \label{appfig:HD 287823}
\end{figure}

\begin{figure}
   \centering
   \includegraphics[width=\linewidth]{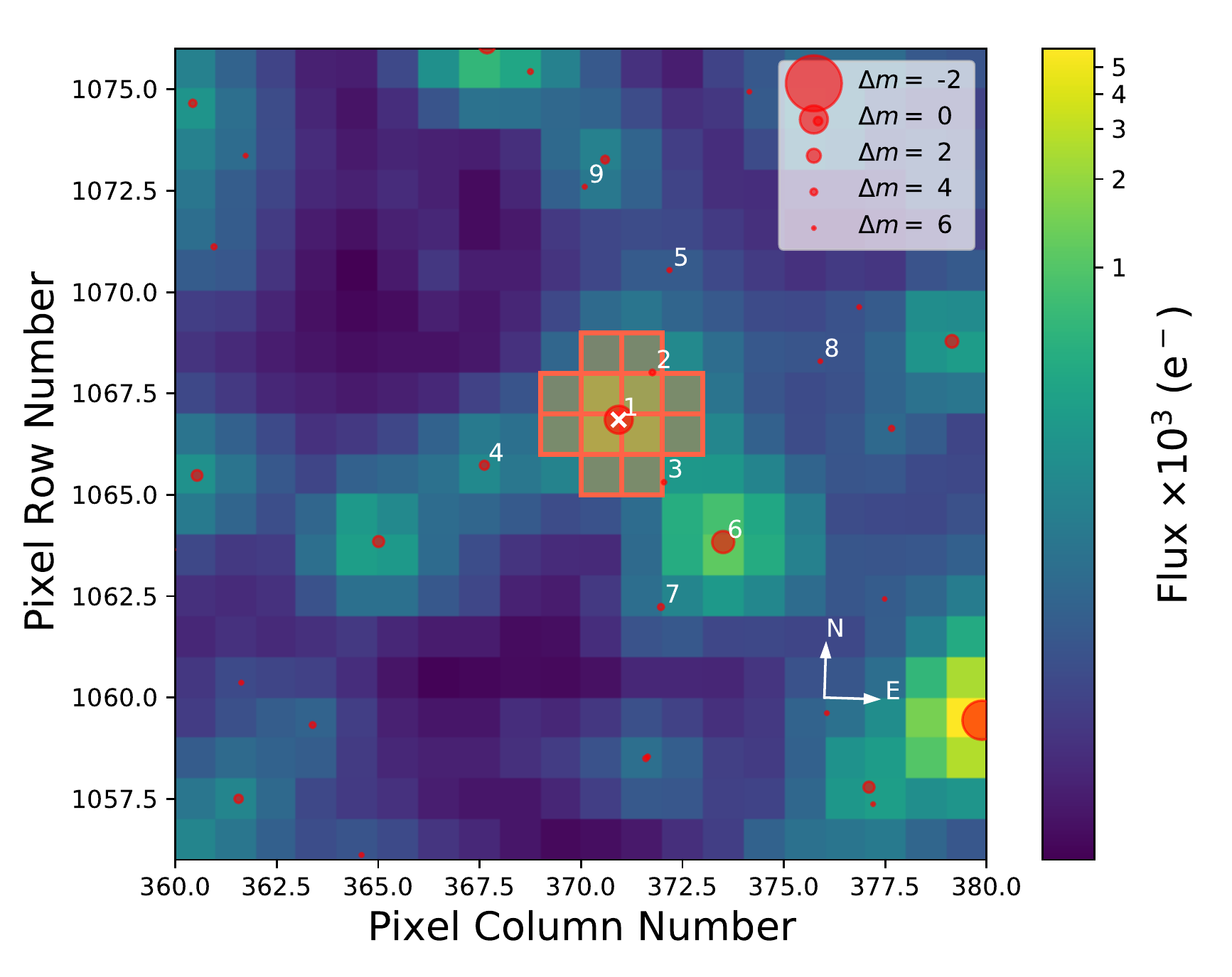}
      \caption{Same as fig. \ref{appfig:HD152799} but for HD 290500 in sector 6. 
              }
         \label{appfig:HD 290500}
\end{figure}

\begin{figure}
   \centering
   \includegraphics[width=\linewidth]{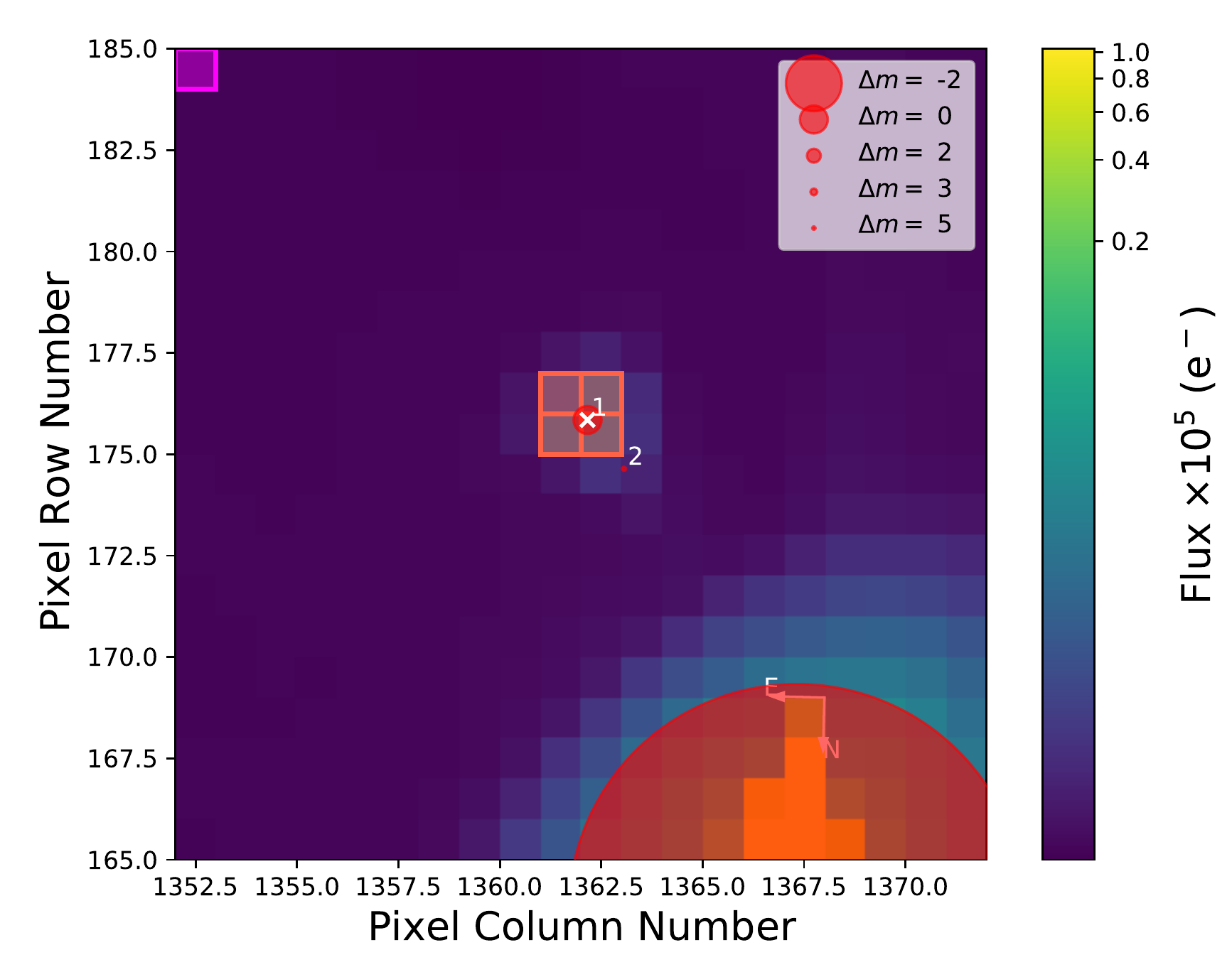}
      \caption{TESS FFI image in a $20 \times 20$ pixel cutout around the position of V599 Ori in sector 6. Red filled area shows the chosen aperture and the magenta area shows the chosen pixel for the simple background flux. The red dots show stars within a GAIA magnitude limit $\propto 6$ with varying size according to their GAIA magnitude and the white cross marks the position of V599 Ori. The influence of the naked eye star 49 Ori is clearly visible.
              }
         \label{appfig:V599 Ori}
\end{figure}

\begin{figure}
   \centering
   \includegraphics[width=\linewidth]{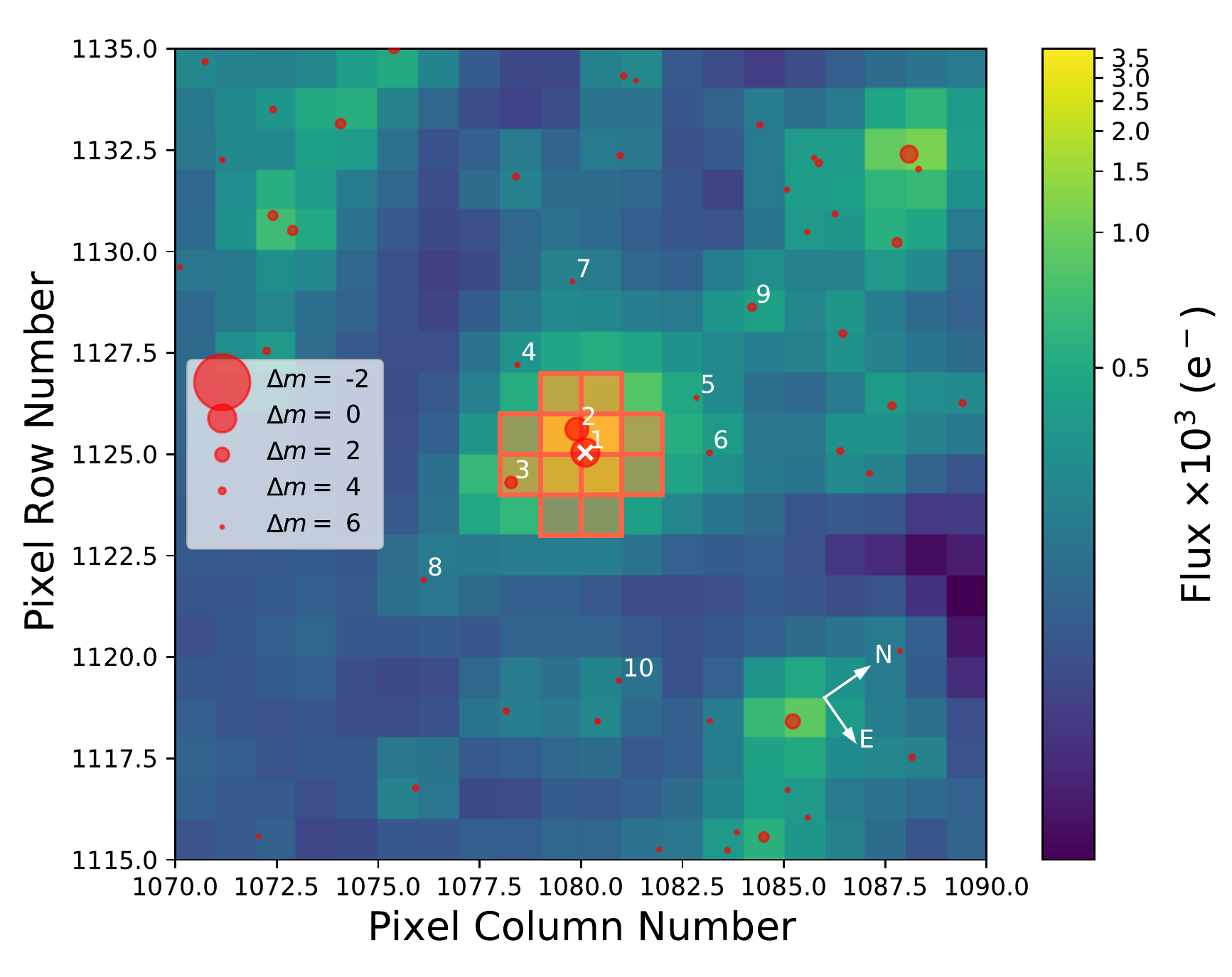}
      \caption{Same as fig. \ref{appfig:HD152799} but for TYC 8581-2002-1 in sector 10. Sectors 8 and 9 have also been observed. 
              }
         \label{appfig:TYC 8581-2002-1}
\end{figure}

\begin{figure}
   \centering
   \includegraphics[width=\linewidth]{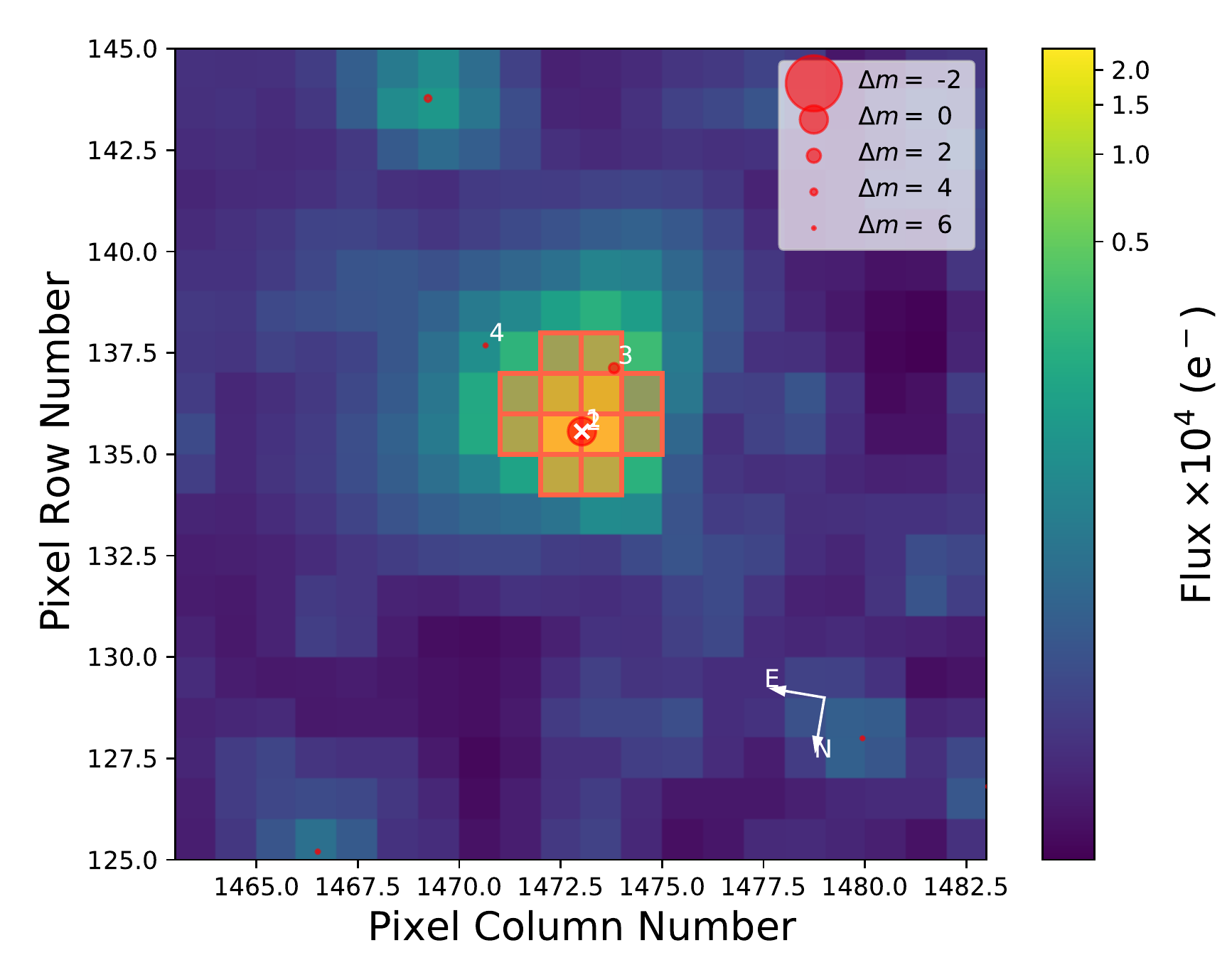}
      \caption{Same as fig. \ref{appfig:HD152799} but for HD 144432 in sector 6. 
              }
         \label{appfig:HD 144432}
\end{figure}

\begin{figure}
   \centering
   \includegraphics[width=\linewidth]{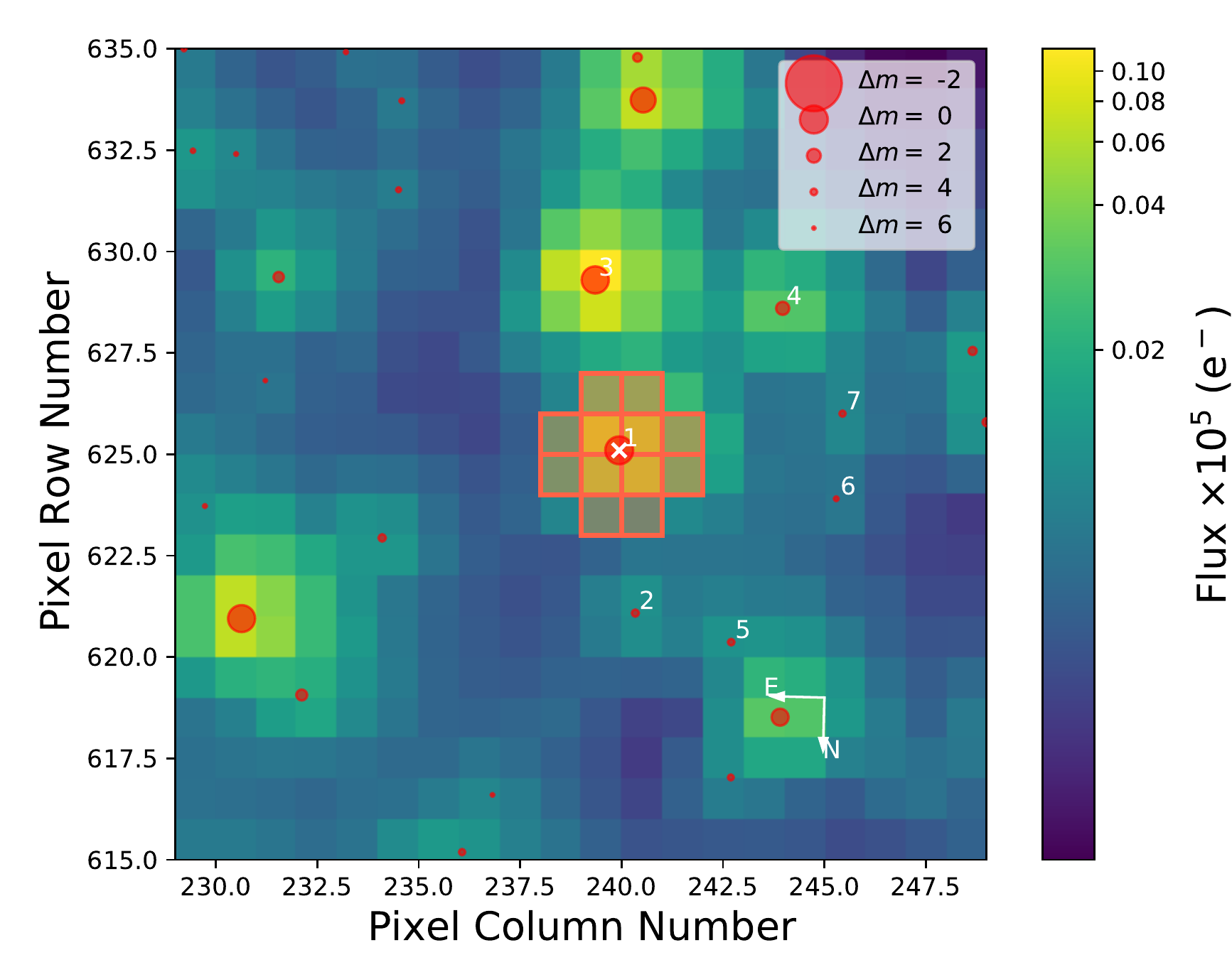}
      \caption{Same as fig. \ref{appfig:HD152799} but for HD 317859 in sector 12.
              }
         \label{appfig:CD-3212910}
\end{figure}

\begin{figure}
   \centering
   \includegraphics[width=\linewidth]{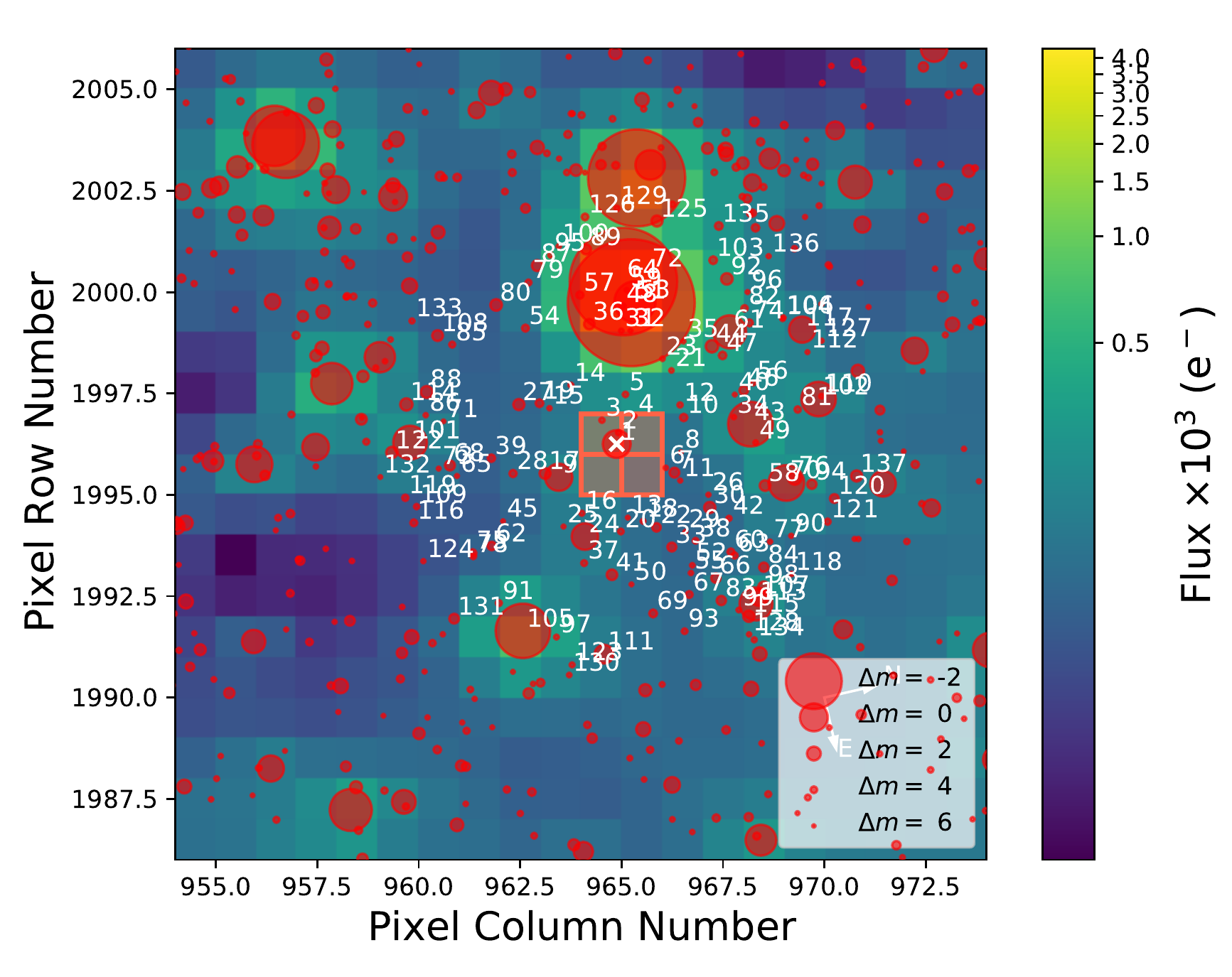}
      \caption{Same as fig. \ref{appfig:HD152799} but for 2MASS 11120327-7637034 in sector 12.
              }
         \label{appfig:2MASS 11120327-7637034}
\end{figure}

In the following we provide the tables of extracted frequencies for the newly discovered pre-main sequence pulsators from TESS data (i.e., Tables \ref{tab:spb} to \ref{tab:dscutican} and the respective Full Frame Images (FFIs) that are described in the main part of the article (i.e., Figures \ref{appfig:HD152822} to \ref{appfig:2MASS 11120327-7637034}). 

It is important to state that the reported uncertainties for frequency, amplitude, and phase correspond only to the uncertainties obtain from the modelling of the adopted light curve. Given that different apertures and and corrections lead to slightly different light curves, we assume that our uncertainties are strongly underestimated. 

  \begin{table}
    \caption[]{Extracted frequencies for the pre-main sequence SPB star BD+65 1637.}
    \label{tab:spb}
    \begin{tabular*}{\linewidth}{lrrrr}
        \hline
        \noalign{\smallskip}
        Designation &\multicolumn{1}{c}{$f$}  &  \multicolumn{1}{c}{$A$}    & \multicolumn{1}{c}{$\phi$} & \multicolumn{1}{c}{SNR}\\
        &\multicolumn{1}{c}{(\cd)} &  \multicolumn{1}{c}{(mmag)}   & \multicolumn{1}{c}{($\frac{\rm rad}{2 \pi}$)} & \\
        \noalign{\smallskip}
        \hline
        \noalign{\smallskip}
        F1                             & $3.076462(4)$  & $1.094(12)$ & $0.854(2)$ & $11.53$\\
        F2                             & $0.507339(4)$  & $1.085(12)$ & $0.692(2)$ & $13.70$\\
        F3                             & $2.151676(4)$  & $1.054(12)$ & $0.284(2)$ & $13.60$\\
        F4                             & $3.018586(6)$  & $0.772(12)$ & $0.555(3)$ & $11.60$\\
        F5                             & $0.199488(12)$  & $0.374(12)$ & $0.522(5)$ & $5.36$\\
        F6                             & $2.905707(14)$  & $0.335(12)$ & $0.592(6)$ & $5.31$\\
        F7                             & $3.190572(14)$  & $0.333(12)$ & $0.731(6)$ & $5.74$\\
        F8                             & $3.261994(14)$  & $0.334(12)$ & $0.744(6)$ & $6.14$\\
        F8                             & $5.228138(15)$  & $0.311(12)$ & $0.806(6)$ & $6.79$\\
        F9                             & $3.09616(2)$  & $0.292(12)$ & $0.500(7)$ & $5.40$\\
        F10                            & $2.98411(2)$  & $0.273(12)$ & $0.798(7)$ & $5.11$\\
        F11                            & $5.74287(2)$  & $0.230(12)$ & $0.433(8)$ & $5.59$\\
        F12                            & $2.54778(2)$  & $0.224(12)$ & $0.425(9)$ & $4.34$\\
        F13                            & $6.09464(2)$  & $0.210(12)$ & $0.975(9)$ & $5.78$\\
        F14                            & $6.16729(2)$  & $0.194(12)$ & $0.791(10)$ & $5.92$\\
        F15                            & $3.08344(2)$  & $0.191(12)$ & $0.846(10)$ & $4.00$\\
        F16                            & $8.59152(3)$  & $0.161(12)$ & $0.292(11)$ & $8.53$\\
        F17                            & $10.51292(3)$  & $0.150(12)$ & $0.139(13)$ & $10.06$\\
        F18                            & $5.41367(3)$  & $0.148(12)$ & $0.870(13)$ & $4.12$\\
        F19                            & $7.32194(3)$  & $0.134(12)$ & $0.647(14)$ & $6.02$\\
        (F20                           & $3.00381(2)$  & $0.191(12)$ & $0.078(10)$ & $3.85$)\\
        (F21                           & $5.17067(3)$  & $0.139(12)$ & $0.843(14)$ & $3.77$)\\
        \noalign{\smallskip}
        \hline

    \end{tabular*}    
    \tablefoot{
    The values in parentheses give the $1\sigma$ uncertainty as reported by the standard error estimates formulated by \citet{montgomery1999}.
    }
 \end{table}

 \begin{table}
    \caption[]{Extracted frequencies for the pre-main sequence SPB candidates found in this work.}
    \label{tab:spbcan}
    \begin{tabular*}{\linewidth}{lrrrr}
        \hline
        \noalign{\smallskip}
        Designation &\multicolumn{1}{c}{$f$}  &  \multicolumn{1}{c}{$A$}    & \multicolumn{1}{c}{$\phi$} & \multicolumn{1}{c}{SNR}\\
        &\multicolumn{1}{c}{(\cd)} &  \multicolumn{1}{c}{(mmag)}   & \multicolumn{1}{c}{($\frac{\rm rad}{2 \pi}$)} & \\
        \noalign{\smallskip}
        \noalign{\smallskip}
        \hline
        \noalign{\smallskip}
        \multicolumn{5}{l}{HD 152822 -- candidate}\\
        \noalign{\smallskip}
        \hline
        \noalign{\smallskip}
        F1                             & $2.25479(2)$  & $1.16(5)$ & $0.135(7)$ & $10.56$\\
        F2                             & $0.84580(4)$  & $0.73(5)$ & $0.310(11)$ & $6.48$\\
        F3                             & $2.31573(5)$  & $0.49(5)$ & $0.333(17)$ & $6.52$\\
        F4                             & $1.97877(7)$  & $0.34(5)$ & $0.40(3)$ & $4.93$\\
        F5                             & $2.47346(8)$  & $0.31(5)$ & $0.13(3)$ & $5.38$\\
        F6                             & $4.84655(12)$  & $0.21(5)$ & $0.31(4)$ & $4.56$\\
        F7                             & $5.15125(13)$  & $0.19(5)$ & $0.09(4)$ & $4.68$\\
        F8                             & $4.51676(14)$  & $0.18(5)$ & $0.55(5)$ & $4.90$\\
        \noalign{\smallskip}
        \hline
         \noalign{\smallskip}
        \multicolumn{5}{l}{HD 152799 -- candidate}\\
        \noalign{\smallskip}
        \hline
        \noalign{\smallskip}          
        F1                             & $1.40530(2)$  & $1.33(5)$ & $0.257(6)$ & $10.04$\\
        F2                             & $0.70984(4)$  & $0.52(5)$ & $0.433(15)$ & $4.75$\\
        F3                             & $1.16862(7)$  & $0.35(5)$ & $0.70(2)$ & $4.41$\\
        F4                             & $2.25479(14)$  & $0.17(5)$ & $0.67(5)$ & $4.55$\\
        F5                             & $4.4988(2)$  & $0.11(5)$ & $0.68(7)$ & $4.64$\\  
        (F6                             & $1.49842(9)$  & $0.25(5)$ & $0.81(3)$ & $3.88$)\\
        (F7                             & $1.0898(1)$  & $0.24(5)$ & $0.92(3)$ & $3.70$)\\
        \noalign{\smallskip}
        \hline
         \noalign{\smallskip}
        \multicolumn{5}{l}{HD 329271 -- candidate}\\
        \noalign{\smallskip}
        \hline
        \noalign{\smallskip}            
        F1                             & $2.473508(2)$  & $1.54(6)$ & $0.301(6)$ & $13.84$\\
        F2                             & $2.51648(8)$  & $0.36(5)$ & $0.92(3)$ & $5.49$\\
        \noalign{\smallskip}
        \hline
         \noalign{\smallskip}
        \multicolumn{5}{l}{CD-32 12908 -- candidate}\\
        \noalign{\smallskip}
        \hline
        \noalign{\smallskip}            
        F1                             & $2.73037(2)$  & $3.22(15)$ & $0.268(8)$ & $9.80$\\
        F2                             & $2.68658(3)$  & $2.59(15)$ & $0.91(1)$ & $11.63$\\
        F3                             & $0.08751(7)$  & $1.11(15)$ & $0.21(2)$ & $4.73$\\
        F4                             & $5.42128(7)$  & $0.95(15)$ & $0.66(3)$ & $6.06$\\
        F5                             & $1.21202(9)$  & $0.88(15)$ & $0.43(3)$ & $5.79$\\  
        F6                             & $5.67506(9)$  & $0.84(15)$ & $0.14(3)$ & $6.60$\\  
        F7                             & $5.37315(11)$  & $0.68(15)$ & $0.36(4)$ & $6.41$\\  
        F8                             & $3.06725(11)$  & $0.66(15)$ & $0.37(4)$ & $6.20$\\  
        F9                             & $1.26015(13)$  & $0.59(15)$ & $0.41(4)$ & $4.52$\\  
        F10                             & $1.17264(15)$  & $0.51(15)$ & $0.04(5)$ & $4.31$\\ 
        F11                            & $2.7785(2)$  & $0.28(15)$ & $0.28(6)$ & $4.54$\\  
        F12                            & $7.2940(3)$  & $0.24(15)$ & $0.16(10)$ & $5.51$\\ 
        (F13                             & $11.4945(3)$  & $0.37(15)$ & $0.137(15)$ & $3.73$)\\  
        \noalign{\smallskip}
        \hline
         \noalign{\smallskip}
        \multicolumn{5}{l}{HD 76534 -- candidate}\\
        \noalign{\smallskip}
        \hline
        \noalign{\smallskip}    
        F1                             & $5.289865(6)$  & $1.131(13)$ & $0.108(2)$ & $13.33$\\
        F2                             & $2.611675(9)$  & $0.761(13)$ & $0.901(3)$ & $8.30$\\
        F3                             & $2.574505(10)$  & $0.645(13)$ & $0.149(3)$ & $8.38$\\
        F4                             & $0.491034(11)$  & $0.583(13)$ & $0.782(4)$ & $6.38$\\
        F5                             & $2.862083(12)$  & $0.573(13)$ & $0.178(4)$ & $10.37$\\  
        F6                             & $1.67265(2)$  & $0.409(13)$ & $0.135(5)$ & $6.31$\\  
        F7                             & $0.43235(2)$  & $0.387(13)$ & $0.195(5)$ & $4.93$\\  
        F8                             & $5.14901(2)$  & $0.312(13)$ & $0.147(7)$ & $5.43$\\  
        F9                             & $4.85361(2)$  & $0.324(13)$ & $0.903(6)$ & $5.95$\\  
        F10                            & $5.090322(2)$ & $0.245(13)$ & $0.114(8)$ & $4.94$\\ 
        F11                            & $2.36518(3)$  & $0.220(13)$ & $0.909(9)$ & $4.44$\\  
        F12                            & $5.22139(3)$  & $0.213(13)$ & $0.834(10)$ & $4.64$\\ 
        F13                            & $2.42974(3)$  & $0.209(13)$ & $0.166(10)$ & $4.56$\\ 
        F14                            & $9.360949(3)$  & $0.204(13)$ & $0.595(11)$ & $8.02$\\ 
        F15                            & $9.45681(4)$  & $0.164(13)$ & $0.457(13)$ & $7.78$\\ 
        F16                            & $2.59798(4)$  & $0.156(13)$ & $0.014(13)$ & $4.00$\\ 
        (F17                            & $4.57386(4)$  & $0.165(13)$ & $0.215(13)$ & $3.77$)\\ 
        (F18                            & $5.35051(5)$  & $0.145(13)$ & $0.043(13)$ & $3.66$)\\ 
        \noalign{\smallskip}
        \hline

    \end{tabular*}    
    \tablefoot{
    The values in parentheses give the $1\sigma$ uncertainty as reported by the standard error estimates formulated by \citet{montgomery1999}.
    }
 \end{table}
 
  \begin{table}
    \caption[]{Extracted frequencies for the pre-main sequence $\gamma$ Doradus candidate HD 317859.}
    \label{tab:gamdor}
    \begin{tabular*}{\linewidth}{lrrrr}
        \hline
        \noalign{\smallskip}
        Designation &\multicolumn{1}{c}{$f$}  &  \multicolumn{1}{c}{$A$}    & \multicolumn{1}{c}{$\phi$} & \multicolumn{1}{c}{SNR}\\
        &\multicolumn{1}{c}{(\cd)} &  \multicolumn{1}{c}{(mmag)}   & \multicolumn{1}{c}{($\frac{\rm rad}{2 \pi}$)} & \\
        \noalign{\smallskip}
        \hline
        \noalign{\smallskip}
        F1                             & $1.21202(3)$  & $2.51(13)$ & $0.912(8)$ & $7.44$\\
        F2                             & $1.260153(3)$  & $2.01(13)$ & $0.915(10)$ & $7.16$\\
        F3                             & $1.17264(4)$  & $1.48(13)$ & $0.568(14)$ & $6.46$\\
        F4                             & $2.73471(8)$  & $0.78(13)$ & $0.66(3)$ & $6.44$\\
        F5                             & $2.68658(10)$  & $0.62(13)$ & $0.43(5)$ & $6.28$\\
        F6                             & $5.4213(2)$  & $0.28(13)$ & $0.16(7)$ & $4.95$\\
        F7                             & $5.3775(3)$  & $0.22(13)$ & $0.68(10)$ & $4.23$\\
        F8                             & $5.6794(4)$  & $0.18(13)$ & $0.47(11)$ & $4.33$\\
        (F9                           & $2.4022(2)$  & $0.28(13)$ & $0.99(7)$ & $3.64$)\\
        \noalign{\smallskip}
        \hline

    \end{tabular*}    
    \tablefoot{
    The values in parentheses give the $1\sigma$ uncertainty as reported by the standard error estimates formulated by \citet{montgomery1999}.
    }
 \end{table}

 \begin{table}
    \caption[]{Extracted frequencies for the pre-main sequence $\delta$ Scuti stars found in this work.}
    \label{tab:dscuti}
    \begin{tabular*}{\linewidth}{lrrrr}
        \hline
        \noalign{\smallskip}
        Designation &\multicolumn{1}{c}{$f$}  &  \multicolumn{1}{c}{$A$}    & \multicolumn{1}{c}{$\phi$} & \multicolumn{1}{c}{SNR}\\
        &\multicolumn{1}{c}{(\cd)} &  \multicolumn{1}{c}{(mmag)}   & \multicolumn{1}{c}{($\frac{\rm rad}{2 \pi}$)} & \\
        \noalign{\smallskip}
        \noalign{\smallskip}
        \hline
        \noalign{\smallskip}
        \multicolumn{5}{l}{HD 287823}\\
        \noalign{\smallskip}
        \hline
        \noalign{\smallskip}
         F1   & $20.32224(5)$  & $0.189(17)$ & $0.713(14)$ & $6.41$  \\
         F2   & $23.74746(5)$  & $0.186(17)$ & $0.428(14)$ & $5.22$  \\
         F3   & $19.69683(7)$  & $0.137(17)$ & $0.803(19)$ & $6.56$  \\
         F4   & $16.41862(8)$  & $0.107(17)$ & $0.39(2)$   & $6.07$  \\
         F5   & $20.99800(9)$  & $0.103(17)$ & $0.17(3)$   & $5.87$  \\
         F6   & $19.06234(11)$ & $0.083(17)$ & $0.69(3)$   & $5.01$  \\
         F7   & $23.22791(12)$ & $0.078(17)$ & $0.89(3)$   & $4.31$  \\
         F8   & $16.14735(13)$ & $0.072(17)$ & $0.09(4)$   & $4.87$  \\
         F9   & $21.40260(13)$ & $0.070(17)$ & $0.63(4)$   & $4.47$  \\
         F10  & $18.65773(14)$ & $0.063(17)$ & $0.08(4)$   & $4.46$  \\
         F11  & $19.44395(18)$ & $0.051(17)$ & $0.52(5)$   & $4.08$  \\
         (F12 & $23.97275(9)$  & $0.100(17)$ & $0.73(3)$   & $3.58$) \\
        \noalign{\smallskip}
        \hline
         \noalign{\smallskip}
        \multicolumn{5}{l}{HD 290500}\\
        \noalign{\smallskip}
        \hline
        \noalign{\smallskip}   
         F1   & $22.72214(3)$  & $0.85(5)$ & $0.995(9)$ & $8.24$  \\
         F2   & $20.92903(7)$  & $0.37(5)$ & $0.70(2)$  & $5.34$  \\
         F3   & $23.46700(8)$  & $0.34(5)$ & $0.01(2)$  & $5.05$  \\
         F4   & $23.75206(8)$  & $0.33(5)$ & $0.83(2)$  & $6.14$  \\
         F5   & $19.38418(8)$  & $0.32(5)$ & $0.23(2)$  & $5.85$  \\
         F6   & $21.76122(9)$  & $0.29(5)$ & $0.17(3)$  & $5.02$  \\
         F7   & $19.22786(10)$ & $0.26(5)$ & $0.70(3)$  & $5.67$  \\
         F8   & $21.03018(11)$ & $0.23(5)$ & $0.21(3)$  & $4.97$  \\
         F9   & $21.68766(13)$ & $0.20(5)$ & $0.72(4)$  & $5.06$  \\
         F10  & $13.14041(13)$ & $0.20(5)$ & $0.43(4)$  & $6.00$  \\
         F11  & $20.22557(13)$ & $0.20(5)$ & $0.50(4)$  & $6.38$  \\
         F12  & $2.84142(15)$  & $0.17(5)$ & $0.04(5)$  & $4.00$  \\
         F13  & $23.21872(17)$ & $0.16(5)$ & $0.68(5)$  & $4.22$  \\
         F14  & $17.61864(18)$ & $0.15(5)$ & $0.50(5)$  & $4.32$  \\
         F15  & $21.3842(2)$   & $0.12(5)$ & $1.00(6)$  & $4.21$  \\
         F16  & $19.8761(3)$   & $0.10(5)$ & $0.40(8)$  & $4.07$  \\
         (F17 & $1.42991(10)$  & $0.25(5)$ & $0.49(3)$  & $3.85$) \\
         (F18 & $21.2831(3)$   & $0.10(5)$ & $0.85(8)$  & $3.67$) \\
        \noalign{\smallskip}
        \hline
         \noalign{\smallskip}
        \multicolumn{5}{l}{HD 144432}\\
        \noalign{\smallskip}
        \hline
        \noalign{\smallskip}   
          F1   & $0.89620(2)$   & $2.08(10)$ & $0.558(8)$ & $8.36$  \\
         F2   & $14.50410(3)$  & $1.77(10)$ & $0.350(9)$ & $14.30$ \\
         F3   & $13.18489(7)$  & $0.67(10)$ & $0.25(2)$  & $11.35$ \\
         F4   & $23.16855(8)$  & $0.63(10)$ & $0.31(2)$  & $11.19$ \\
         F5   & $15.51142(8)$  & $0.58(10)$ & $0.38(3)$  & $9.17$  \\
         F6   & $21.46936(9)$  & $0.57(10)$ & $0.16(3)$  & $12.08$ \\
         F7   & $20.61618(10)$ & $0.47(10)$ & $0.46(3)$  & $11.25$ \\
         F8   & $18.8560(2)$   & $0.25(10)$ & $0.90(6)$  & $6.20$  \\
         F9   & $15.3609(2)$   & $0.24(10)$ & $0.38(7)$  & $6.79$  \\
         F10  & $13.2709(2)$   & $0.19(10)$ & $0.20(8)$  & $6.35$  \\
         F11  & $18.4617(3)$   & $0.18(10)$ & $0.56(9)$  & $5.64$  \\
         F12  & $16.3072(3)$   & $0.16(10)$ & $0.49(10)$ & $4.94$  \\
         F13  & $15.8233(4)$   & $0.11(10)$ & $0.69(14)$ & $4.47$  \\
         (F14 & $23.3693(4)$   & $0.11(10)$ & $0.89(14)$ & $3.84$) \\
         (F15 & $18.5406(5)$   & $0.09(10)$ & $0.31(17)$ & $3.54$) \\
         (F16 & $14.5686(5)$   & $0.09(10)$ & $0.32(17)$ & $3.56$) \\
          \noalign{\smallskip}
         \hline
    \end{tabular*}    
    \tablefoot{
    The values in parentheses give the $1\sigma$ uncertainty as reported by the standard error estimates formulated by \citet{montgomery1999}.
    }
 \end{table}

  \begin{table}
    \caption[]{Extracted frequencies for the pre-main sequence $\delta$ Scuti candidates found in this work.}
    \label{tab:dscutican}
    \begin{tabular*}{\linewidth}{lrrrr}
        \hline
        \noalign{\smallskip}
        Designation &\multicolumn{1}{c}{$f$}  &  \multicolumn{1}{c}{$A$}    & \multicolumn{1}{c}{$\phi$} & \multicolumn{1}{c}{SNR}\\
        &\multicolumn{1}{c}{(\cd)} &  \multicolumn{1}{c}{(mmag)}   & \multicolumn{1}{c}{($\frac{\rm rad}{2 \pi}$)} & \\
        \noalign{\smallskip}
        \noalign{\smallskip}
        \hline
        \noalign{\smallskip}
        \multicolumn{5}{l}{V599 Ori -- candidate}\\
        \noalign{\smallskip}
        \hline
        \noalign{\smallskip}
         F1   & $23.22791(3)$ & $14.7(7)$ & $0.683(8)$  & $11.85$ \\
         F2   & $17.59105(4)$ & $9.4(7)$  & $0.234(12)$ & $8.42$  \\
         F3   & $21.53593(4)$ & $9.3(7)$  & $0.864(13)$ & $12.74$ \\
         F4   & $18.49681(7)$ & $5.6(7)$  & $0.39(2)$   & $8.18$  \\
         F5   & $17.83933(8)$ & $5.0(7)$  & $0.00(2)$   & $11.24$ \\
         F6   & $5.6415(3)$   & $1.3(7)$  & $0.30(9)$   & $8.24$  \\
         F7   & $23.1957(4)$  & $1.1(7)$  & $0.71(11)$  & $6.51$  \\
         F8   & $3.9449(4)$   & $0.9(7)$  & $0.38(12)$  & $4.50$  \\
         F9   & $4.7311(6)$   & $0.7(7)$  & $0.99(16)$  & $5.16$  \\
         F10  & $7.1449(6)$   & $0.6(7)$  & $0.77(18)$  & $4.99$  \\
         F11  & $23.2555(7)$  & $0.6(7)$  & $0.9(2)$    & $4.27$  \\
         F12  & $7.5541(9)$   & $0.5(7)$  & $0.1(2)$    & $4.06$  \\
         F13  & $12.5105(9)$  & $0.4(7)$  & $0.6(3)$    & $4.82$  \\
         F14  & $8.3817(10)$  & $0.4(7)$  & $0.5(3)$    & $4.12$  \\
         (F15 & $17.6324(6)$  & $0.7(7)$  & $0.54(17)$  & $3.61$) \\
         (F16 & $17.0485(7)$  & $0.6(7)$  & $0.4(2)$    & $3.55$) \\
         (F17 & $23.1635(10)$ & $0.4(7)$  & $0.9(3)$    & $3.67$) \\
        \noalign{\smallskip}
        \hline
         \noalign{\smallskip}
        \multicolumn{5}{l}{TYC 8581-2002-1 -- candidate}\\
        \noalign{\smallskip}
        \hline
        \noalign{\smallskip} 
         F1   & $16.69326(5)$  & $0.155(14)$ & $0.915(15)$ & $10.66$ \\
         F2   & $17.50498(6)$  & $0.121(14)$ & $0.400(19)$ & $10.29$ \\
         F3   & $16.27781(6)$  & $0.120(14)$ & $0.993(19)$ & $9.85$  \\
         F4   & $15.60543(7)$  & $0.102(14)$ & $0.20(2)$   & $9.59$  \\
         F5   & $18.56213(7)$  & $0.100(14)$ & $0.81(2)$   & $8.14$  \\
         F6   & $1.75382(7)$   & $0.097(14)$ & $0.73(2)$   & $4.25$  \\
         F7   & $11.15057(7)$  & $0.095(14)$ & $0.33(2)$   & $8.81$  \\
         F8   & $18.74492(8)$  & $0.088(14)$ & $0.81(3)$   & $7.25$  \\
         F9   & $15.17848(11)$ & $0.066(14)$ & $0.07(3)$   & $6.94$  \\
         F10  & $22.23083(11)$ & $0.063(14)$ & $0.64(4)$   & $6.35$  \\
         F11  & $19.65890(11)$ & $0.062(14)$ & $0.30(4)$   & $5.50$  \\
         F12  & $20.30956(14)$ & $0.052(14)$ & $0.16(4)$   & $5.24$  \\
         F13  & $16.62806(17)$ & $0.042(14)$ & $0.58(5)$   & $4.62$  \\
         F14  & $16.98599(17)$ & $0.040(14)$ & $0.14(6)$   & $4.50$  \\
         (F15 & $14.3182(2)$   & $0.033(14)$ & $0.50(7)$   & $3.55$) \\
        \noalign{\smallskip}
        \hline
         \noalign{\smallskip}
        \multicolumn{5}{l}{HD 135344A -- candidate}\\
        \noalign{\smallskip}
        \hline
        \noalign{\smallskip} 
         F1  & $74.26705(3)$   & $0.289(16)$ & $0.299(9)$  & $8.33$  \\
         F2  & $73.94123(6)$   & $0.134(16)$ & $0.927(19)$ & $6.48$  \\
         F3  & $60.53231(7)$   & $0.115(16)$ & $0.42(2)$   & $5.60$  \\
         F4  & $79.93974(7)$   & $0.114(16)$ & $0.62(2)$   & $5.32$  \\
         F5  & $63.05118(9)$   & $0.087(16)$ & $0.98(3)$   & $4.79$  \\
         F6  & $80.81278(10)$  & $0.075(16)$ & $0.15(3)$   & $4.19$  \\
         F7  & $77.27049(11)$  & $0.073(16)$ & $0.42(3)$   & $4.17$  \\
         (F8 & $259.40209(10)$ & $0.078(16)$ & $0.83(3)$   & $3.54$) \\
        \noalign{\smallskip}
        \hline
    \end{tabular*}    
    \tablefoot{
    The values in parentheses give the $1\sigma$ uncertainty as reported by the standard error estimates formulated by \citet{montgomery1999}.
    }
 \end{table}
 
   \begin{table}
    \caption[]{Extracted frequencies for HD 329379.}
    \label{tab:HD 329379}
    \tabcolsep=0.1175cm
    \begin{tabular*}{\linewidth}{lrrrr}
        \hline
        % \columnsep{0.1}
        \noalign{\smallskip}
        Designation &\multicolumn{1}{c}{$f$}  &  \multicolumn{1}{c}{$A$}    & \multicolumn{1}{c}{$\phi$} & \multicolumn{1}{c}{SNR}\\
        &\multicolumn{1}{c}{(\cd)} &  \multicolumn{1}{c}{(mmag)}   & \multicolumn{1}{c}{($\frac{\rm rad}{2 \pi}$)} & \\
        \noalign{\smallskip}
        \hline
        \noalign{\smallskip}
         F1 = 2$f_{\rm orb}$  & $0.889013(15)$ & $32.3(2)$ & $0.283(5)$  & $13.13$ \\
         F2 = $f_{\rm orb}$  & $0.44451(3)$   & $14.2(2)$ & $0.283(11)$ & $11.25$ \\
         F3   & $6.30912(14)$  & $3.5(2)$  & $0.91(5)$   & $13.41$ \\
         F4 = 4$f_{\rm orb}$  & $1.7816(2)$    & $2.2(2)$  & $0.45(7)$   & $5.71$  \\
         F5 = 6$f_{\rm orb}$  & $2.6670(4)$    & $1.1(2)$  & $0.35(15)$  & $5.99$  \\
         F6 = F3 - 2$f_{\rm orb}$ & $5.4165(5)$    & $1.0(2)$  & $0.72(16)$  & $7.15$  \\
         F7 = 8$f_{\rm orb}$  & $3.5632(5)$    & $1.0(2)$  & $0.62(16)$  & $6.91$  \\
         F8 = 14$f_{\rm orb}$  & $6.2123(6)$    & $0.9(2)$  & $0.79(19)$  & $8.35$  \\
         F9 = 10$f_{\rm orb}$  & $4.4558(8)$    & $0.6(2)$  & $0.8(3)$    & $6.38$  \\
         F10 = 13$f_{\rm orb}$ & $5.7642(10)$   & $0.5(2)$  & $0.7(3)$    & $6.71$  \\
         F11  & $7.4885(13)$   & $0.4(2)$  & $0.8(4)$    & $5.58$  \\
         F12 = 12$f_{\rm orb}$ & $5.3305(15)$   & $0.3(2)$  & $0.4(5)$    & $4.92$  \\
         F13 = F3 - 4$f_{\rm orb}$ & $4.5275(16)$   & $0.3(2)$  & $0.7(5)$    & $4.29$  \\
         F14 = 16$f_{\rm orb}$ & $7.1013(17)$   & $0.3(2)$  & $0.9(6)$    & $4.77$  \\
         F15 = 17$f_{\rm orb}$ & $7.5459(19)$   & $0.3(2)$  & $0.4(6)$    & $4.82$  \\
         F16 = F3 + 2 $f_{\rm orb}$ & $7.198(2)$     & $0.2(2)$  & $0.5(7)$    & $4.72$  \\
         F17 = 15$f_{\rm orb}$ & $6.653(3)$     & $0.2(2)$  & $0.7(9)$    & $4.36$  \\
        \noalign{\smallskip}
        \hline

    \end{tabular*}    
    \tablefoot{
    The values in parentheses give the $1\sigma$ uncertainty as reported by the standard error estimates formulated by \citet{montgomery1999}.
    }
 \end{table}
 
    \begin{table}
    \caption[]{Extracted frequencies for $\beta$ Cephei star HD 96042.}
    \label{tab:HD 96042}
    \begin{tabular*}{\linewidth}{lrrrr}
        \hline
        \noalign{\smallskip}
        Designation &\multicolumn{1}{c}{$f$}  &  \multicolumn{1}{c}{$A$}    & \multicolumn{1}{c}{$\phi$} & \multicolumn{1}{c}{SNR}\\
        &\multicolumn{1}{c}{(\cd)} &  \multicolumn{1}{c}{(mmag)}   & \multicolumn{1}{c}{($\frac{\rm rad}{2 \pi}$)} & \\
        \noalign{\smallskip}
        \hline
        \noalign{\smallskip}
         F1  & $2.036134(15)$ & $3.10(9)$ & $0.079(5)$  & $12.55$ \\
         F2  & $1.94242(3)$   & $1.38(9)$ & $0.456(11)$ & $7.01$  \\
         F3   & $6.88504(7)$   & $0.64(9)$ & $1.06(2)$  & $13.93$ \\
         F4   & $4.41531(8)$   & $0.59(9)$ & $0.17(2)$  & $7.96$  \\
         F5   & $10.13048(8)$  & $0.55(9)$ & $0.53(3)$  & $15.11$ \\
         F6   & $8.85513(9)$   & $0.52(9)$ & $0.17(3)$  & $13.11$ \\
         F7   & $11.46647(13)$ & $0.35(9)$ & $0.33(4)$  & $13.09$ \\
         F8   & $11.2018(3)$   & $0.15(9)$ & $0.31(10)$ & $7.24$  \\
         F9   & $8.4343(4)$    & $0.13(9)$ & $0.98(12)$ & $4.58$  \\
         F10  & $9.6968(4)$    & $0.13(9)$ & $0.24(12)$ & $5.18$  \\
         F11  & $9.6214(4)$    & $0.12(9)$ & $0.46(12)$ & $5.32$  \\

         (F12 & $2.14640(7)$   & $0.67(9)$ & $0.60(2)$   & $3.80$) \\
         (F13 & $1.90015(7)$   & $0.65(9)$ & $0.24(2)$   & $3.65$) \\
         (F14 & $2.07473(8)$   & $0.56(9)$ & $0.19(3)$   & $3.63$) \\
         (F15 & $1.95896(8)$   & $0.55(9)$ & $0.84(3)$   & $3.52$) \\
         (F16 & $2.18867(9)$   & $0.53(9)$ & $0.49(3)$   & $3.78$) \\
         (F17 & $4.8490(2)$    & $0.22(9)$ & $0.57(7)$  & $3.76$) \\
         (F18 & $7.5375(3)$    & $0.14(9)$ & $0.15(11)$ & $3.95$) \\

        \noalign{\smallskip}
        \hline

    \end{tabular*}    
    \tablefoot{
    The values in parentheses give the $1\sigma$ uncertainty as reported by the standard error estimates formulated by \citet{montgomery1999}.
    }
 \end{table}

\begin{table}
    \caption[]{Extracted frequencies for HD 53367.}
    \label{tab:HD 53367}
    \begin{tabular*}{\linewidth}{lrrrr}
        \hline
        \noalign{\smallskip}
        Designation &\multicolumn{1}{c}{$f$}  &  \multicolumn{1}{c}{$A$}    & \multicolumn{1}{c}{$\phi$} & \multicolumn{1}{c}{SNR}\\
        &\multicolumn{1}{c}{(\cd)} &  \multicolumn{1}{c}{(mmag)}   & \multicolumn{1}{c}{($\frac{\rm rad}{2 \pi}$)} & \\
        \noalign{\smallskip}
        \hline
        \noalign{\smallskip}
         F1 & $2.245051(8)$  & $0.755(12)$ & $0.422(2)$ & $7.19$ \\
         F2 & $2.204158(12)$ & $0.542(12)$ & $0.323(4)$ & $6.41$ \\
         F3 & $2.281855(19)$ & $0.336(12)$ & $0.544(6)$ & $5.04$ \\
         F4 & $12.11828(14)$ & $0.047(12)$ & $0.15(4)$ & $4.65$ \\
         
        \noalign{\smallskip}
        \hline

    \end{tabular*}    
    \tablefoot{
    The values in parentheses give the $1\sigma$ uncertainty as reported by the standard error estimates formulated by \citet{montgomery1999}.
    }
 \end{table}

\begin{table}
    \caption[]{Extracted frequencies for HD 152743.}
    \label{tab:HD 152743}
    \begin{tabular*}{\linewidth}{lrrrr}
        \hline
        \noalign{\smallskip}
        Designation &\multicolumn{1}{c}{$f$}  &  \multicolumn{1}{c}{$A$}    & \multicolumn{1}{c}{$\phi$} & \multicolumn{1}{c}{SNR}\\
        &\multicolumn{1}{c}{(\cd)} &  \multicolumn{1}{c}{(mmag)}   & \multicolumn{1}{c}{($\frac{\rm rad}{2 \pi}$)} & \\
        \noalign{\smallskip}
        \hline
        \noalign{\smallskip}
         F1  & $0.39791(3)$  & $1.05(7)$ & $0.722(10)$ & $4.40$  \\
         F2  & $2.19027(3)$  & $0.96(7)$ & $0.073(11)$ & $7.08$  \\
         F3  & $0.99297(4)$  & $0.84(7)$ & $0.215(13)$ & $4.78$  \\
         F4  & $2.44837(5)$  & $0.65(7)$ & $0.629(17)$ & $7.12$  \\
         F5  & $3.20475(13)$ & $0.26(7)$ & $0.96(4)$   & $4.36$  \\
         F6  & $4.45223(14)$ & $0.23(7)$ & $0.66(5)$   & $4.75$  \\
         F7  & $2.92155(16)$ & $0.20(7)$ & $0.41(5)$   & $4.27$  \\
         (F8 & $1.24032(6)$  & $0.56(7)$ & $0.801(19)$ & $3.95$) \\
         
        \noalign{\smallskip}
        \hline

    \end{tabular*}    
    \tablefoot{
    The values in parentheses give the $1\sigma$ uncertainty as reported by the standard error estimates formulated by \citet{montgomery1999}.
    }
 \end{table}
 


%KZ copies

\subsection{Other variable stars}
\begin{figure}
   \centering
   \includegraphics[width=\linewidth]{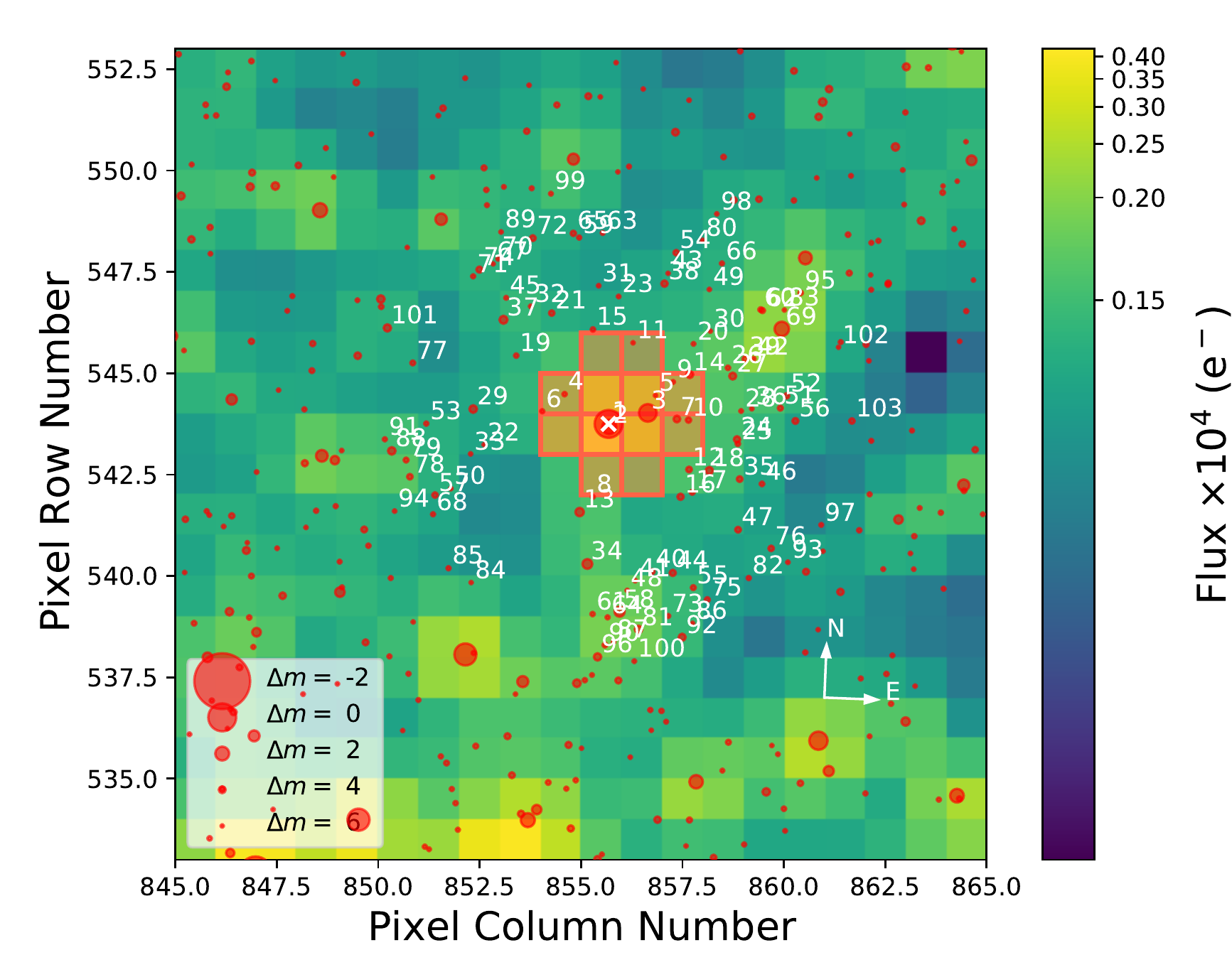}
      \caption{Same as fig. \ref{appfig:HD152799} but for CD-4911096 in sector 12.
              }
         \label{appfig:CD-4911096}
\end{figure}

\textit{\underline{CD-49 11096}} is a B9V \citep{Aidelman2018} star that has been observed in sector 12. No short cadence data is available for this star, hence we use the light curve obtained using the \texttt{RegressionCorrector}. CD-4911096 has a TESS magnitude of $10.584$. Multiple other stars lie within the chosen aperture as the sky region is very crowded. Most notably are GAIA DR2 5938324042340525696 with a TESS magnitude of $12.277$ and GAIA DR2 5938324038039512192 with a TESS magnitude of $11.533$ which do not have an entry in the SIMBAD database. We extract two significant frequencies at $5.45237$ and $5.71764$ \cd\, and the amplitude spectrum shows additional signal around $15$ and $21$\cd\, although no frequency is significant in this range. The TIC reports $T_{\rm eff}  = 6501$~K for one of the two close-by objects, GAIA DR2 5938324038039512192; an effective temperature as would be expected for $\delta$ Scuti stars. Given the frequency ranges, we might see the signal of a pre-main sequence $\delta$ Scuti star. However, more data will be needed to verify this preliminary identification and we restrain from adding CD-4911096 with $T_{\rm eff} = 11450 \pm 308$~K and $\log(g) = 4.26 \pm 0.12$ to our sample. The FFI and chosen aperture are shown in Fig. \ref{appfig:CD-4911096}.

\begin{figure}
   \centering
   \includegraphics[width=\linewidth]{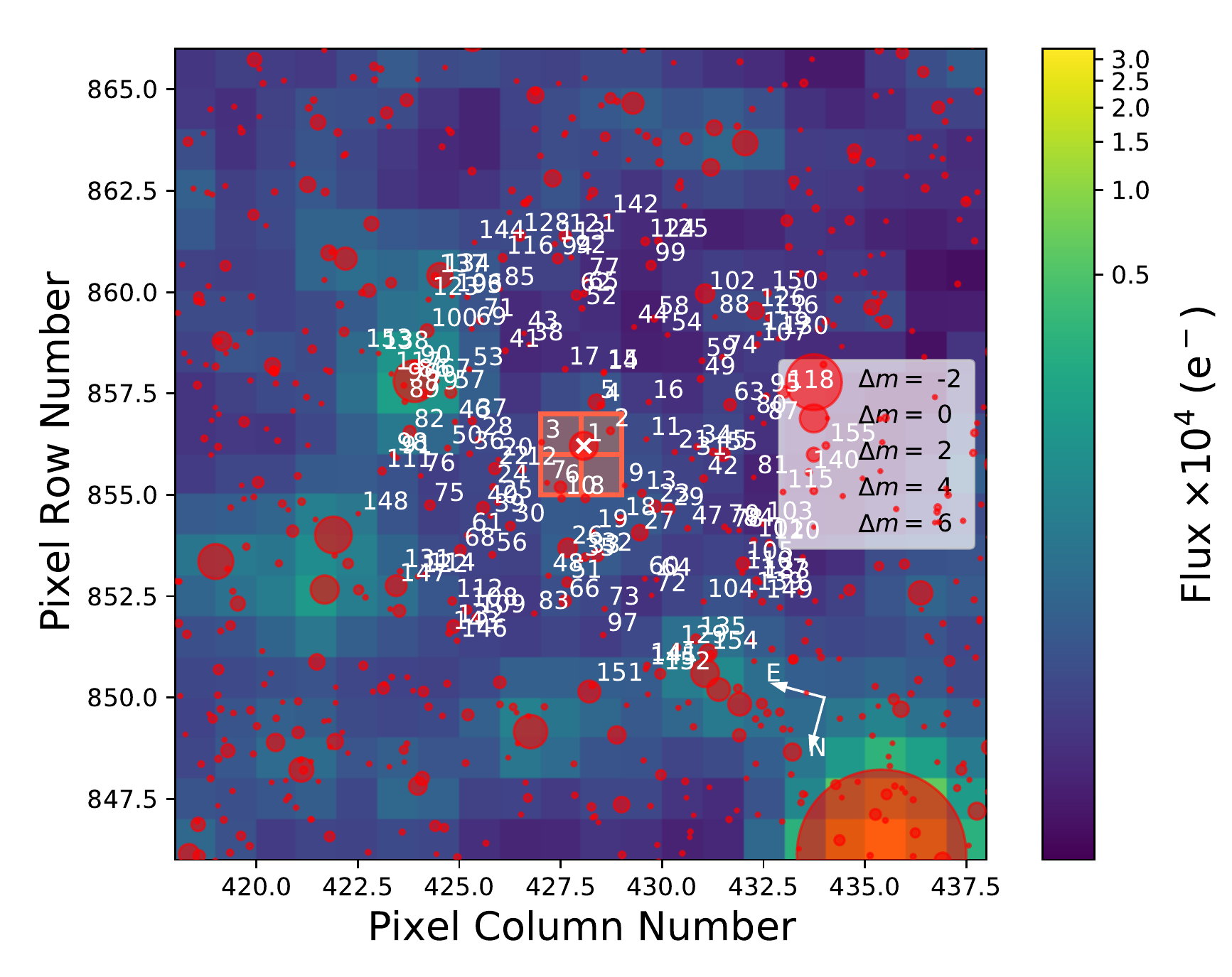}
      \caption{Same as fig. \ref{appfig:HD152799} but for PDS 344 in sector 10. The aperture was shifted in sector 11 which leads to having stars 4 and 5 within the aperture.
              }
         \label{appfig:PDS344}
\end{figure}

\textit{\underline{PDS 344}} is most likely a B5 star \citep{Vieira2003} that has been observed in sector 10 and 11. No short cadence data is available for this star, hence we use the light curve obtained using the \texttt{RegressionCorrector}. PDS 344 has a TESS magnitude of $13.071$. The apertures for sector 10 and sector 11 are slightly shifted with respect to each other, resulting in different stars lying inside the aperture of each sector. Most notably GAIA DR2 5332670397352388608 with a TESS magnitude of $14.374$ lies within the aperture in sector 11. We extract one frequency at $6.96842(3)$~\cd with an amplitude of $0.69(4)$~mmag and a SNR of $16.18$. It is a clear detection of the frequency, as it is evident in both sectors. However, given the crowded nature of the FFIs we restrain from adding PDS 344 to our sample of pulsating stars. More observations are needed for an identification. The FFI and chosen aperture for sector 10 are shown in Fig. \ref{appfig:PDS344}.

\begin{figure}
   \centering
   \includegraphics[width=\linewidth]{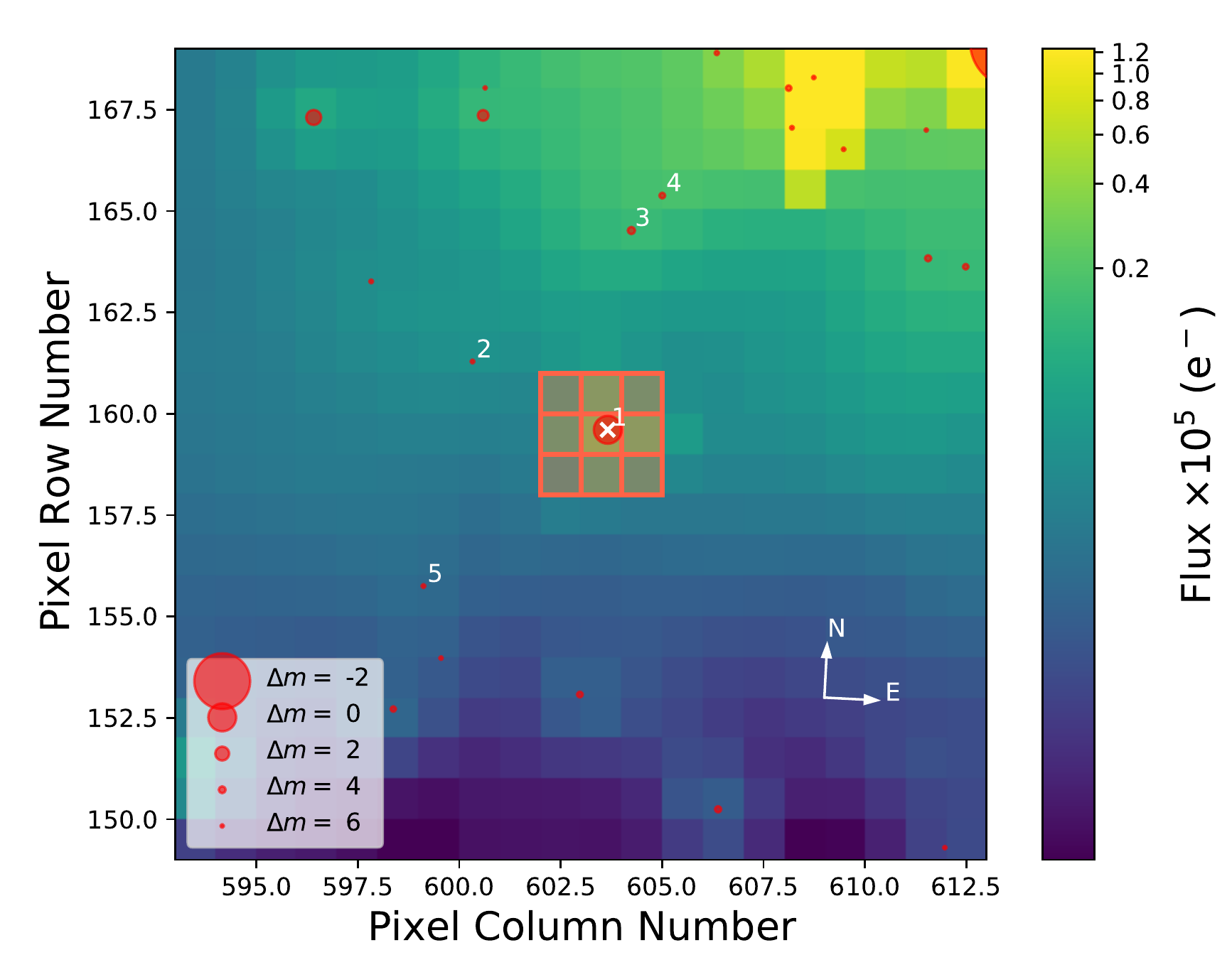}
      \caption{Same as fig. \ref{appfig:HD152799} but for HD 36982 in sector 6. The aperture was shifted in sector 11 which leads to having stars 4 and 5 within the aperture.
              }
         \label{appfig:HD36982}
\end{figure}

\textit{\underline{HD 36982}} is a B1.5Vp star according to SIMBAD that has been observed in sector 6. No short cadence data is available for this star, hence we use the light curve obtained using the \texttt{RegressionCorrector}. There are no other stars in the aperture, but strong background light from nearby O-type stars can be detected. We extract one frequency at $6.08744(3)$~\cd\, with an amplitude of $0.41(3)$~mmag and a SNR of $11.11$. This is similar to the case of PDS 344. More observations are needed for an identification. The FFI and chosen aperture for sector 10 are shown in Fig. \ref{appfig:HD36982}.

\begin{figure}
   \centering
   \includegraphics[width=\linewidth]{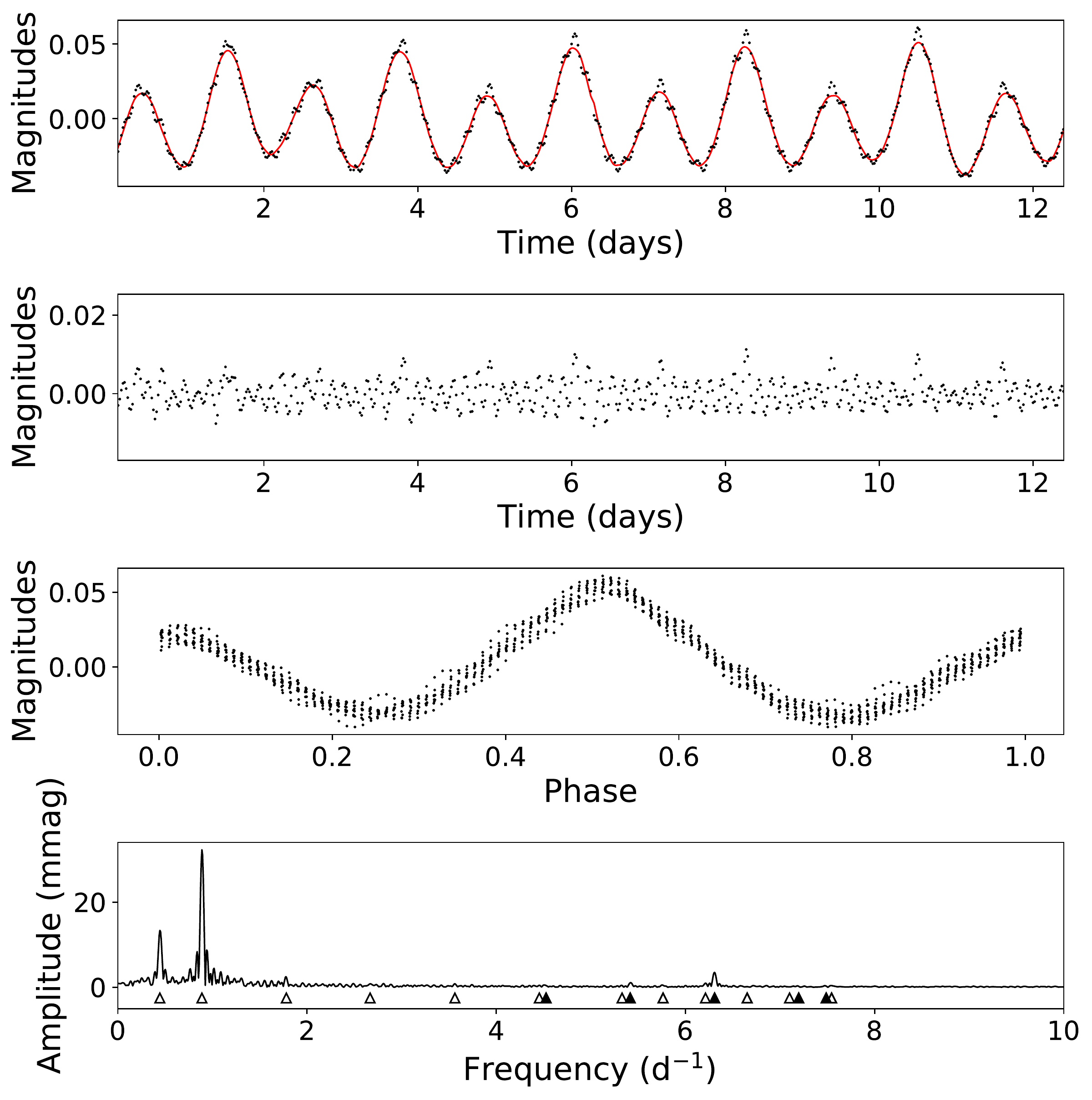}
      \caption{TESS observation of HD 329379. The top panel shows the the first orbit of the TESS observation (black dots) and a simple model for the binary light curve (red line). The second panel shows the residual of the top panel. The third panel shows the folded TESS light curve. The bottom panel shows the amplitude spectrum (black line). Significant frequencies are marked as open triangles for multiples of the orbital frequency and filled triangles otherwise.
              }
         \label{appfig:HD 329379_lcnstuff}
\end{figure}

\begin{figure}
   \centering
   \includegraphics[width=\linewidth]{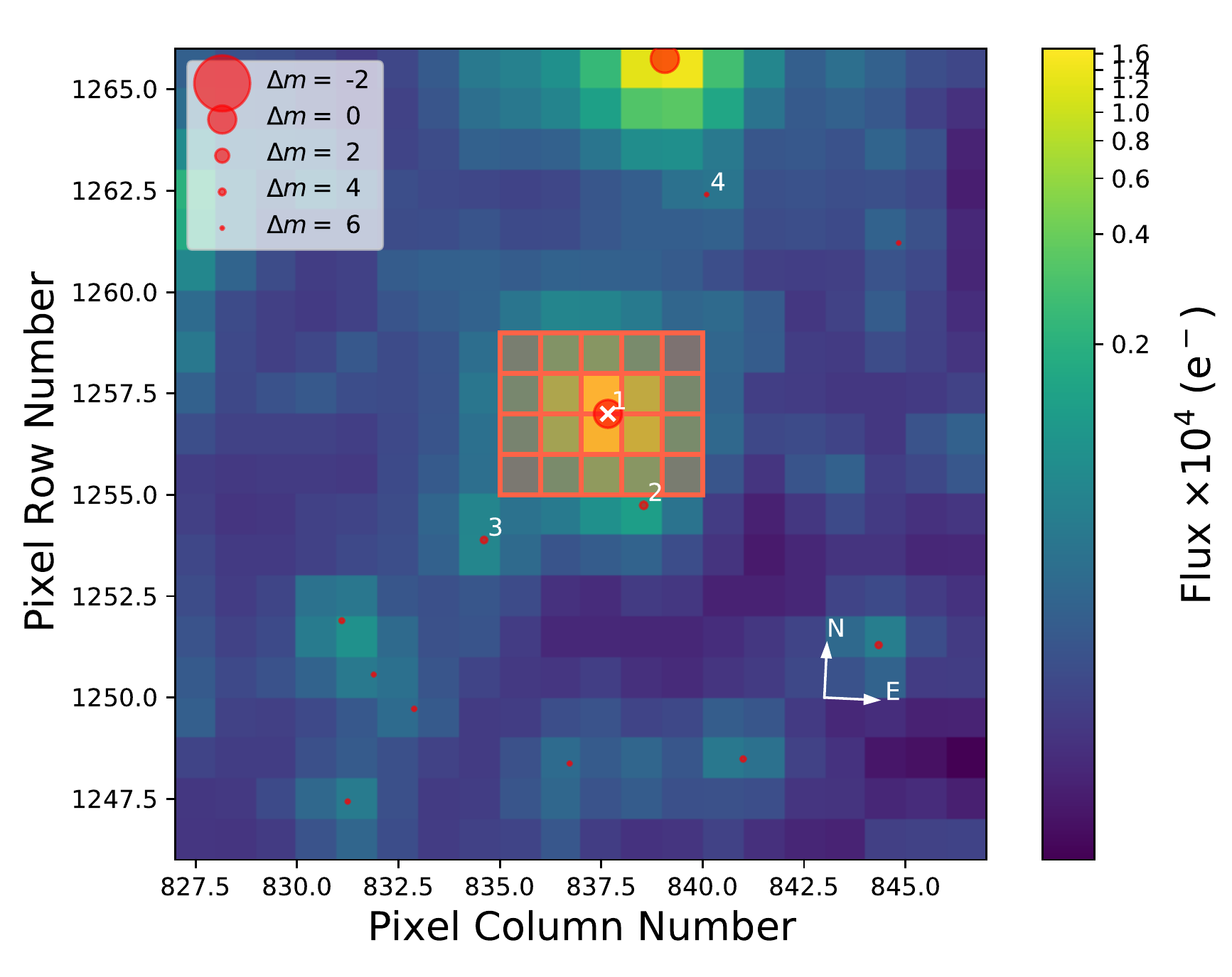}
      \caption{Same as fig. \ref{appfig:HD152799} but for HD 329379 in sector 12.
              }
         \label{appfig:HD 329379}
\end{figure}

\textit{\underline{HD 329379}} is a B0II star \citep{Aidelman2018} that has been observed in sector 12. No short cadence data is available for this star, hence we use the light curve obtained using the \texttt{RegressionCorrector}. There are no other stars in the aperture, but two darker stars in the vicinity from which we expect no contamination. We extract $17$ significant frequencies and interpret the light curve and the frequency spectrum as ellipsoidal variability of a close binary system with an orbital period of about $2.25$~d. All but five of the extracted frequencies are multiples of the orbital frequency F2. The others constitute a quartet (F3, F6, F13, F16) and a singlet F11 (see Table \ref{tab:HD 329379}). The frequencies F3 and F16 and F6 and F13, respectively are each split by \textit{exactly} the orbital frequency while there is a small additional offset of 0.008 $f_{\rm orb}$ between F3 and F6. Thus, this could also be two doublets that coincide to a quartet. 
The light curve (see Fig. \ref{appfig:HD 329379_lcnstuff}) clearly shows additional variability that resembles pulsations. We constructed a simple model for the binary light curve using a Savgol filter with a window length of 31. The residuals show an envelope that rises whenever the light curve obtains a maximum. It is difficult to speak on the origin of the observed light curve and frequencies. Multiple explanations are possible that consist of, e.g. tidally trapped pulsations \citep{Handler2020, Kurtz2020, Fuller2020}, tidally perturbed pulsations \citep{reyniers2003a, reyniers2003b, Steindl2020}, and tidally induced pulsation including heartbeat effects \citep{Fuller2017}. HD 36982 should therefore be studied in more detail. The FFI and chosen aperture for sector 12 are shown in Fig. \ref{appfig:HD 329379}.

\begin{figure}
   \centering
   \includegraphics[width=\linewidth]{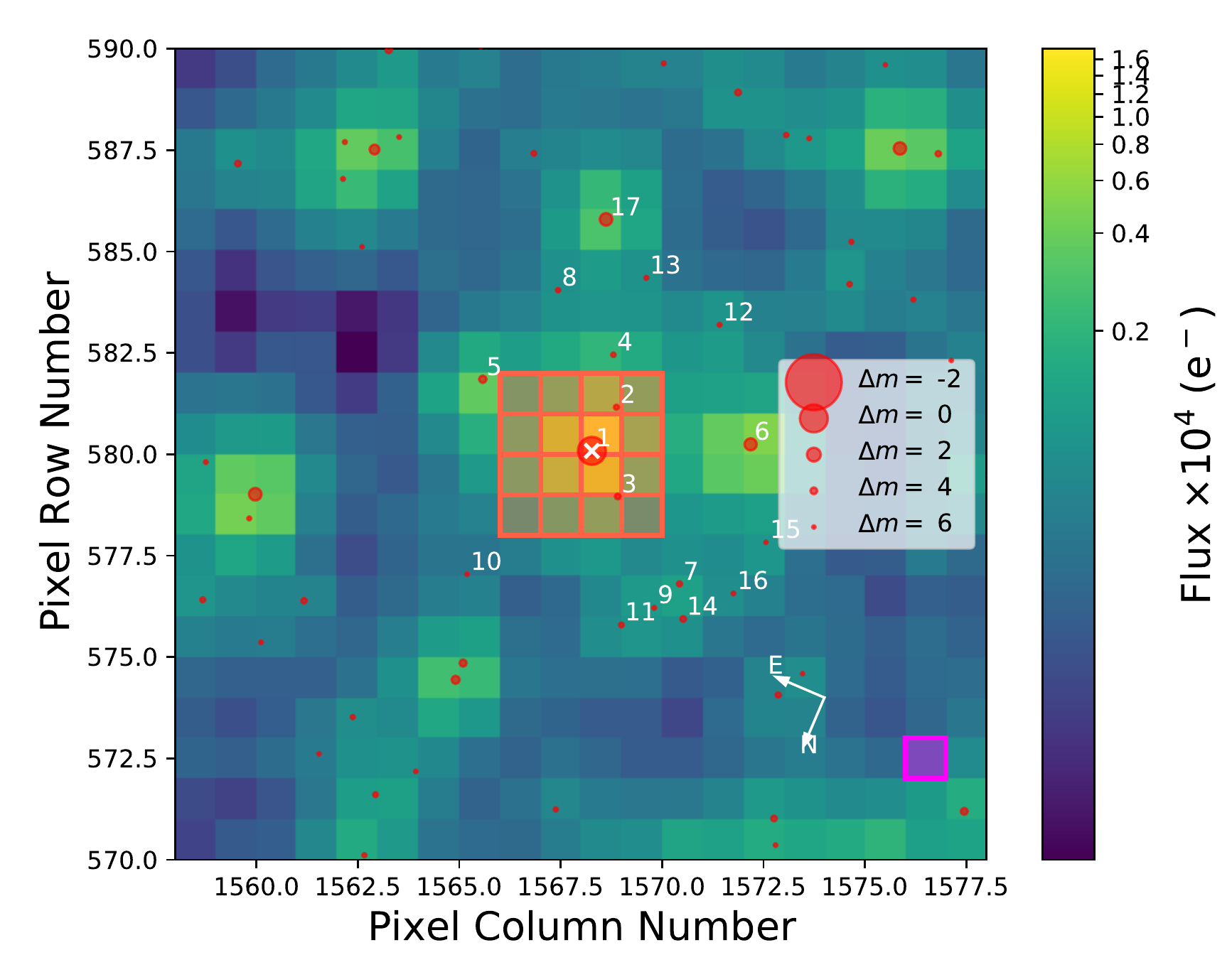}
      \caption{Same as fig. \ref{appfig:HD152799} but for HD 96042 in sector 10. The star has also been observed in sector 11.
              }
         \label{appfig:HD 96042}
\end{figure}

\textit{\underline{HD 96042}} is a B1Vne star \citep{Houk1975} that has been observed in sectors 10 and 11. No short cadence data is available for this star, hence we use the light curve obtained using the \texttt{RegressionCorrector}. Two other stars lie within the chosen aperture, but HD 96042 outshines them by four TESS magnitudes. We extract three significant frequencies in the g-mode ($<5$~\cd) and eight in the p-mode regime ($>6.5$~\cd). Six more frequencies in the g-mode regime are just below our significance criterion. This indicates that multiple g-modes are excited in HD 96042 and a period spacing might be inferred with more observational data. Given the stellar parameters of $T_{\rm eff} = 25000 \pm 1500$~K and $\log(g) = 3.8 \pm 0.2$ \citep{Fairlamb2015}, we classify HD 96042 as $\beta$ Cephei star on the main sequence. The FFI and chosen aperture for sector 10 are shown in Fig. \ref{appfig:HD 96042} and the frequencies are listed in Table \ref{tab:HD 96042}.

\textit{\underline{HD 53367}} is a B0IV/Ve star \citep{Djie2001} that has been observed in sector 7. Short cadence data is available, hence we downloaded the PDCSAP flux from the MAST portal. We extract three frequencies in the g-mode and one additional frequency in the p-mode regime. The residuals show more variability in the low frequency range, hence further observations of HD 53367 might enable the identification of further modes, similar to HD 96042.  Given the stellar parameters of $T_{\rm eff} = 29000 \pm 2000$~K and $\log(g) = 4$ \citep{Alecian2013}, we classify HD 53367 as $\beta$ Cephei star on the main sequence. The frequencies are listed in Table \ref{tab:HD 53367}.

\begin{figure}
   \centering
   \includegraphics[width=\linewidth]{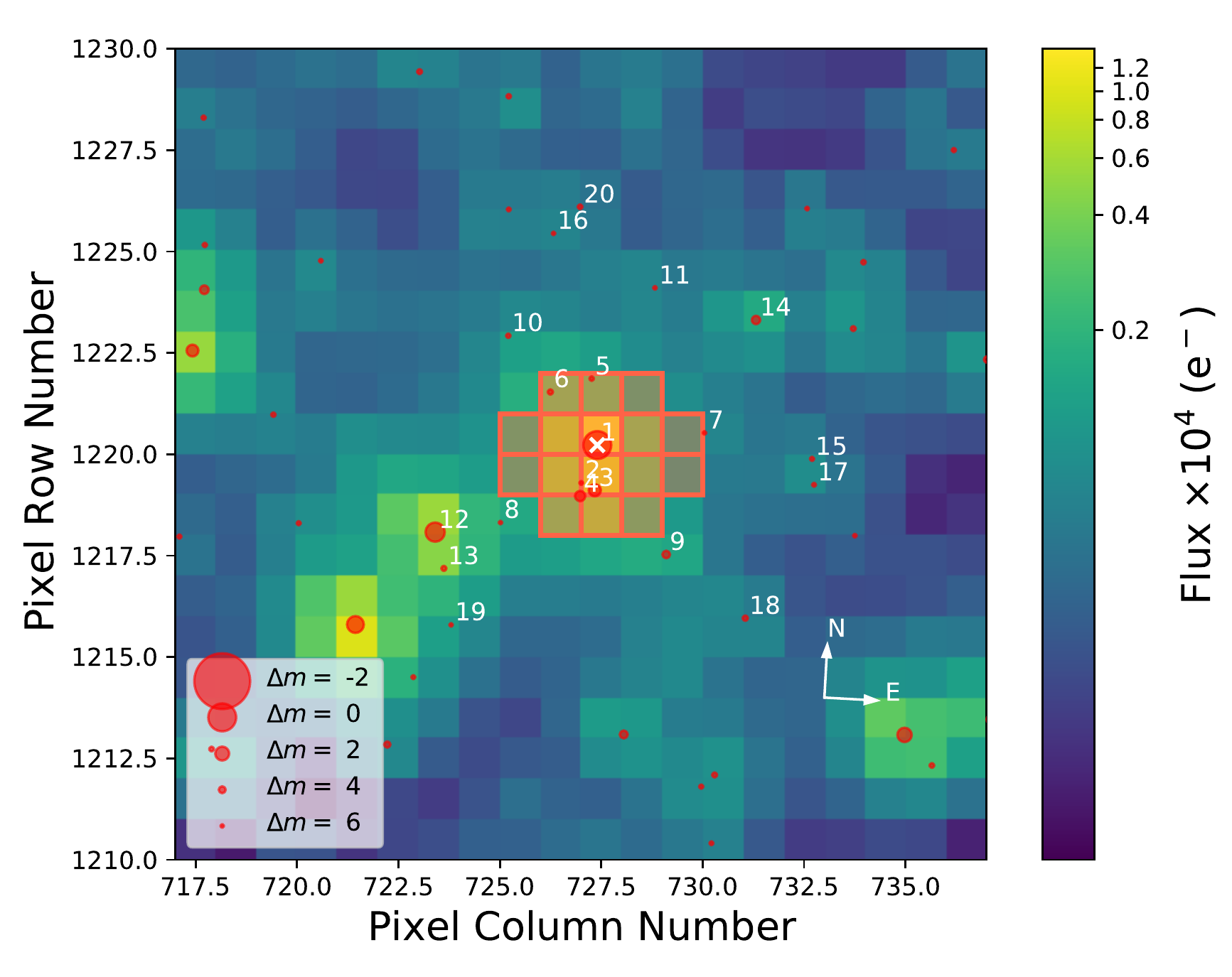}
      \caption{Same as fig. \ref{appfig:HD152799} but for HD 152743 in sector 12.
              }
         \label{appfig:HD 152743}
\end{figure}

\textit{\underline{HD 152743}} is a B2IV star \citep{Houk1978} that has been observed in sector 12. No short cadence data is available for this star, hence we use the light curve obtained using the \texttt{RegressionCorrector}. Five other stars lie within the chosen aperture with up to only one magnitude less than HD 152743. We extract seven significant frequencies clearly indicating g-mode pulsations. With $T_{\rm eff} = 24600 \pm 2371$~K and $\log(g) = 3.76 \pm 0.34$ \citep{Aidelman2018}, HD 152743 could very well be the source of these g-mode pulsations. However, some of the other stars also have effective temperatures expected for g-mode pulsators according to the TIC. Hence, additional observations are needed for a clear identification. The FFI and chosen aperture for sector 10 are shown in Fig. \ref{appfig:HD 152743} and the frequencies are listed in Table \ref{tab:HD 152743}.


\clearpage

\section{Pre-main sequence $\gamma$ Doradus and hybrid stars discovered from CoRoT data: Frequency tables and spectroscopic data}
\label{app:corot}

\subsection{Frequency tables}
Tables \ref{app:tab_CoID1870}, \ref{app:tab_CoID7565} and \ref{app:tab_CoID9101} list the frequencies extracted in the two pre-main sequence $\gamma$ Doradus stars and the pre-main sequence $\delta$ Scuti -- $\gamma$ Doradus hybrid discovered from CoRoT data.

\begin{table}
\begin{scriptsize}
            \caption[]{Extracted frequencies for Cl$^{\star}$ NGC 2264 VAS 196.}
    \label{app:tab_CoID1870}
    \begin{tabular*}{\linewidth}{lrrrrr}
        \hline
        \noalign{\smallskip}
        Designation &\multicolumn{1}{c}{$f$}  &  \multicolumn{1}{c}{$A_{\rm SRa05}$}    & \multicolumn{1}{c}{$\phi$} & \multicolumn{1}{c}{SNR} & \multicolumn{1}{c}{cross-ref }\\
        &\multicolumn{1}{c}{(\cd)} &  \multicolumn{1}{c}{(mmag)}   & \multicolumn{1}{c}{($\frac{\rm rad}{2 \pi}$)} & &
        \multicolumn{1}{c}{ }  \\
        \noalign{\smallskip}
        \hline
        \noalign{\smallskip}
F1	&	0.88674(8)	&	1.344(7)	&	0.4615(9)	&	7.83	&	SRa01, TESS	\\
F2	&	1.3564(1)	&	1.029(7)	&	0.434(1)	&	9.75	&	SRa01, TESS	\\
F3	&	1.3217(1)	&	0.762(7)	&	0.353(2)	&	10.99	&	SRa01, TESS	\\
F4	&	1.4819(2)	&	0.519(7)	&	0.621(2)	&	10.99	&	SRa01	\\
F5	&	0.9083(2)	&	0.451(7)	&	0.062(2)	&	5.84	&	SRa01	\\
F6	&	0.1569(4)	&	0.289(7)	&	0.750(4)	&	4.06	&		\\
F7	&	1.2874(4)	&	0.248(7)	&	0.657(5)	&	5.07	&		\\
F8	&	2.7511(5)	&	0.212(7)	&	0.660(6)	&	7.14	&	SRa01	\\
F9	&	1.4420(5)	&	0.233(7)	&	0.558(5)	&	6.22	&		\\
        \noalign{\smallskip}
        \hline
    \end{tabular*}    
    \tablefoot{
    The values in parentheses give the $1\sigma$ uncertainty as reported by the standard error estimates formulated by \citet{montgomery1999}. The amplitudes ($A_{\rm SRa05}$) are determined from the CoRoT SRa05 data. The last column (cross-ref) marks the frequencies that can also be detected in the CoRoT SRa01 and TESS data.
    }
\end{scriptsize}
 \end{table}

\begin{table}
\begin{scriptsize}
            \caption[]{Extracted frequencies for Cl$^{\star}$ NGC 2264 VAS 219.}
    \label{app:tab_CoID7565}
    \begin{tabular*}{\linewidth}{lrrrrr}
        \hline
        \noalign{\smallskip}
        Designation &\multicolumn{1}{c}{$f$}  &  \multicolumn{1}{c}{$A_{\rm SRa05}$}    & \multicolumn{1}{c}{$\phi$} & \multicolumn{1}{c}{SNR} & \multicolumn{1}{c}{cross-ref }\\
        &\multicolumn{1}{c}{(\cd)} &  \multicolumn{1}{c}{(mmag)}   & \multicolumn{1}{c}{($\frac{\rm rad}{2 \pi}$)} & &
        \multicolumn{1}{c}{ }  \\
        \noalign{\smallskip}
        \hline
        \noalign{\smallskip}
F1	&	2.09926(5)	&	4.07(1)	&	0.2272(5)	&	7.37	&	SRa01, TESS	\\
F2	&	1.92554(6)	&	3.34(1)	&	0.9149(6)	&	7.05	&	SRa01, TESS	\\
F3	&	1.36574(7)	&	2.77(1)	&	0.3421(7)	&	6.44	&	SRa01, TESS	\\
F4	&	0.04759(7)	&	2.51(1)	&	0.8480(8)	&	5.93	&		\\
F5	&	0.1993(1)	&	1.88(1)	&	0.126(1)	&	4.57	&		\\
F6	&	2.0562(1)	&	1.96(1)	&	0.103(1)	&	5.22	&	SRa01, TESS	\\
F7	&	1.3392(1)	&	1.88(1)	&	0.290(1)	&	4.94	&	SRa01	\\
F8	&	2.1481(1)	&	1.88(1)	&	0.521(1)	&	6.10	&	SRa01	\\
F9	&	3.4281(1)	&	1.62(1)	&	0.416(1)	&	9.91	&	SRa01	\\
F10	&	1.8049(1)	&	1.58(1)	&	0.302(1)	&	5.32	&	SRa01, TESS	\\
F11	&	3.7810(1)	&	1.50(1)	&	0.301(1)	&	13.95	&	SRa01	\\
F12	&	0.1087(1)   &	1.57(1)	&	0.929(1)	&	5.00	&		\\
F13	&	1.6359(1)	&	1.45(1)	&	0.303(1)	&	5.01	&	SRa01	\\
F14	&	0.5601(1)	&	1.49(1)	&	0.277(1)	&	4.19	&	SRa01	\\
F15	&	2.4539(1)	&	1.34(1)	&	0.327(1)	&	5.88	&	SRa01	\\
F16	&	1.5451(2)	&	1.24(1)	&	0.416(2)	&	5.10	&	SRa01	\\
F17	&	0.1662(2)	&	1.00(1)	&	0.329(2)	&	4.26	&		\\
F18	&	1.0608(2)	&	1.11(1)	&	0.554(2)	&	3.74	&	SRa01	\\
F19	&	0.3389(2)	&	1.03(1)	&	0.218(2)	&	4.10	&	SRa01	\\
F20	&	2.3782(2)	&	1.01(1)	&	0.961(2)	&	5.07	&		\\
F21	&	0.0773(2)	&	1.77(1)	&	0.278(2)	&	8.98	&		\\
F22	&	1.8715(2)	&	1.00(1)	&	0.119(2)	&	4.93	&		\\
F23	&	1.9070(2)	&	0.89(1)	&	0.316(2)	&	5.51	&		\\
F24	&	1.6848(3)	&	0.72(1)	&	0.751(3)	&	4.26	&		\\         
        \noalign{\smallskip}
        \hline

    \end{tabular*}    
    \tablefoot{
    The values in parentheses give the $1\sigma$ uncertainty as reported by the standard error estimates formulated by \citet{montgomery1999}. The amplitudes ($A_{\rm SRa05}$) are determined from the CoRoT SRa05 data. The last column (cross-ref) marks the frequencies that can also be detected in the CoRoT SRa01 and TESS data.
    }
\end{scriptsize}
 \end{table}
 
\begin{table}
\begin{scriptsize}
            \caption[]{Extracted frequencies for Cl$^{\star}$ NGC 2264 VAS 230.}
    \label{app:tab_CoID9101}
    \begin{tabular*}{\linewidth}{lrrrrr}
        \hline
        \noalign{\smallskip}
        Designation &\multicolumn{1}{c}{$f$}  &  \multicolumn{1}{c}{$A_{\rm SRa01}$}    & \multicolumn{1}{c}{$\phi$} & \multicolumn{1}{c}{SNR} & \multicolumn{1}{c}{cross ref }\\
        &\multicolumn{1}{c}{(\cd)} &  \multicolumn{1}{c}{(mmag)}   & \multicolumn{1}{c}{($\frac{\rm rad}{2 \pi}$)} & &
        \multicolumn{1}{c}{ }  \\
        \noalign{\smallskip}
        \hline
        \noalign{\smallskip}
F1	&	8.1866(4)	&	0.77(1)	&	0.336(3)	&	6.71	&	SRa05, TESS  \\
F2	&	8.4660(6)	&	0.62(1)	&	0.730(3)	&	9.10	&	SRa05, TESS  \\
F3	&	8.5591(8)	&	0.42(1)	&	0.891(4)	&	7.98	&	SRa05, TESS  \\
F4  &   11.8419(9)  &   0.35(1) &   0.175(6)    &   10.87   &   SRa05, TESS  \\
F5	&	8.385(1)	&	0.29(1)	&	0.464(6) 	&	6.33	&	TESS  \\
F6	&	2.3203(7)	&	0.58(1)	&	0.282(4)	&	4.18	&	SRa05  \\
F7	&	2.2407(6)	&	0.54(1)	&	0.650(4)	&	5.07	&	SRa05  \\
F8	&	2.1643(6)	&	0.49(1)	&	0.718(3)	&	4.15	&	SRa05  \\
F9	&	3.3006(8)	&	0.46(1)	&	0.275(5)	&	5.48	&	SRa05  \\
        \noalign{\smallskip}
        \hline
    \end{tabular*}    
    \tablefoot{
    The values in parentheses give the $1\sigma$ uncertainty as reported by the standard error estimates formulated by \citet{montgomery1999}. The amplitudes ($A_{\rm SRa01}$) are determined from the CoRoT SRa01 data. The last column (cross ref) marks the frequencies that can also be detected in the CoRoT SRa05 and TESS data. The first five frequencies are the p-modes and the second four frequencies are the g-modes; both are sorted by decreasing amplitude.
    }
\end{scriptsize}
 \end{table}

\subsection{High-resolution spectroscopy: observations, data reduction, and analysis}
We obtained high-resolution spectra with the Robert G. Tull Coud\'e spectrograph (TS) on the 2.7-m telescope of Mc Donald Observatory for Cl$^{\star}$ NGC 2264 VAS 196 (CoID0223991870) and Cl$^{\star}$ NGC 2264 VAS 219 (CoID0223997565). The cross-dispersed \'echelle spectrograph has a resolving power of 60\,000 in the configuration adopted by us. Each spectrum covers the wavelength range from 3633 to 10\,849\,\AA\, with gaps between the \'echelle orders at wavelengths longer than 5880\,\AA. The spectra for the two stars have signal-to-noise (S/N) ratios of 84 and 80, respectively, calculated at $\sim$5240\,\AA.

Bias and Flat Field frames were obtained at the beginning of each night, while several Th-Ar comparison lamp spectra were obtained each night for wavelength calibration purposes. The reduction was performed using the Image Reduction and Analysis Facility\footnote{IRAF (http://iraf.noao.edu) is distributed by the National Optical Astronomy Observatory, which is operated by the Association of Universities for Research in Astronomy (AURA) under cooperative agreement with the National Science Foundation.} (IRAF). The spectra were normalized by fitting a low order polynomial to carefully selected continuum points. 

For Cl$^{\star}$ NGC 2264 VAS 230 (CoID0223999101) and Cl$^{\star}$ NGC 2264 VAS 196 (CoID 0223991870), a spectrum obtained with the cross-dispersed \'echelle spectrograph UVES at the ESO Very Large Telescope (VLT) was available in the ESO archive\footnote{http://archive.eso.org}. UVES has a resolving power of 40\,000 in the standard mode. All reduction steps were performed within the UVES pipeline (version 5.2.0) and the Reflex software\footnote{http://www.eso.org/sci/software/reflex}. The UVES spectra for Cl$^{\star}$ NGC 2264 VAS 230 (CoID0223999101) and Cl$^{\star}$ NGC 2264 VAS 196 have S/N values of 139 and 157 calculated at $\sim$5845\,\AA, respectively.

The spectroscopic analysis was performed using the {\sc LLmodels} model atmosphere code \citep{Shulyak2004}, the VALD database for atomic line parameters\footnote{http://vald.astro.uu.se} \citep{Kupka1999}, SYNTH3 \citep{Kochukhov2007} for the computation of synthetic spectra and an updated version of the Spectroscopy Made Easy (SME, version 474) software package \citep{Valenti1996,Piskunov2017}. Using these methods for each of the three pre-main sequence pulsators, the effective temperature, surface gravity, projected rotational velocity, and metallicity were determined. A detailed description of the adopted procedure can be found for example in \citet{Zwintz2017c}.

\begin{table}
\begin{footnotesize}
        \caption[]{Fundamental parameters for the six pre-main sequence $\gamma$ Doradus stars discovered from CoRoT data.}
    \label{tab:gamDor-specdata}
    \begin{tabular*}{\linewidth}{lrrrr}
        \hline
        \noalign{\smallskip}
        Identifier &\multicolumn{1}{c}{$T_{\rm eff}$}  &  \multicolumn{1}{c}{log(g)}    & \multicolumn{1}{c}{$v$\,sin\,$i$} & \multicolumn{1}{c}{[Fe/H]}\\
        &\multicolumn{1}{c}{(K)} &  \multicolumn{1}{c}{(dex)}   & \multicolumn{1}{c}{(km\,s$^{-1}$)} & \multicolumn{1}{c}{(dex)} \\
        \noalign{\smallskip}
        \hline
        \noalign{\smallskip}
         Cl$^{\star}$ NGC 2264 VAS 196  & 6784(150)  & 4.0(2) & 77(5) & -0.072  \\
         Cl$^{\star}$ NGC 2264 VAS 219  & 7170(150)  & 4.3(3) & 74(5) & -0.286  \\
         Cl$^{\star}$ NGC 2264 VAS 230  & 7520(150)  & 4.3(4) & 205(25) & -0.200  \\
        \noalign{\smallskip}
        \hline

    \end{tabular*}    
    %\tablefoot{...    }
 \end{footnotesize}
\end{table}

\begin{figure}
   \centering
   \includegraphics[width=\linewidth]{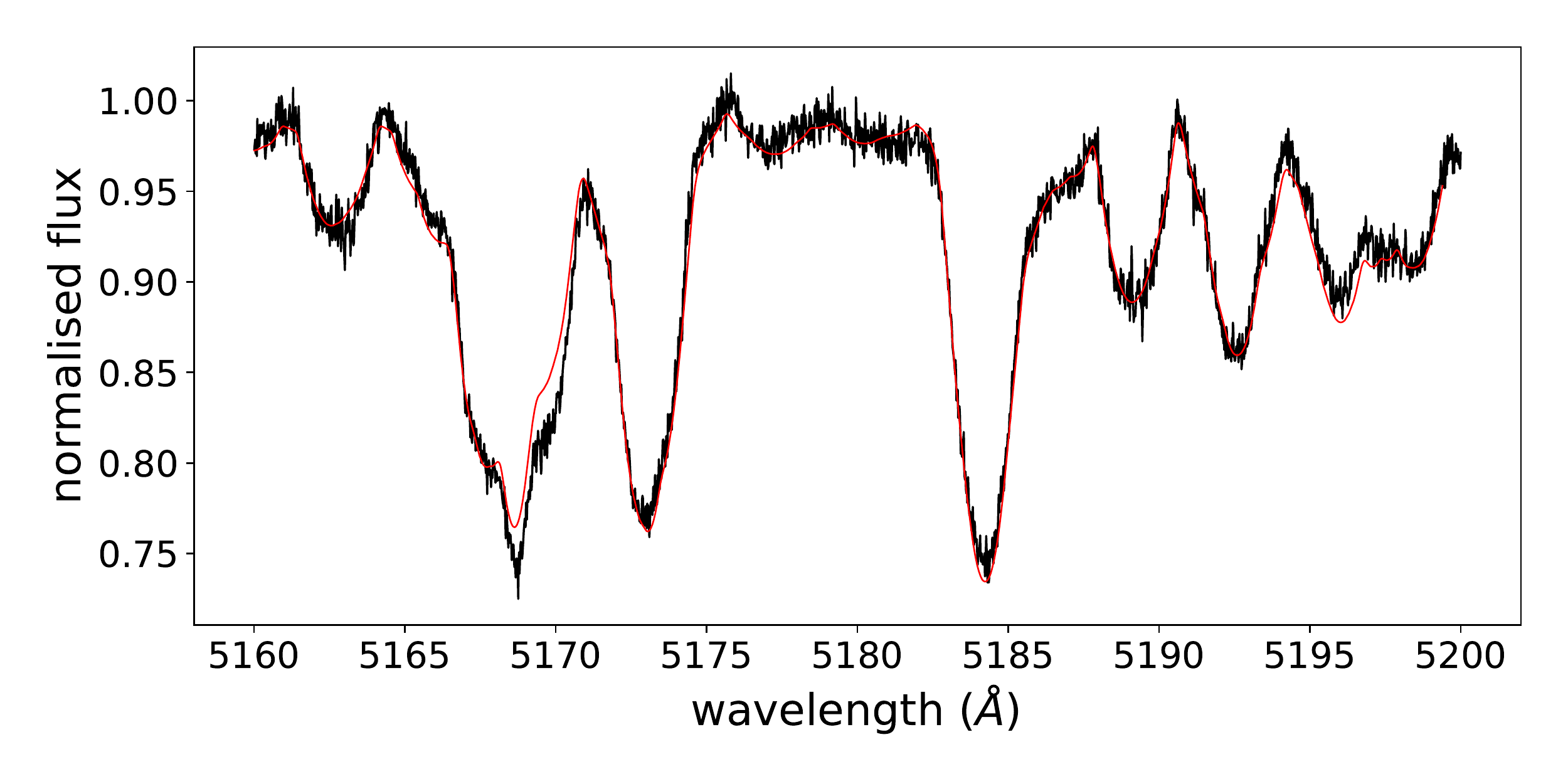}
      \caption{Observed (black line) and synthetic (red line) spectrum of Cl$^{\star}$ NGC 2264 VAS 196 in the region around the \ion{Mg}{i}b triplet. The synthetic spectrum was calculated with $T_{\rm eff}$ = 6784\,K, log(g) of 4.0 and a $v$\,sin\,$i$ of 77\,km\,s$^{-1}$.
              }
         \label{fig:spectrum_VAS196}
\end{figure}

\textit{\underline{Cl$^{\star}$ NGC 2264 VAS 196}} (CoID 0223991870): For this star, two spectra are available: one obtained at Mc Donald Observatory and one observed with the UVES spectrograph at the VLT that we retrieved from the ESO archive. As the S/N value of the UVES spectrum (i.e., 157) is significantly higher than that of the spectrum from Mc Donald observatory (i.e., 50), we base our analysis on the UVES spectrum. We determine a $T_{\rm eff}$ of 6784 $\pm$ 150\,K and a log(g) of 4.0 $\pm$ 0.2. The star rotates moderately fast with a $v$\,sin\,$i$, of 77 km\,s$^{-1}$ and it shows basically solar chemical abundances. The observed and synthetic spectra are shown in Fig. \ref{fig:spectrum_VAS196}.

\begin{figure}
   \centering
   \includegraphics[width=\linewidth]{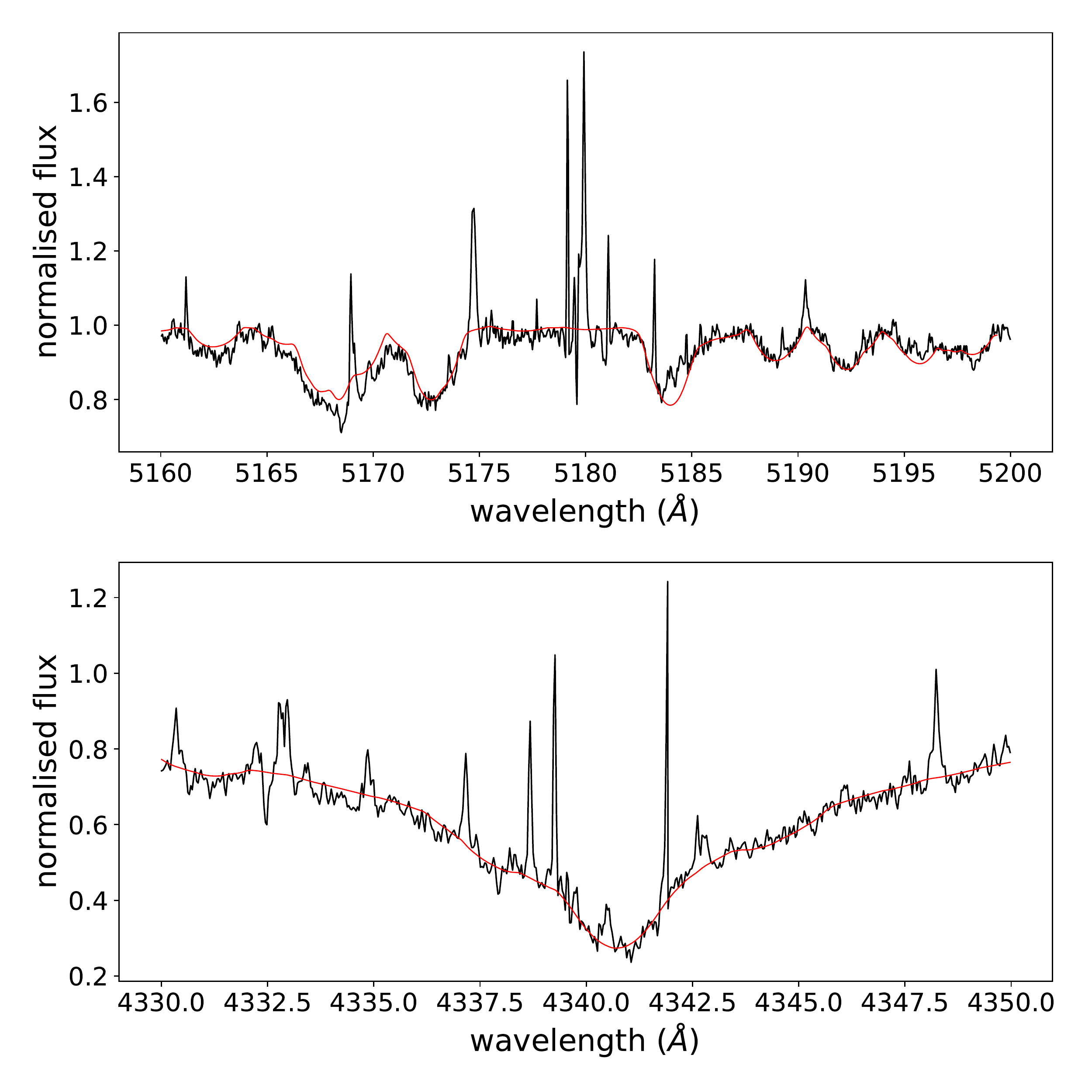}
      \caption{Observed (black line) and synthetic (red line) spectrum of Cl$^{\star}$ NGC 2264 VAS 219: the top panel shows the region around the \ion{Mg}{i}b triplet, the bottom panel the H$\gamma$ line. The synthetic spectrum was calculated with $T_{\rm eff}$ = 7170\,K, log(g) of 4.3 and a $v$\,sin\,$i$ of 74\,km\,s$^{-1}$. 
              }
         \label{fig:spectrum_VAS219}
\end{figure}

\textit{\underline{Cl$^{\star}$ NGC 2264 VAS 219}} (CoID 0223997565): The spectrum obtained at Mc Donald observatory for Cl$^{\star}$ NGC 2264 VAS 219 shows many emission lines (as can be seen in Fig. \ref{fig:spectrum_VAS219}). Our analysis yielded a $T_{\rm eff}$ of 7170 $\pm$ 150\,K, a log(g) of 4.3 $\pm$ 0.3, a $v$\,sin\,$i$, of 74 $\pm$ 5\,km\,s$^{-1}$, and slightly less than solar abundance. The observed and synthetic spectra are shown in Fig. \ref{fig:spectrum_VAS219}.

\begin{figure}
   \centering
   \includegraphics[width=\linewidth]{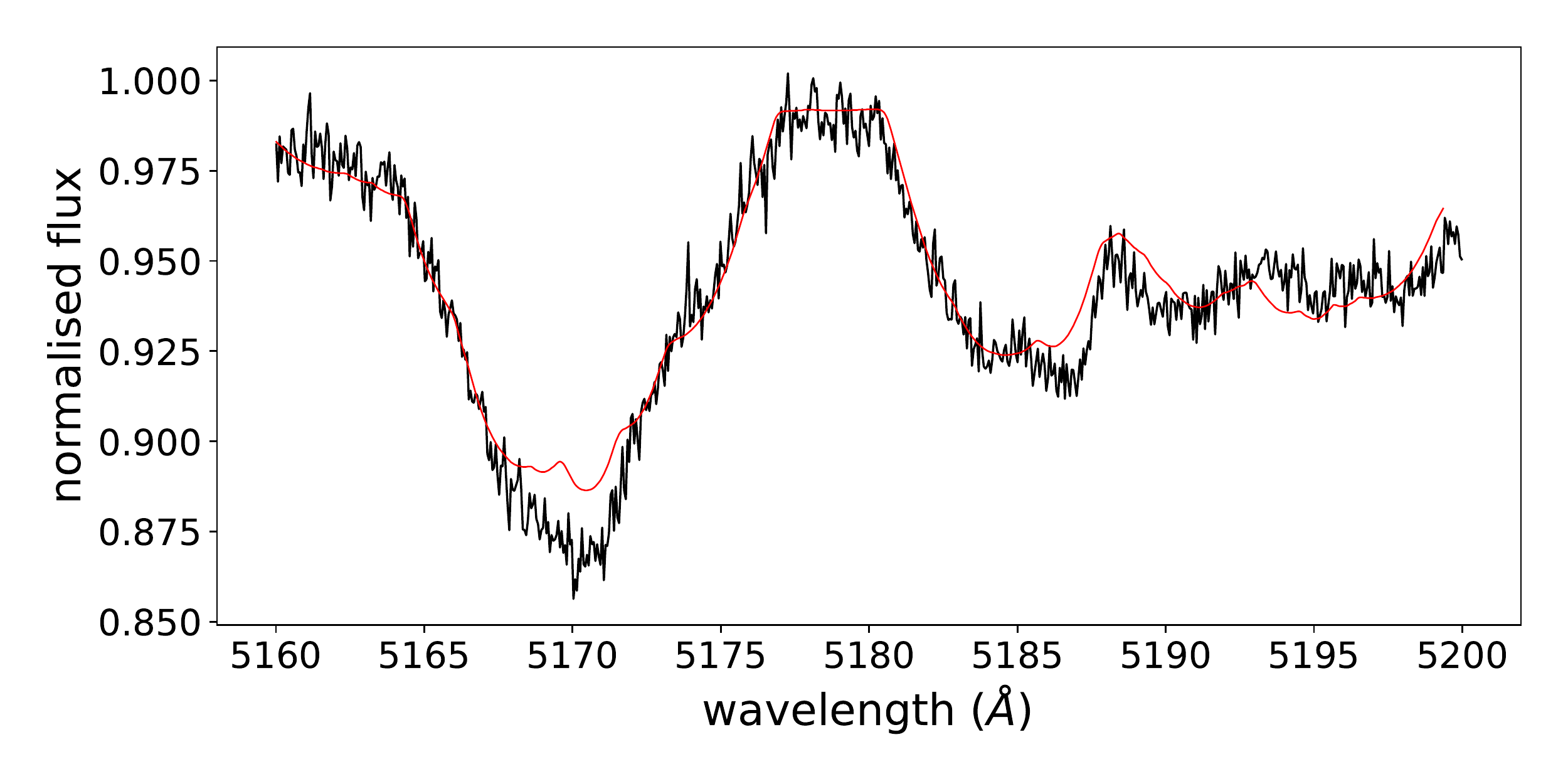}
      \caption{Observed (black line) and synthetic (red line) spectrum of Cl$^{\star}$ NGC 2264 VAS 230 in the region around the \ion{Mg}{i}b triplet. The synthetic spectrum was calculated with $T_{\rm eff}$ = 7520\,K, log(g) of 4.3 and a $v$\,sin\,$i$ of 205\,km\,s$^{-1}$. 
              }
         \label{fig:spectrum_VAS230}
\end{figure}

\textit{\underline{Cl$^{\star}$ NGC 2264 VAS 230}} (CoID 0223999101): The UVES spectrum retrieved from the ESO archive shows that Cl$^{\star}$ NGC 2264 VAS 230 has an extremely high $v$\,sin\,$i$ value.
This makes it challenging to determine reliable values for the star's fundamental parameters. Our best fitting spectrum synthesis model yields a $T_{\rm eff}$ of 7520 $\pm$ 150\,K, a log(g) of 4.3 $\pm$ 0.4, and a $v$\,sin\,$i$ of 205 $\pm$ 25\,km\,s$^{-1}$ with slightly less than solar abundance. The observed and synthetic spectra are shown in Fig. \ref{fig:spectrum_VAS230}.

\section{MESA microphysics}
\label{app:mesaphysics}
The MESA EOS is a blend of the OPAL \citet{Rogers2002}, SCVH
\citet{Saumon1995}, PTEH \citet{Pols1995}, HELM
\citet{Timmes2000}, and PC \citet{Potekhin2010} EOSes.

Radiative opacities are primarily from OPAL \citep{Iglesias1993,
Iglesias1996}, with low-temperature data from \citet{Ferguson2005}
and the high-temperature, Compton-scattering dominated regime by
\citet{Buchler1976}.  Electron conduction opacities are from
\citet{Cassisi2007}.

Nuclear reaction rates are a combination of rates from
NACRE \citep{Angulo1999}, JINA REACLIB \citep{Cyburt2010}, plus
additional tabulated weak reaction rates \citet{Fuller1985, Oda1994,
Langanke2000}. Screening
is included via the prescription of \citet{Chugunov2007}.  Thermal
neutrino loss rates are from \citet{Itoh1996}.

\section{Instability regions}
\begin{figure}
   \centering
   \includegraphics[width=\linewidth]{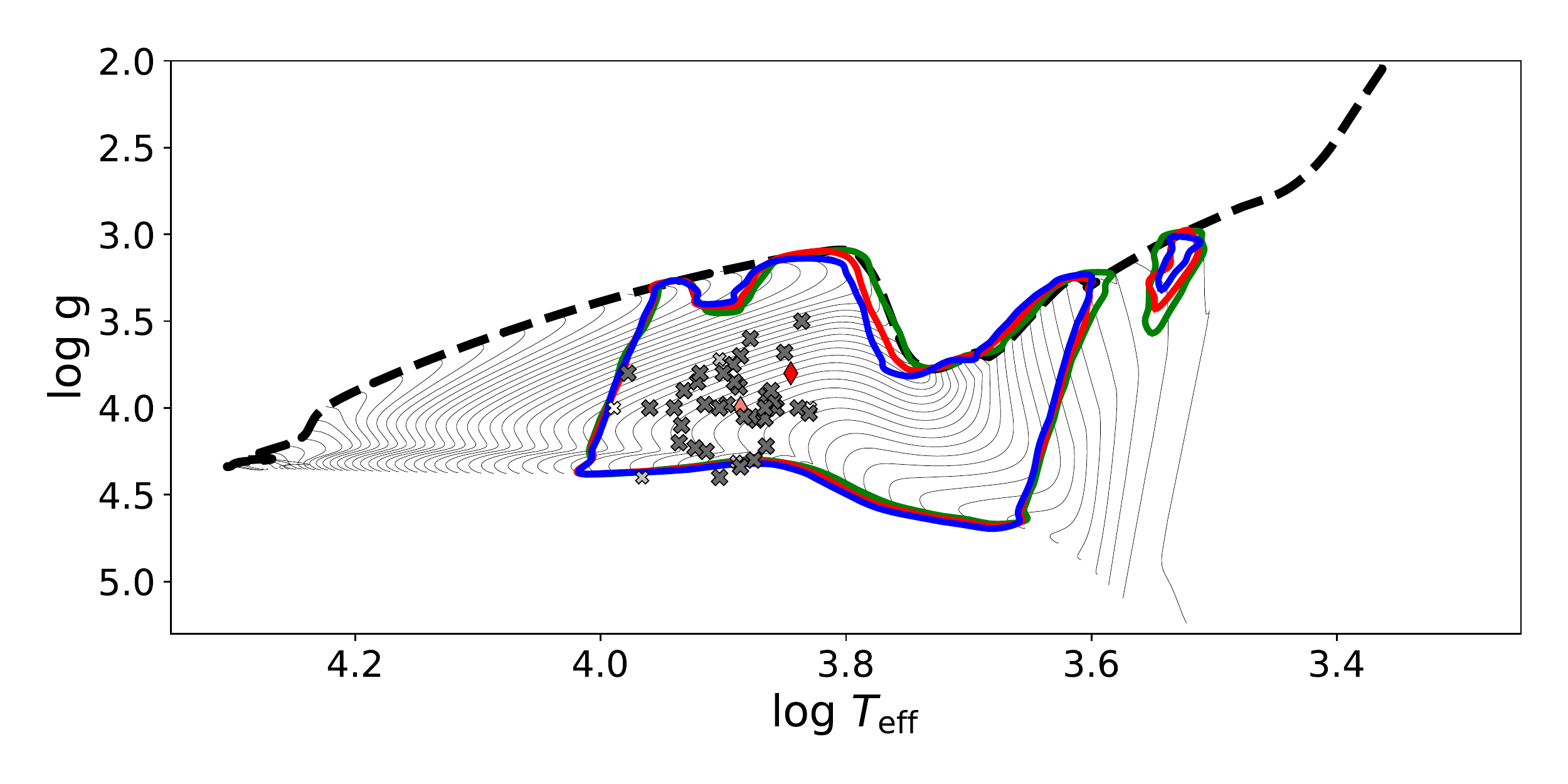}
      \caption{P-mode instability regions for different mixing lengths. The green, red, and blue lines show the boundary for $\alpha_{\rm MLT} = 1.8, 2.0$, and $2.2$, respectively.
              }
         \label{fig:mix_pmodes}
\end{figure}

\begin{figure}
   \centering
   \includegraphics[width=\linewidth]{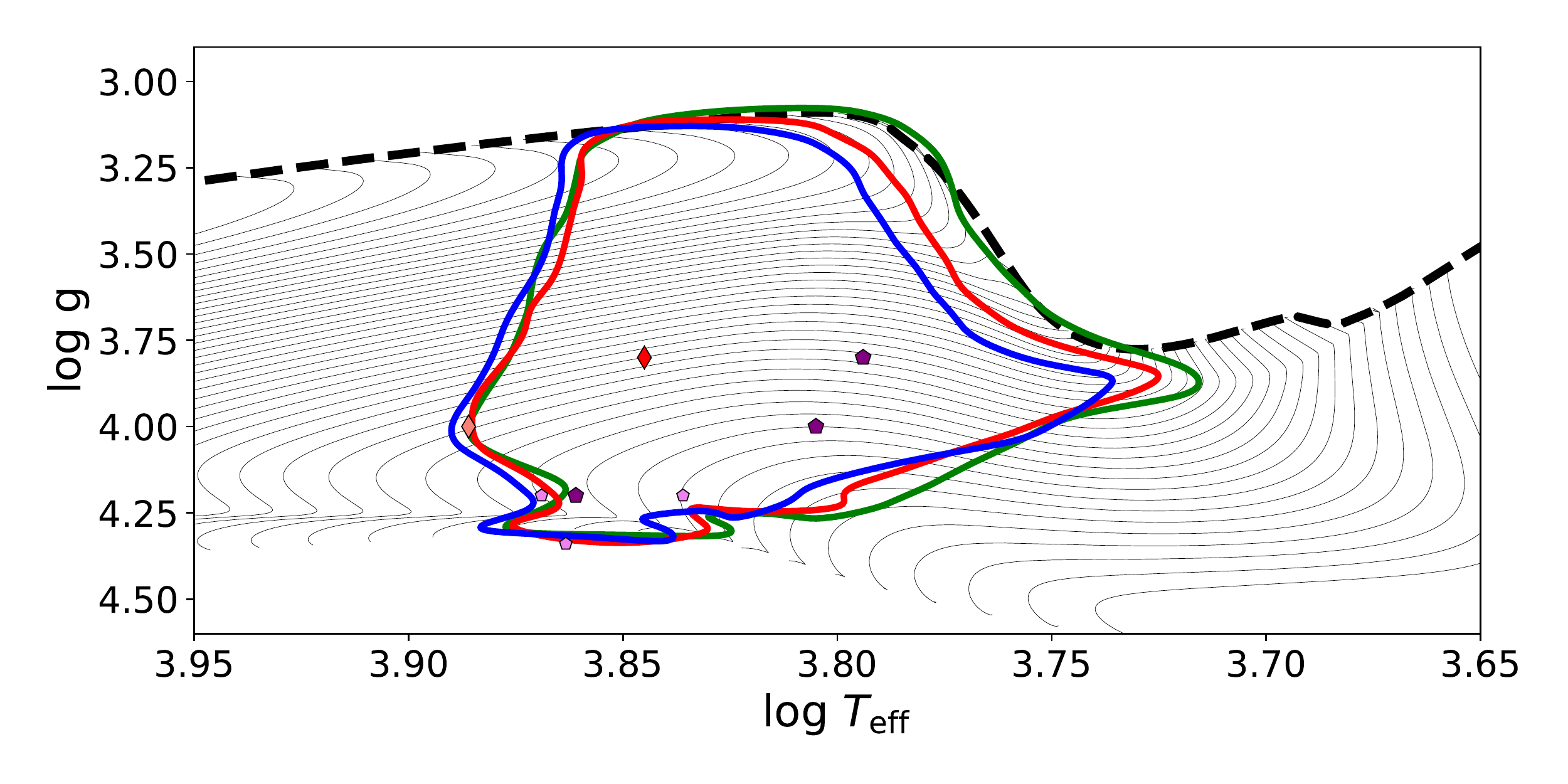}
      \caption{Same as Fig. \ref{fig:mix_pmodes} but for g-modes and the $\gamma$ Doradus instability region.
              }
         \label{fig:mix_gamma}
\end{figure}

\begin{figure*}
   \centering
   \includegraphics[width=\linewidth]{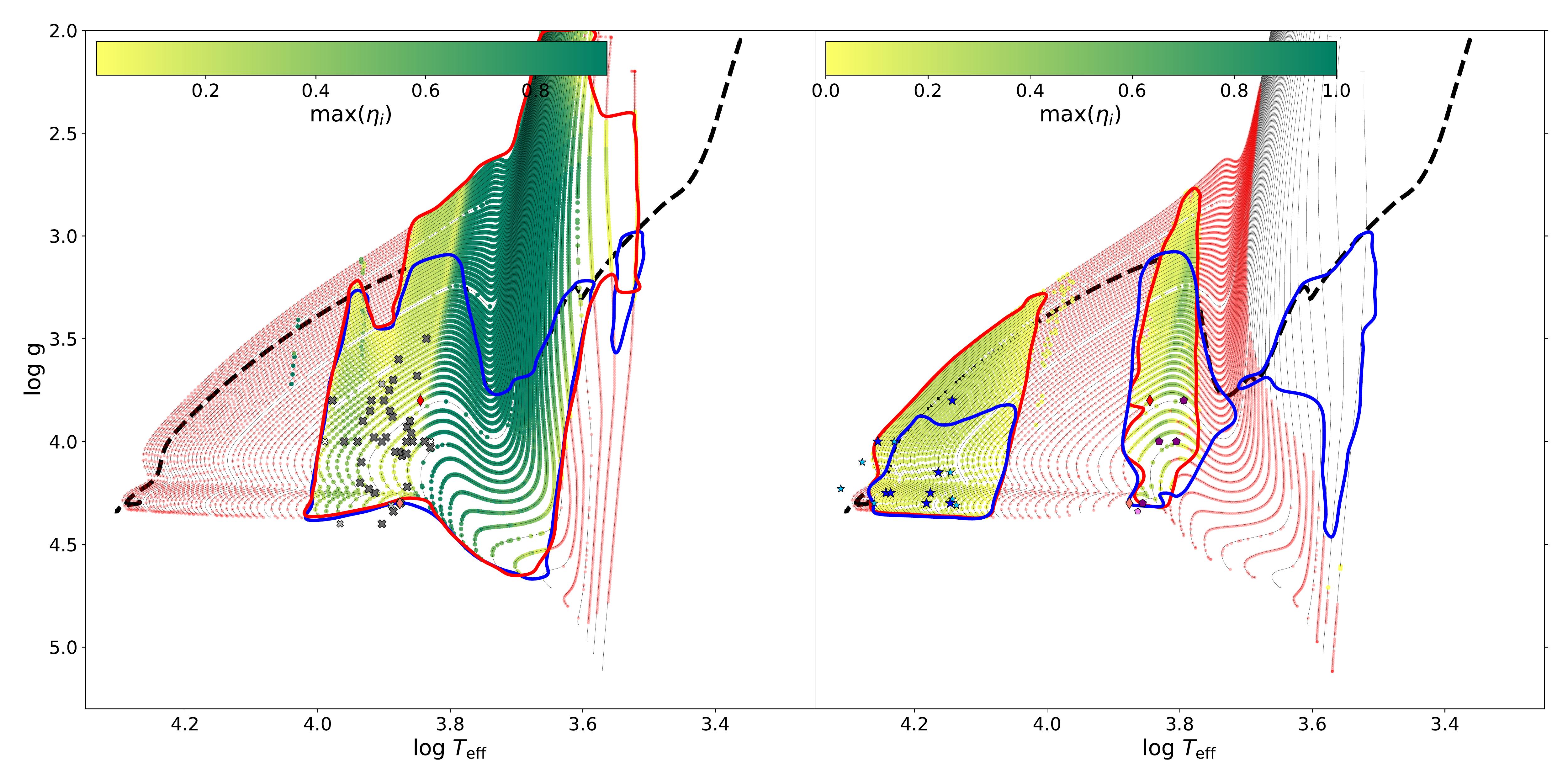}
      \caption{Comparision of the instability regions from accreting models versus classical models. The dashed line is the evolutionary track from the accreting model. Black lines show the evolutionary tracks of classical models and the colour code is the same as in Fig. \ref{fig:instab_main}. The red lines show the border of instability regions of the classical model while the blue lines show the same borders as in Fig. \ref{fig:instab_main}.}
         \label{fig:instab_compare_classic}
\end{figure*}

In the following we show instability regions resulting from our calculations. Figures \ref{fig:mix_pmodes} and \ref{fig:mix_gamma} show the influence of the mixing length on the instability regions of p-modes and g-modes, respectively. Figure \ref{fig:instab_compare_classic} shows the comparison between classical models and the models from accreting protostars.

\end{document}